\documentclass[12pt]{article}

\usepackage{axodraw}
\usepackage{epsfig}

\makeatletter

\long\def\@caption#1[#2]#3{%
  \par
  \addcontentsline{\csname ext@#1\endcsname}{#1}%
    {\protect\numberline{\csname the#1\endcsname}{\ignorespaces #2}}%
  \begingroup
    \@parboxrestore
    \if@minipage
      \@setminipage
    \fi
    \normalsize
    \@makecaption{\csname fnum@#1\endcsname}{\it \ignorespaces #3}\par
  \endgroup}

\def\paragraph{\@startsection{paragraph}{4}{\z@}{+2.00ex plus
 +1ex minus +.2ex}{1.5ex plus .2ex}{\it\normalsize}}

\def\section{\@startsection {section}{1}{\z@}{+3.0ex plus +1ex minus
  +.2ex}{2.3ex plus .2ex}{\normalsize\bf\boldmath}}
\def\subsection{\@startsection{subsection}{2}{\z@}{+2.5ex plus +1ex
minus +.2ex}{1.5ex plus .2ex}{\normalsize\bf\boldmath}}
\def\subsubsection{\@startsection{subsubsection}{3}{\z@}{+3.25ex plus
 +1ex minus +.2ex}{1.5ex plus .2ex}{\normalsize\bf\boldmath}}

\expandafter\ifx\csname mathrm\endcsname\relax\def\mathrm#1{{\rm #1}}\fi

\newcounter{saveeqn}

\@addtoreset{equation}{section}

\newcount\@tempcntc
\def\@citex[#1]#2{\if@filesw\immediate\write\@auxout{\string\citation{#2}}\fi
  \@tempcnta\z@\@tempcntb\m@ne\def\@citea{}\@cite{\@for\@citeb:=#2\do
    {\@ifundefined
       {b@\@citeb}{\@citeo\@tempcntb\m@ne\@citea
        \def\@citea{,\penalty\@m\ }{\bf ?}\@warning
       {Citation `\@citeb' on page \thepage \space undefined}}%
    {\setbox\z@\hbox{\global\@tempcntc0\csname
b@\@citeb\endcsname\relax}%
     \ifnum\@tempcntc=\z@ \@citeo\@tempcntb\m@ne
       \@citea\def\@citea{,\penalty\@m}
       \hbox{\csname b@\@citeb\endcsname}%
     \else
      \advance\@tempcntb\@ne
      \ifnum\@tempcntb=\@tempcntc
      \else\advance\@tempcntb\m@ne\@citeo
      \@tempcnta\@tempcntc\@tempcntb\@tempcntc\fi\fi}}\@citeo}{#1}}

\def\@citeo{\ifnum\@tempcnta>\@tempcntb\else\@citea
  \def\@citea{,\penalty\@m}%
  \ifnum\@tempcnta=\@tempcntb\the\@tempcnta\else
   {\advance\@tempcnta\@ne\ifnum\@tempcnta=\@tempcntb \else
\def\@citea{--}\fi
    \advance\@tempcnta\m@ne\the\@tempcnta\@citea\the\@tempcntb}\fi\fi}

\newcommand{\lsim}
{\mathrel{\raisebox{-.3em}{$\stackrel{\displaystyle <}{\sim}$}}}
\newcommand{\gsim}
{\mathrel{\raisebox{-.3em}{$\stackrel{\displaystyle >}{\sim}$}}}
\def\asymp#1%
{\mathrel{\raisebox{-.4em}{$\widetilde{\scriptstyle #1}$}}}

\def\Nequal#1%
{\mathrel{\raisebox{-.5em}{$\stackrel{=}{\scriptstyle\rm#1}$}}}
\newcommand{\dsl}[1]{\not \hspace{-0.7mm}#1}
\def\dsl{\mathpalette\make@slash}
\def\make@slash#1#2{\setbox\z@\hbox{$#1#2$}%
  \hbox to 0pt{\hss$#1/$\hss\kern-\wd0}\box0}

\def\beq{\begin{equation}}
\def\eeq{\end{equation}}
\def\beqar{\begin{eqnarray}}
\def\eeqar{\end{eqnarray}}
\def\barr#1{\begin{array}{#1}}
\def\earr{\end{array}}
\def\bfi{\begin{figure}}
\def\efi{\end{figure}}
\def\btab{\begin{table}}
\def\etab{\end{table}}
\def\bce{\begin{center}}
\def\ece{\end{center}}
\def\nn{\nonumber}

\def\text{\textstyle}

\def\al{\alpha}
\def\be{\beta}

\def\ga{\gamma}
\def\de{\delta}
\def\De{\Delta}
\def\eps{\epsilon}
\def\veps{\varepsilon}

\def\la{\lambda}

\def\si{\sigma}

\def\refeq#1{\mbox{(\ref{#1})}}
\def\refeqs#1{\mbox{(\ref{#1})}}

\def\reffi#1{\mbox{Fig.~\ref{#1}}}
\def\reffis#1{\mbox{Figs.~\ref{#1}}}
\def\refta#1{\mbox{Table~\ref{#1}}}

\def\refse#1{\mbox{Section~\ref{#1}}}

\def\refpar#1{\mbox{\it\ref{#1}}}
\def\refapp#1{\mbox{Appendix~\ref{#1}}}
\def\citere#1{\mbox{Ref.~\cite{#1}}}
\def\citeres#1{\mbox{Refs.~\cite{#1}}}

\newcommand{\GeV}{\unskip\,\mathrm{GeV}}
\newcommand{\MeV}{\unskip\,\mathrm{MeV}}

\newcommand{\fb}{\unskip\,\mathrm{fb}}

\newcommand{\ri}{{\mathrm{i}}}
\newcommand{\rd}{{\mathrm{d}}}

\newcommand{\M}{{\cal{M}}}

\def\mathswitchr#1{\relax\ifmmode{\mathrm{#1}}\else$\mathrm{#1}$\fi}
\def\mathswitch#1{\relax\ifmmode#1\else$#1$\fi}

\newcommand{\Pg}{\mathswitchr g}
\newcommand{\PH}{\mathswitch {H}}

\newcommand{\Pt}{\mathswitch {t}}

\newcommand{\MH}{\mathswitch {M_\PH}}

\newcommand{\Mt}{\mathswitch {m_\Pt}}

\def\ie{i.e.\ }
\def\eg{e.g.\ }

\newcommand{\onel}{{\mathrm{1-loop}}}

\newcommand{\cut}{{\mathrm{cut}}}

\newcommand{\soft}{{\mathrm{soft}}}
\newcommand{\coll}{{\mathrm{coll}}}

\newcommand{\Coul}{{\mathrm{Coul}}}

\newcommand{\UV}{\mathrm{UV}}
\newcommand{\IR}{\mathrm{IR}}
\newcommand{\LO}{\mathrm{LO}}
\newcommand{\NLO}{\mathrm{NLO}}
\newcommand{\ISR}{\mathrm{ISR}}
\newcommand{\FSR}{\mathrm{FSR}}
\newcommand{\ren}{\mathrm{ren}}
\newcommand{\MSbar}{\overline{\mathrm{MS}}}

\def\Li{\mathop{\mathrm{Li}_2}\nolimits}

\def\cLi{\mathop{{\cal L}i_2}\nolimits}
\def\Re{\mathop{\mathrm{Re}}\nolimits}

\renewcommand{\O}{{\cal O}}
\newcommand{\A}{{\cal A}}
\renewcommand{\C}{{\cal C}}

\def\draftdate{\relax}
\def\mda{\relax}
\def\mua{\relax}
\def\mla{\relax}
\def\draft{
\def\thtystars{******************************}
\def\sixtystars{\thtystars\thtystars}
\typeout{}
\typeout{\sixtystars**}
\typeout{* Draft mode!
         For final version remove \protect\draft\space in source file *}
\typeout{\sixtystars**}
\typeout{}
\def\draftdate{\today}
\def\mua{\marginpar[\boldmath\hfil$\uparrow$]%
                   {\boldmath$\uparrow$\hfil}%
                    \typeout{marginpar: $\uparrow$}\ignorespaces}
\def\mda{\marginpar[\boldmath\hfil$\downarrow$]%
                   {\boldmath$\downarrow$\hfil}%
                    \typeout{marginpar: $\downarrow$}\ignorespaces}
\def\mla{\marginpar[\boldmath\hfil$\rightarrow$]%
                   {\boldmath$\leftarrow $\hfil}%
                    \typeout{marginpar: $\leftrightarrow$}\ignorespaces}
\def\Mua{\marginpar[\boldmath\hfil$\Uparrow$]%
                   {\boldmath$\Uparrow$\hfil}%
                    \typeout{marginpar: $\uparrow$}\ignorespaces}
\def\Mda{\marginpar[\boldmath\hfil$\Downarrow$]%
                   {\boldmath$\Downarrow$\hfil}%
                    \typeout{marginpar: $\downarrow$}\ignorespaces}
\def\Mla{\marginpar[\boldmath\hfil$\Rightarrow$]%
                   {\boldmath$\Leftarrow $\hfil}%
                    \typeout{marginpar: $\leftrightarrow$}\ignorespaces}
\overfullrule 5pt
\oddsidemargin -15mm
\marginparwidth 29mm
}

\def\stars{\strut\leaders\hbox{*}\hfill\strut}
\def\starline{\hfil\strut\hfil\hbox to \textwidth {\stars}\hfil}

\catcode`\@=11
\ifx\@Hxfloat\@Hundef\else\expandafter\endinput\fi
\let\@Hxfloat\@xfloat
\def\@xfloat#1[{\@ifnextchar{H}{\@HHfloat{#1}[}{\@Hxfloat{#1}[}}
\def\@HHfloat#1[H]{%
\expandafter\let\csname end#1\endcsname\end@Hfloat
\vskip\intextsep\vbox\bgroup\def\@captype{#1}\parindent\z@
\ignorespaces}
\def\end@Hfloat{\egroup\vskip \intextsep}
\relax

\begin{document}
\thispagestyle{empty}
\def\thefootnote{\fnsymbol{footnote}}
\setcounter{footnote}{1}
\null
\draftdate\hfill DESY 02-177 \\
\strut\hfill Edinburgh 2002/18 \\
\strut\hfill MPI-PhT/2002-70\\
\strut\hfill PSI-PR-02-22\\
\strut\hfill hep-ph/0211352

\vfill
\begin{center}

{\Large \bf  NLO QCD corrections to \boldmath{$t\bar tH$} production 
\\[.5em]
in hadron collisions}%
\footnote{This work has been supported in part by the Swiss Bundesamt f\"ur
          Bildung und Wissenschaft and by the European Union under
          contract HPRN-CT-2000-00149.}

\vspace*{8mm}

{\sc W.~Beenakker$^1$, S.~Dittmaier$^{2,3}$
M.~Kr\"amer$^4$, B.~Pl\"umper$^2$, \\[.3em]
M.~Spira$^5$ and P.M.~Zerwas$^2$}

\vspace*{8mm}

{\normalsize \it
$^1$ Theoretical Physics, University of Nijmegen, 
NL-6500 GL Nijmegen, The Netherlands\\[.2cm]
$^2$ Deutsches Elektronen-Synchrotron DESY, D-22603 Hamburg, Germany \\[.2cm]
$^3$ Max-Planck-Institut f\"ur Physik (Werner-Heisenberg-Institut), 
D-80805 M\"unchen, Germany\\[.2cm]
$^4$ School of Physics, The University~of~Edinburgh,
Edinburgh EH9 3JZ, Scotland \\[.2cm]
$^5$ Paul Scherrer Institut PSI, CH-5232 Villigen PSI, Switzerland}
\par
\end{center}
\vskip 5mm
\begin{center}
\bf Abstract
\end{center} 
The Higgs boson $\,H\,$ of the Standard Model can be searched for in
the channels $\,p\bar p/pp\to t\bar tH+X\,$ at the Tevatron and the
LHC. The cross sections for these processes and the final-state
distributions of the Higgs boson and top quarks are presented at
next-to-leading order QCD. To calculate these QCD corrections, a
special calculational technique for pentagon diagrams has been
developed and the dipole subtraction formalism has been adopted for
massive particles. The impact of the corrections on the total cross
sections is characterized by $K$ factors, the ratios of the cross
sections in next-to-leading order over leading order QCD. At the
central scale $\,\mu_0=(2\Mt+\MH)/2\,$ the $K$ factors are found to be
slightly below unity for the Tevatron ($K\sim 0.8$) and slightly above
unity for the LHC ($K\sim 1.2$). Including the corrections
significantly stabilizes the theoretical predictions for total cross
sections and for the distributions in rapidity and transverse momentum
of the Higgs boson and top quarks.

\par
\vskip 1cm
\vfill
\noindent
November 2002
\null
\setcounter{page}{0}

\clearpage

\def\thefootnote{\arabic{footnote}}
\setcounter{footnote}{0}

\section{Introduction}
\label{se:intro}
  
Even though the Standard Model (SM) has been tremendously successful
in describing matter and forces in particle physics, the third
component of the model, the Higgs mechanism \cite{Higgs:1964ia}, has
not been established yet experimentally. This component, however, is
crucial for the closure of the SM in a mathematically consistent form,
thereby allowing for systematic perturbative expansions and precise
predictions in the electroweak sector \cite{'tHooft:rn}.  The search
for Higgs bosons \cite{Carena:2000yx,atlas_cms_tdrs} is therefore one
of the most important experimental programs in present-day high-energy
physics \cite{Carena:2002rm}.

For a light Higgs boson the electroweak interactions can be extended
perturbatively up to the grand unification scale. The validity of this
extrapolation is backed by the fact that it provides a qualitatively
correct estimate of the electroweak mixing angle. Demanding,
therefore, the perturbative extrapolation to hold up to the grand
unification scale, the mass of the Higgs boson is bounded to be below
180 GeV in the SM~\cite{Cabibbo:1979ay}. An upper limit in the same
range is also obtained experimentally by evaluating radiative
corrections to electroweak precision observables, \ie $\MH \le 196$
GeV at the 95\% confidence level~\cite{Abbaneo:2001ix}. A lower
experimental mass limit of 114.4 GeV has been set by the LEP
experiments~\cite{:2001xw}.

In the near future, the search for Higgs bosons will continue at
hadron colliders, the proton--antiproton collider
Tevatron~\cite{Carena:2000yx} with a centre-of-mass energy of 2 TeV,
followed by the proton--proton collider LHC~\cite{atlas_cms_tdrs} with
a centre-of-mass energy of 14 TeV.  Various channels can be exploited
at hadron colliders to search for Higgs bosons in the intermediate
mass range. Among these channels, Higgs-boson radiation off heavy top
quarks~\cite{Kunszt:1984ri,Beenakker:2001rj,Reina:2001sf} plays an
important r\^ole:
\beq
\label{ttHreactions}
p\bar p/pp\to t\bar tH+X.  
\eeq 
Although the expected rate is low at the Tevatron, a sample of a few
but very clean events could be observed for Higgs masses below
140~GeV~\cite{Goldstein:2000bp}. At the LHC, Higgs-boson radiation off
the top quarks is an important search channel for Higgs masses below
$\sim 125$~GeV~\cite{Richter-Was:sa,Drollinger:2001ym}.  Moreover,
analyzing the $t\bar tH$ production rate at the LHC can provide
valuable information on the top--Higgs Yukawa coupling in relation to
other Yukawa couplings~\cite{Maltoni:2002jr,Belyaev:2002ua}. Precision
measurements of the absolute value of the $t\bar tH$ coupling can be
completed later at $e^+e^-$ colliders~\cite{Djouadi:1992tk,%
Dittmaier:1998dz,Dawson:1999ej,Abe:2001np,Juste:1999af,Baer:1999ge}.

Theoretical predictions for cross sections that are based merely on
leading-order (LO) QCD are notoriously imprecise. They are plagued by
considerable uncertainties owing to the strong dependence on the
renormalization and factorization scales, introduced by the QCD
coupling and the parton densities.  Therefore, higher-order QCD
corrections are needed for satisfactory theoretical predictions.

The cross sections for the processes (\ref{ttHreactions}) are found in
next-to-leading order (NLO) QCD by convoluting the cross sections
$d\sigma^{ab}$ of the parton subprocesses $ab \rightarrow t\bar t H
(+k)$ with the parton distributions $f_a$ and $f_b$ of the
initial-state hadrons, both evaluated in next-to-leading order:
\begin{equation}
\label{eq:hadxs}
{\rm d}\sigma(h_1h_2\to t\bar tH+X)= \!\!\sum_{a,b=q,\bar{q},g} 
\int {\rm d}\xi_a \,
f_a^{h_1}(\xi_a; \mu_{\rm F}^2) \int {\rm d}\xi_b \, 
f_b^{h_2}(\xi_b;\mu_{\rm F}^2) \;
{\rm d}\sigma^{ab}(\hat{s}=\xi_a \xi_b s; \mu_{\rm F}^2, \mu_{\rm R}^2).
\end{equation}
In this convolution the partons $a$ and $b$ carry fractions $\xi_a$
and $\xi_b$ of the original momenta of the hadrons $h_1$ (=\,$p$) and
$h_2$ (=\,$\bar p/p$), respectively. The total hadronic centre-of-mass
energy is denoted by $\sqrt{s}$, whereas the partonic subprocess
energy is given by $\sqrt{\hat{s}}$.  The factorization and
renormalization scales are denoted by $\mu_F$ and $\mu_R$,
respectively. In general these two scales will be identified.
 
A brief report on the radiative QCD corrections for the processes
(\ref{ttHreactions}) in NLO QCD has been presented for both the
Tevatron and the LHC in \citere{Beenakker:2001rj}. The results of that
study were found to be in agreement with a parallel investigation for
the Tevatron in \citere{Reina:2001sf}, which was based exclusively on
the dominant partonic process $q\bar q \to t\bar tH$. As expected, in
complete calculations of the NLO QCD corrections the scale dependence
is reduced significantly, and stable theoretical predictions can be
derived for the cross sections.

At the technical level there are two main obstacles in calculating the
NLO QCD corrections. First, the one-loop pentagon (5-point) diagrams,
involving both soft/collinear singularities and massive particles,
have to be calculated in dimensional regularization. To this end, we
have developed a calculational technique, exploiting the fact that the
singularity structure of the pentagon diagrams can be derived in a
universal way. This enables us to switch to regularization schemes
that are more suitable for analyzing pentagon diagrams, without losing
regularization-scheme-dependent constant terms. Second, the extraction
of the singularities in the real part of the NLO QCD corrections has
to be performed in a numerically stable way. For this purpose we have
adopted the dipole subtraction formalism \cite{Catani:1996jh} for
massive particles \cite{Catani:2002hc}. [This is the first application
of this particular method to a complex NLO QCD calculation with
massive quarks.] As an additional cross-check, the partial results for
$q\bar{q}\to t\bar{t}H$ have been recalculated by means of the
phase-space slicing method.

Besides the total cross sections, final-state distributions of the
Higgs particle and the top quarks in rapidity $y$ and transverse
momentum $p_T$ will be presented. As in the case of the total cross sections,
including the NLO corrections significantly reduces the scale
dependence of these distributions. It should be noted that based on
this complete calculation, the NLO QCD corrections to arbitrary
distributions can be investigated.

The report is divided into four sections. In the next section we
describe the calculation of the NLO QCD corrections at the partonic
level.  In the third section the hadronic results for the SM reactions
$p\bar p/pp\to t\bar tH$ are presented, comprising the total hadronic
cross sections as well as the distributions in rapidity and transverse
momentum. The analytical and numerical results of our study are
summarized in the last section.  Finally, the appendices provide some
useful formulas, such as scalar one-loop integrals, results on colour
algebra, and splitting functions.

\section{Calculation of the NLO corrections}

\subsection{LO cross sections and conventions}
\label{se:LOconv}

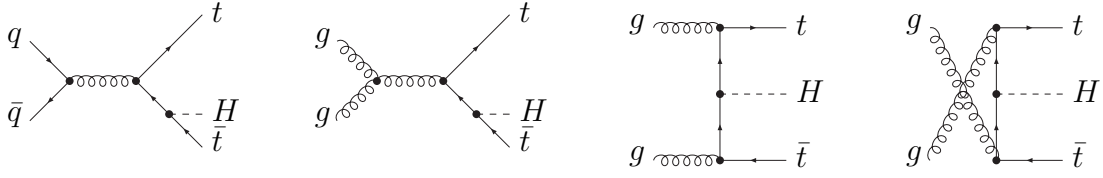
\begin{figure}
\SetScale{0.5}
\noindent
{\unitlength 0.5pt 
\begin{picture}(170,120)(-50,-10)
\ArrowLine(20, 80)(50,50)
\ArrowLine(50, 50)(20,20)
\Gluon(50,50)(100,50){4}{5}
\Vertex(50,50){3}
\Vertex(100,50){3}
\Vertex(125,25){3}
\DashLine(125, 25)(150,25){5}
\ArrowLine(100,50)(150,100)
\ArrowLine(125,25)(100, 50)
\ArrowLine(150, 0)(125, 25)
\put( 3,80){$q$}
\put( 3,20){$\bar q$}
\put(158, 95){$t$}
\put(158,20){$H$}
\put(158,-3){$\bar{t}$}
\end{picture}
\hspace*{2em}
\begin{picture}(170,120)(-50,-10)
\Gluon(20, 80)(50,50){4}{4}
\Gluon(20,20)(50, 50){4}{4}
\Gluon(50,50)(100,50){4}{5}
\Vertex(50,50){3}
\Vertex(100,50){3}
\Vertex(125,25){3}
\DashLine(125, 25)(150,25){5}
\ArrowLine(100,50)(150,100)
\ArrowLine(125,25)(100, 50)
\ArrowLine(150, 0)(125, 25)
\put( 3,80){$g$}
\put( 3,20){$g$}
\put(158, 95){$t$}
\put(158,20){$H$}
\put(158,-3){$\bar{t}$}
\end{picture}
\hspace*{1em}
\begin{picture}(170,120)(-50,0)
\Gluon(50,100)(100,100){4}{5}
\Gluon(50,  0)(100,  0){4}{5}
\Vertex(100,100){3}
\Vertex(100,  0){3}
\Vertex(100, 50){3}
\DashLine(100, 50)(150,50){5}
\ArrowLine(100,100)(150,100)
\ArrowLine(100, 50)(100,100)
\ArrowLine(100,  0)(100, 50)
\ArrowLine(150,  0)(100,  0)
\put(158, 95){$t$}
\put(158,44){$H$}
\put(158,-5){$\bar{t}$}
\put(32,100){$g$}
\put(32,  0){$g$}
\end{picture}
\hspace*{1em}
\begin{picture}(170,120)(-50,0)
\Gluon(50,100)(100,  0){4}{11}
\Gluon(50,  0)(100,100){4}{11}
\Vertex(100,100){3}
\Vertex(100,  0){3}
\Vertex(100, 50){3}
\DashLine(100, 50)(150,50){5}
\ArrowLine(100,100)(150,100)
\ArrowLine(100, 50)(100,100)
\ArrowLine(100,  0)(100, 50)
\ArrowLine(150,  0)(100,  0)
\put(158, 95){$t$}
\put(158,44){$H$}
\put(158,-5){$\bar{t}$}
\put(32,100){$g$}
\put(32,  0){$g$}
\end{picture}
}
\caption{The basic LO diagrams for the partonic processes $q\bar q, 
    gg\to t\bar tH$.}
\label{fig:LOdiags}
\end{figure}

The LO hadronic processes 
$p\bar p/pp\to t\bar tH+X$ proceed at the
parton level via $q\bar q$ annihilation and $gg$ fusion:
\beqar
q(p_1)+\bar q(p_2) &\to& t(p_3)+\bar t(p_4)+H(p_5), \\
g(p_1)+g(p_2) &\to& t(p_3)+\bar t(p_4)+H(p_5), 
\eeqar 
where the momenta of the particles are given in brackets.  The momenta
obey the on-shell conditions $p_1^2=p_2^2=0$, $p_3^2=p_4^2=\Mt^2$, and
$p_5^2=\MH^2$. For later use, the following set of kinematical
invariants is defined:
\beqar
\hat{s} &=& (p_1+p_2)^2, \nn\\
s_{ij} &=& (p_i+p_j)^2, \qquad i,j=3,4,5, \nn\\
t_{ij} &=& (p_i-p_j)^2, \qquad i=1,2, \quad j=3,4,5.  
\eeqar 
A generic set of LO diagrams is shown in \reffi{fig:LOdiags}; all LO
diagrams not shown differ only in the point where the Higgs line is
attached to the top-quark line.

We express the amplitudes $\M^{q\bar q,gg}$ of the LO and one-loop
diagrams for the $q\bar q$ annihilation and the $gg$ fusion channel in
terms of colour factors $\C^{q\bar q,gg}_j$, standard matrix elements
(SME) $\M^{q\bar q,gg}_i$, and invariant functions $F^{q\bar
  q,gg}_{ij}$.  The SME include the Dirac structure, spinors, and
polarization vectors in a standard form, while the $F^{q\bar
  q,gg}_{ij}$ comprise all remaining factors, such as propagators,
couplings, and standard one-loop integrals.

To introduce a compact notation for the SME, the tensors
\beqar
\Gamma^{q\bar q}_{\{1,\mu,\mu\nu,\dots\}} &=&
\bar v_{\bar q}(p_2)\left\{{\bf 1},\ga_\mu,\ga_\mu\ga_\nu,\dots\right\}
u_q(p_1),
\nn\\
\Gamma^{\Pt\bar\Pt}_{\{1,\mu,\mu\nu,\dots\}} &=&
\bar u_\Pt(p_3)\left\{{\bf 1},\ga_\mu,\ga_\mu\ga_\nu,\dots\right\}
v_{\bar\Pt}(p_4),
\eeqar
are defined with obvious notations for the Dirac spinors $u_q(p_1)$,
etc. Furthermore, symbols like $\Gamma_p$ are used as shorthand for
the contraction $\Gamma_\mu\, p^\mu$.  For the $q\bar q$ channel we
introduce the following set of SME,
\beqar
\M^{q\bar q}_{\{1,2,3,4\}} &=& \Gamma^{q\bar q,\mu} \;
\Gamma^{\Pt\bar\Pt}_{\{\mu,\mu p_1,\mu p_2,\mu p_1 p_2\}},
\nn\\
\M^{q\bar q}_{\{5,6,7,8\}} &=& \Gamma^{q\bar q,\mu p_3 p_4} \;
\Gamma^{\Pt\bar\Pt}_{\{\mu,\mu p_1,\mu p_2,\mu p_1 p_2\}},
\nn\\
\M^{q\bar q}_{\{9,10,11,12\}} &=& \Gamma^{q\bar q,p_3} \;
\Gamma^{\Pt\bar\Pt}_{\{1,p_1,p_2,p_1 p_2\}},
\nn\\
\M^{q\bar q}_{\{13,14,15,16\}} &=& \Gamma^{q\bar q,p_4} \;
\Gamma^{\Pt\bar\Pt}_{\{1,p_1,p_2,p_1 p_2\}},
\nn\\
\M^{q\bar q}_{\{17,18,19,20\}} &=& \Gamma^{q\bar q,\mu\nu p_3} \;
\Gamma^{\Pt\bar\Pt}_{\{\mu\nu,\mu\nu p_1,\mu\nu p_2,\mu\nu p_1 p_2\}},
\nn\\
\M^{q\bar q}_{\{21,22,23,24\}} &=& \Gamma^{q\bar q,\mu\nu p_4} \;
\Gamma^{\Pt\bar\Pt}_{\{\mu\nu,\mu\nu p_1,\mu\nu p_2,\mu\nu p_1 p_2\}},
\nn\\
\M^{q\bar q}_{\{25,26,27,28\}} &=& \Gamma^{q\bar q,\mu\nu\rho} \;
\Gamma^{\Pt\bar\Pt}_{\{\mu\nu\rho,\mu\nu\rho p_1,\mu\nu\rho p_2,
\mu\nu\rho p_1 p_2\}}.
\eeqar

The LO amplitude $\M^{q\bar q}_{\LO}$ of the $q\bar q$ channel is 
proportional to the colour factor
\beq
\C^{q\bar q} = \la^c \otimes \la^c,
\label{eq:Cqq}
\eeq
where the first Gell--Mann matrix $\la^c$ belongs to the $q\bar q$
state, the second matrix to the $t\bar t$ state. The one-loop amplitudes
$\M^{q\bar q}_{\onel}$ also involve other colour operators, such as
${\bf 1}\otimes{\bf 1}$ and $(\la^a\la^b)\otimes(\la^a\la^b)$, but
they can be reduced to multiples of $\C^{q\bar q}$ and tensors that are
orthogonal to $\C^{q\bar q}$ and therefore do not contribute in the 
interference 
$\sum_{\mathrm{colours}}(\M^{q\bar q}_{\LO})^*\M^{q\bar q}_{\onel}$. 
Thus, $\C^{q\bar q}$ is the only relevant colour structure at one loop, 
and the produced $t\bar t$ pair is in a pure colour-octet state in the 
$q\bar{q}$ channel.  Both the LO and the (relevant part of the) 
one-loop amplitude can be written as
\beq
\M^{q\bar q} = \C^{q\bar q} \A^{q\bar q}, \qquad
\A^{q\bar q} = \sum_{i=1}^{28} F^{q\bar q}_i \M^{q\bar q}_i,
\label{eq:Mqqdecomp}
\eeq
leading in particular to
\beq
\sum_{\mathrm{spins,\,colours}}|\M^{q\bar q}_{\LO}|^2 
= 
32 \sum_{\mathrm{spins}}|\A^{q\bar q}_{\LO}|^2.
\eeq

Explicitly, the LO amplitude can be cast in the form
\beqar
\M^{q\bar q}_{\LO} &=& {}-\C^{q\bar q} \, 
\frac{\pi\alpha_{\mathrm{s}} g_{ttH}}{\hat{s}}
\left[ \phantom{+} \frac{1}{s_{35}-\Mt^2}
        \left(\M^{q\bar q}_2+\M^{q\bar q}_3+2\M^{q\bar q}_{13}\right)
\right.
\nn\\
&& \left. \phantom{-\C^{q\bar q} \, 
        \frac{\pi\alpha_{\mathrm{s}} g_{ttH}}{\hat{s}}\Bigg\{} {}
      +\frac{1}{s_{45}-\Mt^2}
        \left(\M^{q\bar q}_2+\M^{q\bar q}_3-2\M^{q\bar q}_{9}\right)
\ \right],
\eeqar
from which the invariant functions $F^{q\bar q}_{\LO,i}$ can be read
off easily.  The factor $g_{ttH}=\Mt/v$ denotes the SM Yukawa
coupling strength, where $v$ is the vacuum-expectation value of the
Higgs field, and $\alpha_{\mathrm{s}} = g^2_{\mathrm{s}}/(4\pi)$ is the
QCD coupling constant.

To define the SME for the $gg$ channel, the polarization
vector $\veps^\mu_i$ is introduced for each gluon with momentum
$p^\mu_i$.  Apart from the transversality condition $\veps_i p_i=0$,
the gauge conditions
\beq
\veps_1 p_2=\veps_2 p_1=0
\label{eq:gauge}
\eeq
are used to simplify the algebra in the calculation of the amplitudes.
Within this gauge, the following set of SME is complete,
\beqar
\M^{gg}_1    &=& \text\frac{1}{4} \left[
 \Gamma^{\Pt\bar\Pt}_{p_1 p_2\veps_1\veps_2}
-\Gamma^{\Pt\bar\Pt}_{p_2 p_1\veps_1\veps_2}
-\Gamma^{\Pt\bar\Pt}_{p_1 p_2\veps_2\veps_1}
+\Gamma^{\Pt\bar\Pt}_{p_2 p_1\veps_2\veps_1} \right],
\nn\\
\M^{gg}_{\{2,3,4\}} &=& \Gamma^{\Pt\bar\Pt}_1 \times
\left\{ (\veps_1\veps_2), (\veps_1 p_3)(\veps_2 p_3),
(\veps_1 p_4)(\veps_2 p_4) \right\},
\nn\\
\M^{gg}_5    &=& \text\frac{1}{2} \left[
 \Gamma^{\Pt\bar\Pt}_{\veps_1\veps_2}
-\Gamma^{\Pt\bar\Pt}_{\veps_2\veps_1} \right],
\nn\\
\M^{gg}_{\{6,7,8\}} &=& \text\frac{1}{2} \left[
 \Gamma^{\Pt\bar\Pt}_{p_1 p_2}
-\Gamma^{\Pt\bar\Pt}_{p_2 p_1} \right] \times
\left\{ (\veps_1\veps_2), (\veps_1 p_3)(\veps_2 p_3),
(\veps_1 p_4)(\veps_2 p_4) \right\},
\nn\\
\M^{gg}_9    &=& \text\frac{1}{2} \left[
 \Gamma^{\Pt\bar\Pt}_{p_1\veps_1\veps_2}
-\Gamma^{\Pt\bar\Pt}_{p_1\veps_2\veps_1} \right],
\nn\\
\M^{gg}_{\{10,11\}} &=& \text\frac{1}{2} \left[
 \Gamma^{\Pt\bar\Pt}_{p_1 p_2\veps_1}
-\Gamma^{\Pt\bar\Pt}_{p_2 p_1\veps_1} \right] \times
\left\{ (\veps_2 p_3), (\veps_2 p_4) \right\},
\nn\\
\M^{gg}_{\{12,13\}} &=& \Gamma^{\Pt\bar\Pt}_{p_1\veps_1} \times
\left\{ (\veps_2 p_3), (\veps_2 p_4) \right\},
\nn\\
\M^{gg}_{\{14,15\}} &=& \Gamma^{\Pt\bar\Pt}_{p_2\veps_1} \times
\left\{ (\veps_2 p_3), (\veps_2 p_4) \right\},
\nn\\
\M^{gg}_{16} &=& \text\frac{1}{2} \left[
 \Gamma^{\Pt\bar\Pt}_{p_1 p_2}
-\Gamma^{\Pt\bar\Pt}_{p_2 p_1} \right] (\veps_1 p_3)(\veps_2 p_4),
\nn\\
\M^{gg}_{\{17,18\}} &=& \Gamma^{\Pt\bar\Pt}_{\veps_1} \times
\left\{ (\veps_2 p_3), (\veps_2 p_4) \right\},
\nn\\
\M^{gg}_{\{19,20,21,22,23\}} &=& \Gamma^{\Pt\bar\Pt}_{p_1} \times
\left\{ (\veps_1 \veps_2), (\veps_1 p_3)(\veps_2 p_3), 
(\veps_1 p_3)(\veps_2 p_4), (\veps_1 p_4)(\veps_2 p_3), 
(\veps_1 p_4)(\veps_2 p_4) \right\},
\nn\\
\M^{gg}_{24} &=& \Gamma^{\Pt\bar\Pt}_1
(\veps_1 p_3)(\veps_2 p_4), \nn\\[.5em]
\hat\M^{gg}_i &=& \M^{gg}_i\Big|_{p_1\leftrightarrow p_2,
                                \veps_1\leftrightarrow\veps_2},
\qquad i=1,\dots,24.
\eeqar
In order to exhibit the Bose symmetry, the SME have been divided into the
two subsets $\M^{gg}_i$ and $\hat\M^{gg}_i$ which transform into each
other by interchanging the incoming gluons. Obviously the first
eight SME of the two sets are Bose symmetric or antisymmetric,
\beqar
\hat\M^{gg}_i &=& {}+\M^{gg}_i, \qquad i=1,2,3,4, \nn\\
\hat\M^{gg}_i &=& {}-\M^{gg}_i, \qquad i=5,6,7,8.
\eeqar
Alternatively we may write
\beq
\M^{gg}_{24+i} = \hat\M^{gg}_{8+i} = 
\M^{gg}_{8+i}\Big|_{p_1\leftrightarrow p_2,
                    \veps_1\leftrightarrow\veps_2}, \qquad
i=1,\dots,16,
\eeq
for the independent crossed SME $\hat\M^{gg}_i$ with $i>8$.

For the $gg$ channel it is convenient to introduce the three colour
operators
\beq
\C^{gg}_1 = \de^{c_1c_2} {\bf 1}, \qquad
\C^{gg}_2 = \ri f^{c_1c_2c}\lambda^c, \qquad
\C^{gg}_3 = d^{c_1c_2c}\lambda^c,
\label{eq:Cgg}
\eeq
where $c_n$ is the colour index of gluon $n$, and the matrices ${\bf
  1}$ and $\la^c$ act on the $t\bar t$ state. The totally
antisymmetric and symmetric SU(3) structure constants $f^{abc}$ and
$d^{abc}$ are defined in the usual form.  Parts in amplitudes that are
proportional to different $\C^{gg}_j$ do not interfere with each other
owing to the orthogonality relations:
\beq
{\mathrm{Tr}}\left( \C^{gg\dagger}_j \C^{gg}_k \right) = 
c^{gg}_j \de_{jk} \qquad \mbox{with} \quad
c^{gg}_1=24, \quad 
c^{gg}_2=48, \quad 
c^{gg}_3=\frac{80}{3}.
\eeq
Obviously $\C^{gg}_1$ corresponds to a colour-singlet state of the
$t\bar t$ system, while $\C^{gg}_2$ and $\C^{gg}_3$ describe the two
different octet states. Both the LO and the one-loop amplitude can be 
written as
\beq
\M^{gg} = \sum_{j=1}^{3} \C^{gg}_j \A^{gg}_j, \qquad
\A^{gg}_j = \sum_{i=1}^{40} F^{gg}_{ij} \M^{gg}_i,
\label{eq:Mggdecomp}
\eeq
leading in particular to
\beq
\sum_{\mathrm{spins,\,colours}}|\M^{gg}_{\LO}|^2 
= \sum_{j=1}^{3} c^{gg}_j \sum_{\mathrm{spins}}|\A^{gg}_{\LO,j}|^2.
\eeq
Explicitly, the LO amplitude reads
\beqar
\M^{gg}_{\LO} &=& \phantom{{}+{}}
\left(\frac{2}{3}\C^{gg}_1+\C^{gg}_2+\C^{gg}_3\right) 
\left(M_{\mathrm{direct}}+\frac{1}{2}M_{\mathrm{fusion}}\right)
\nn\\ && {}
+\left(\frac{2}{3}\C^{gg}_1-\C^{gg}_2+\C^{gg}_3\right) 
\left(M_{\mathrm{crossed}}-\frac{1}{2}M_{\mathrm{fusion}}\right)
\label{eq:LOgg}
\eeqar
with
\beqar
M_{\mathrm{direct}} &=& 
\frac{\pi\alpha_{\mathrm{s}} g_{ttH}}{(s_{45}-\Mt^2)(t_{13}-\Mt^2)} 
\biggl[ \M^{gg}_1+\left(\frac{\hat{s}}{2}+t_{13}-\Mt^2\right)
                       (\M^{gg}_2+\M^{gg}_5)
\nn\\ && 
\hphantom{\frac{\pi\alpha_{\mathrm{s}} g_{ttH}}{(s_{45}-\Mt^2)(t_{13}-\Mt^2)} 
          \biggl[ } {}
      -4\M^{gg}_3+\M^{gg}_6-2\M^{gg}_{12}-2\hat\M^{gg}_{12} \biggr]
\nn\\ && 
{}+\frac{\pi\alpha_{\mathrm{s}} g_{ttH}}{(t_{13}-\Mt^2)(t_{24}-\Mt^2)}
\biggl[ \M^{gg}_1+\frac{\hat{s}}{2}(\M^{gg}_2+\M^{gg}_5)+\M^{gg}_6
        +2\M^{gg}_{13}
\nn\\ && 
\hphantom{{}+\frac{\pi\alpha_{\mathrm{s}}g_{ttH}}{(t_{13}-\Mt^2)(t_{24}-\Mt^2)}
          \biggl[} {} 
      +4\M^{gg}_{24}-2\hat\M^{gg}_{12} \biggr]
\nn\\ && 
{}+\frac{\pi\alpha_{\mathrm{s}} g_{ttH}}{(s_{35}-\Mt^2)(t_{24}-\Mt^2)}
\biggl[ \M^{gg}_1+\left(\frac{\hat{s}}{2}+t_{24}-\Mt^2\right)
                       (\M^{gg}_2+\M^{gg}_5)
\nn\\ &&
\hphantom{{}+\frac{\pi\alpha_{\mathrm{s}} g_{ttH}}{(s_{35}-\Mt^2)(t_{24}-\Mt^2)}
          \biggl[} {}
      -4\M^{gg}_4+\M^{gg}_6+2\M^{gg}_{13}+2\hat\M^{gg}_{13} \biggr],
\nn\\[.5em]
M_{\mathrm{crossed}} &=& M_{\mathrm{direct}}
\Big|_{t_{1i}\leftrightarrow t_{2i}, \,
       \M^{gg}_i\leftrightarrow \hat\M^{gg}_i} \ ,
\nn\\[.5em]
M_{\mathrm{fusion}} &=&
\frac{2\pi\alpha_{\mathrm{s}} g_{ttH}}{\hat{s}(s_{45}-\Mt^2)} 
\biggl[ (t_{13}-t_{23})\M^{gg}_2+2\M^{gg}_6 \biggr]
\nn\\ && {}
+\frac{2\pi\alpha_{\mathrm{s}} g_{ttH}}{\hat{s}(s_{35}-\Mt^2)} 
\biggl[ (t_{24}-t_{14})\M^{gg}_2+2\M^{gg}_6 \biggr]
\ =\ -\,M_{\mathrm{fusion}}\Big|_{t_{1i}\leftrightarrow t_{2i}, \,
        \M^{gg}_i\leftrightarrow \hat\M^{gg}_i}\ .\qquad\qquad
\label{eq:Mggauc}
\eeqar
The terms in Eq.~\refeq{eq:LOgg} proportional to $M_{\mathrm{direct}}$
correspond to the three ``direct graphs'', where two parallel gluons
couple to the top-quark line (see 3rd graph in \reffi{fig:LOdiags}).
The terms proportional to $M_{\mathrm{crossed}}$ result from the
direct graphs by crossing, \ie by interchanging the two gluons (see
4th graph in \reffi{fig:LOdiags}).  The terms proportional to
$M_{\mathrm{fusion}}$ correspond to the gluon-fusion graphs (see 2nd
graph in \reffi{fig:LOdiags}) which do not receive crossed
counterparts.

Some comments on the SME and their evaluation ought to be added.  The
$\M^{q\bar q,gg}_i$ defined above lead to a unique representation of
all LO and one-loop diagrams, as indicated in
Eqs.~\refeq{eq:Mqqdecomp} and \refeq{eq:Mggdecomp}, if only the Dirac
equation, the transversality of the polarization vectors, and the
gauge conditions \refeq{eq:gauge} are used. The number of SME could
actually be reduced further by exploiting discrete symmetries and by
using the four-dimensionality of space-time, inducing relations among
the SME given above. However, eliminating some of the SME does not
simplify the numerical evaluation significantly.  The explicit
calculation of the SME was carried out in two different ways.  First
all interference terms $\sum_{\mathrm{spins}}(\M^{q\bar
q,gg}_i)^*\M^{q\bar q,gg}_k$ were calculated by making use of the
general polarization sums for Dirac spinors and polarization
vectors. In a second independent approach we directly expressed the
$\M^{q\bar q,gg}_i$ in terms of Weyl--van der Waerden spinor products
\cite{Dittmaier:1999nn} for each helicity configuration.

The final step in the calculation of the partonic cross sections and 
distributions involves the 4-dimensional integration over the final-state 
three-particle phase space. By introducing an intermediate 
``$t\bar t$-state'' with virtuality $s_{34}$, one obtains
\beqar
  \label{eq:LOxsection}
  \rd \sigma_{\LO}^{ab} &=& 
         \frac{1}{2\hat{s}}\,k^{ab}_{\mathrm{av}}\,\frac{1}{(2\pi)^4}\,
         \frac{\kappa(s_{34},\Mt^2,\Mt^2)}{8\,s_{34}}\,
         \frac{\kappa(\hat{s},s_{34},\MH^2)}{8\,\hat{s}}\,
         \Theta(\sqrt{\hat{s}}-2\Mt-\MH)\,\Theta(s_{34}-4\Mt^2)\times
         \nn\\[1mm]
                   & & 
         \times\,\Theta([\sqrt{\hat{s}}-\MH]^2-s_{34})\!
         \sum_{\mathrm{spins,\,colours}}|\M^{ab}_{\LO}|^2 
         \,\,\rd s_{34}\,\,\rd\Omega_t^*\,\,\rd\cos(\theta_H^{\mathrm{CM}})\,,
\eeqar
with 
\beq
  \kappa(x,y,z) = \sqrt{x^2+y^2+z^2-2xy-2xz-2yz}\,.
\label{eq:kappa}
\eeq
The labels $ab$ indicate the specific initial state of the partonic process. 
The solid angle $\Omega_t^*$ of the top quark is defined in the rest frame of
the ``$t\bar t$-state'', using the direction of flight of the 
``$t\bar t$-state'' in the initial-state centre-of-mass frame as $z$-axis.
The angle $\theta_H^{\mathrm{CM}}$ denotes the polar angle of the Higgs 
boson in the initial-state centre-of-mass frame, using the direction of the 
initial-state beams as $z$-axis. The corresponding azimuthal angle is 
trivially integrated to $2\pi$ in view of rotational invariance about the 
initial-state beams. The factor $k^{ab}_{\mathrm{av}}$ results from the 
average over the initial-state spins and colours:
\beq
  k_{\mathrm{av}}^{q\bar q} = \frac{1}{36}, \qquad
  k_{\mathrm{av}}^{gg} = \frac{1}{256}, \qquad
  k_{\mathrm{av}}^{qg} = k_{\mathrm{av}}^{g\bar q} = \frac{1}{96}\,,
\eeq 
where $k_{\mathrm{av}}^{qg}$ and $k_{\mathrm{av}}^{g\bar q}$ are 
given for future use.
The differential cross section including virtual NLO corrections can be
obtained in a straightforward way from Eq.~\refeq{eq:LOxsection} by 
substituting the correct matrix element.

\subsection{Virtual corrections\label{subsec:virtual}}

\subsubsection{One-loop diagrams and calculational framework}

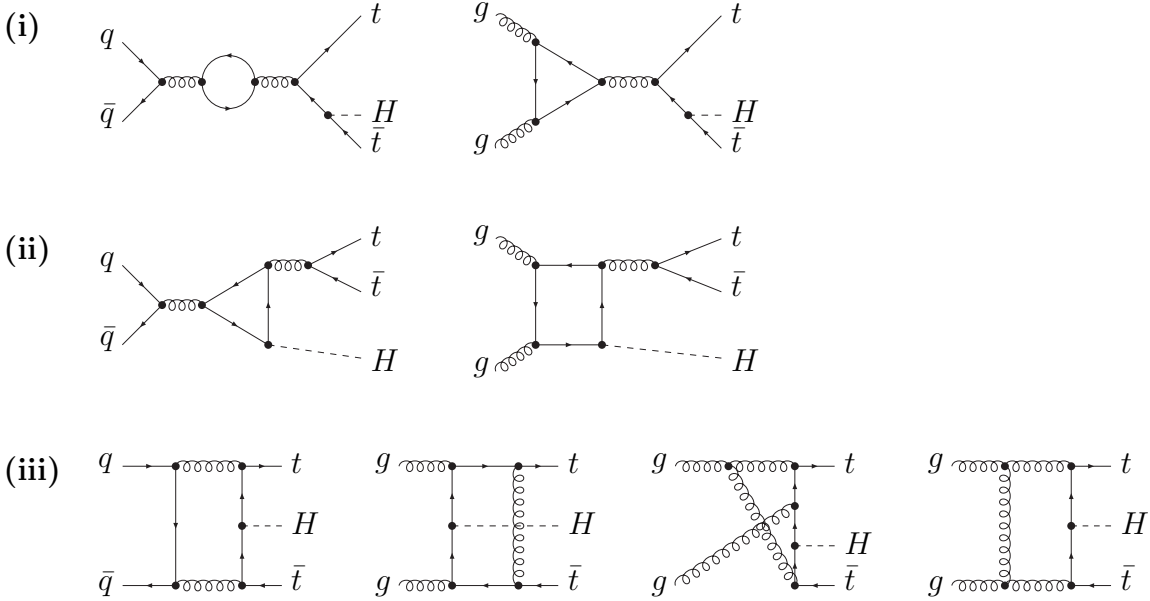
\begin{figure}
\SetScale{0.5}
{\unitlength 0.5pt 
\begin{picture}(220,120)(-80,-10)
\ArrowLine(20, 80)(50,50)
\ArrowLine(50, 50)(20,20)
\ArrowArc(100, 50)(20,0,180)
\ArrowArc(100, 50)(20,180,0)
\Gluon(50,50)(80,50){4}{3}
\Gluon(120,50)(150,50){4}{3}
\Vertex(50,50){3}
\Vertex(80,50){3}
\Vertex(120,50){3}
\Vertex(150,50){3}
\Vertex(175,25){3}
\DashLine(175, 25)(200,25){5}
\ArrowLine(150,50)(200,100)
\ArrowLine(175,25)(150, 50)
\ArrowLine(200, 0)(175, 25)
\put( 3,80){$q$}
\put( 3,20){$\bar q$}
\put(208, 95){$t$}
\put(208,20){$H$}
\put(208,-3){$\bar{t}$}
\put(-70, 85){{\bf (i)}}
\end{picture} 
\hspace*{2em}
\begin{picture}(220,120)(-80,-10)
\Gluon(20,  0)(50,20){4}{4}
\Gluon(20,100)(50,80){4}{4}
\ArrowLine(50,80)(50,20)
\ArrowLine(50,20)(100,50)
\ArrowLine(100,50)(50,80)
\Gluon(100,50)(140,50){4}{4}
\Vertex(50,20){3}
\Vertex(50,80){3}
\Vertex(100,50){3}
\Vertex(140,50){3}
\Vertex(165,25){3}
\DashLine(165, 25)(190,25){5}
\ArrowLine(140,50)(190,100)
\ArrowLine(165,25)(140, 50)
\ArrowLine(190, 0)(165, 25)
\put( 3,100){$g$}
\put( 3,  0){$g$}
\put(198, 95){$t$}
\put(198,20){$H$}
\put(198,-3){$\bar{t}$}
\end{picture} 
}
\\[2.0em]
\noindent
{\unitlength 0.5pt 
\begin{picture}(220,120)(-80,-10)
\ArrowLine(20, 80)(50,50)
\ArrowLine(50, 50)(20,20)
\Gluon(50,50)(80,50){4}{3}
\ArrowLine(80,50)(130,20)
\ArrowLine(130,20)(130,80)
\ArrowLine(130,80)(80,50)
\Gluon(130,80)(160,80){4}{3}
\Vertex(50,50){3}
\Vertex(80,50){3}
\Vertex(130,20){3}
\Vertex(130,80){3}
\Vertex(160,80){3}
\DashLine(130, 20)(200,10){5}
\ArrowLine(160,80)(200,100)
\ArrowLine(200,60)(160, 80)
\put( 3,80){$q$}
\put( 3,20){$\bar q$}
\put(208, 95){$t$}
\put(208,57){$\bar{t}$}
\put(208, 0){$H$}
\put(-70, 85){{\bf (ii)}}
\end{picture} 
\hspace*{2em}
\begin{picture}(220,120)(-80,-10)
\Gluon(20,  0)(50,20){4}{4}
\Gluon(20,100)(50,80){4}{4}
\ArrowLine(50,80)(50,20)
\ArrowLine(50,20)(100,20)
\ArrowLine(100,20)(100,80)
\ArrowLine(100,80)(50,80)
\Gluon(100,80)(140,80){4}{4}
\Vertex(50,20){3}
\Vertex(50,80){3}
\Vertex(100,20){3}
\Vertex(100,80){3}
\Vertex(140,80){3}
\DashLine(100, 20)(190,10){5}
\ArrowLine(140,80)(190,100)
\ArrowLine(190,60)(140, 80)
\put( 3,100){$g$}
\put( 3,  0){$g$}
\put(198, 95){$t$}
\put(198,57){$\bar{t}$}
\put(198, 0){$H$}
\end{picture} 
}
\\[1.5em]
\noindent
{\unitlength 0.5pt 
\begin{picture}(170,120)(-50,0)
\ArrowLine(50, 95)( 90, 95)
\ArrowLine(90, 95)( 90,  5)
\ArrowLine( 90,  5)(50,  5)
\Gluon(90, 95)(140, 95){4}{5}
\Gluon(90,  5)(140,  5){4}{5}
\Vertex( 90, 95){3}
\Vertex( 90,  5){3}
\Vertex(140,  5){3}
\Vertex(140, 50){3}
\Vertex(140, 95){3}
\DashLine(140,50)(170,50){5}
\ArrowLine(140, 95)(170, 95)
\ArrowLine(140, 50)(140, 95)
\ArrowLine(140,  5)(140, 50)
\ArrowLine(170,  5)(140,  5)
\put(178, 90){$t$}
\put(178,44){$H$}
\put(178, 2){$\bar{t}$}
\put(32, 95){$q$}
\put(32,  2){$\bar q$}
\put(-40, 85){{\bf (iii)}}
\end{picture}
\hspace*{1em}
\begin{picture}(170,120)(-50,0)
\Gluon(50, 95)( 90, 95){4}{4}
\Gluon(50,  5)( 90,  5){4}{4}
\ArrowLine(170,  5)(140,  5)
\ArrowLine(140,  5)(90,  5)
\ArrowLine( 90,  5)(90, 50)
\ArrowLine( 90, 50)(90, 95)
\ArrowLine(90, 95)(140, 95)
\ArrowLine(140, 95)(170, 95)
\Vertex( 90, 95){3}
\Vertex( 90,  5){3}
\Vertex(140,  5){3}
\Vertex( 90, 50){3}
\Vertex(140, 95){3}
\DashLine(90,50)(170,50){5}
\Gluon(140,  5)(140, 95){4}{9}
\put(178, 90){$t$}
\put(178,44){$H$}
\put(178, 2){$\bar{t}$}
\put(32, 95){$g$}
\put(32,  2){$g$}
\end{picture}
\hspace*{1em}
\begin{picture}(170,120)(-50,0)
\Gluon(50, 95)( 90, 95){4}{4}
\Gluon(50,  5)(140, 65){4}{11}
\Gluon(90, 95)(140, 95){4}{5}
\Gluon(90, 95)(140,  5){4}{11}
\Vertex( 90, 95){3}
\Vertex(140, 35){3}
\Vertex(140,  5){3}
\Vertex(140, 65){3}
\Vertex(140, 95){3}
\DashLine(140,35)(170,35){5}
\ArrowLine(140, 95)(170, 95)
\ArrowLine(140, 65)(140, 95)
\ArrowLine(140, 35)(140, 65)
\ArrowLine(140,  5)(140, 35)
\ArrowLine(170,  5)(140,  5)
\put(178, 90){$t$}
\put(178,29){$H$}
\put(178, 2){$\bar{t}$}
\put(32, 95){$g$}
\put(32,  2){$g$}
\end{picture}
\hspace*{1em}
\begin{picture}(170,120)(-50,0)
\Gluon(50, 95)( 90, 95){4}{4}
\Gluon(50,  5)( 90,  5){4}{4}
\Gluon(90, 95)(140, 95){4}{5}
\Gluon(90,  5)(140,  5){4}{5}
\Gluon(90, 95)( 90,  5){4}{9}
\Vertex( 90, 95){3}
\Vertex( 90,  5){3}
\Vertex(140,  5){3}
\Vertex(140, 50){3}
\Vertex(140, 95){3}
\DashLine(140,50)(170,50){5}
\ArrowLine(140, 95)(170, 95)
\ArrowLine(140, 50)(140, 95)
\ArrowLine(140,  5)(140, 50)
\ArrowLine(170,  5)(140,  5)
\put(178, 90){$t$}
\put(178,44){$H$}
\put(178, 2){$\bar{t}$}
\put(32, 95){$g$}
\put(32,  2){$g$}
\end{picture}
}
\caption{Representative diagrams of the gauge-invariant subsets 
(i)-(iii) of one-loop QCD diagrams, as described in the text.}
\label{fig:NLOdiags}
\end{figure}

Among the one-loop QCD diagrams, three different gauge-invariant
subsets can be distinguished, for both the $q\bar q$ and $gg$ channels.
Representative diagrams of these subsets are shown in
\reffi{fig:NLOdiags}.
Subset \textbf{(i)} comprises all closed-quark-loop graphs
where the Higgs boson couples to the external top-quark line, 
\textbf{(ii)} includes all
graphs where the Higgs boson couples to a closed quark loop, and
\textbf{(iii)} is formed by all other diagrams, which include at least
one gluon in the loop.  Counting closed quark loops only once per
quark flavour, we get 2, 2, and 25 graphs of the respective subsets in
the $q\bar q$ channel; and 6, 20, and 108 graphs in the $gg$ channel.
The counting is performed in the Feynman gauge, omitting tadpole-like
graphs and external self-energy corrections consistently. Of course,
the subsets \textbf{(i)} and \textbf{(iii)} receive counterterm
contributions from renormalization, which are considered in
\refse{se:ren} in more detail; the second subset is not renormalized
in NLO, since $Hgg$ and $Hggg$ couplings do not exist in LO.
 
In NLO the one-loop amplitudes $\M^{q\bar q,gg}_{\onel}$ contribute
via interference with the corresponding LO amplitudes 
$\M^{q\bar q,gg}_{\LO}$. Using the colour decomposition rules of 
\refse{se:LOconv}, the square of the sum of the LO amplitude and 
its virtual NLO correction reduces to
\beqar
\sum_{\mathrm{spins,\,colours}}|\M^{q\bar q}_{\LO+(\onel)}|^2 &=& 
32\sum_{\mathrm{spins}}\left( |\A^{q\bar q}_{\LO}|^2 
+ 2\Re\left\{ (\A^{q\bar q}_{\LO})^*\A^{q\bar q}_{\onel} \right\} \right),
\nn\\
\sum_{\mathrm{spins,\,colours}}|\M^{gg}_{\LO+(\onel)}|^2 &=& 
\sum_{j=1}^{3} c^{gg}_j \sum_{\mathrm{spins}}\left(
|\A^{gg}_{\LO,j}|^2 +2\Re\left\{ (\A^{gg}_{\LO,j})^*\A^{gg}_{\onel,j} \right\}
\right).
\eeqar

The evaluation of the virtual NLO correction was performed in two
completely independent calculations, leading, in particular, to two
independent computer codes.  In the first evaluation, the Feynman
graphs have been generated by using {\sl Feyn\-Arts}
\cite{Kublbeck:1990xc}. These amplitudes have subsequently been
reduced in terms of SME with the help of {\sl Mathematica}.  Following
standard techniques of one-loop calculations
\cite{'tHooft:1979xw,Passarino:1979jh,Beenakker:1990jr,Denner:1991qq,Denner:1993kt},
the tensor integrals are algebraically reduced to scalar integrals.
The treatment of the so-called ``pentagon'' diagrams, involving five
propagators in the loop, requires particular care and extensions of
the standard methods given in
\citeres{'tHooft:1979xw,Passarino:1979jh,Beenakker:1990jr,Denner:1991qq,Denner:1993kt}.
The treatment of the pentagons is described in \refse{se:pentagons} in
detail, where the results on the scalar 5-point functions are listed
explicitly.  The results on the divergent scalar 3- and 4-point
functions are collected in \refapp{app:C0D0}.  Both ultraviolet (UV)
and soft/collinear infrared (IR) divergences are treated in
dimensional regularization. The renormalization, which removes the UV
divergences, is discussed in
\refse{se:ren}. The explicit structure of the IR singularities occurring
in the virtual corrections is given in \refse{se:UVIRdivs}, and the
cancellation against their counterparts in the real corrections is
described in \refse{se:realcorr}. The algebraic output of the first
calculation has been implemented in {\sl Fortran} for numerical
evaluation. The second calculation basically follows the same
strategy, with a conceptual difference that is described in
\refse{se:pentagons}~\refpar{par:Epract}.  The amplitudes have been
generated by hand and reduced with {\sl
Form}~\cite{Vermaseren:2000nd}, and the numerical evaluation is
carried out in {\sl C++}.

For later use, we introduce the following definitions for the scalar
and tensor integrals in analogy to
\citeres{'tHooft:1979xw,Passarino:1979jh,Beenakker:1990jr,Denner:1991qq,Denner:1993kt},
\beqar
B_{\{0,\mu,\mu\nu,\dots\}}(q_1,m_0,m_1) &=& 
\frac{(2\pi\mu)^{2\eps}}{\ri\pi^2} \int\rd^D q\, 
\frac{\{1,q_\mu,q_\mu q_\nu,\dots\}}{d_0d_1},
\nn\\
C_{\{0,\mu,\mu\nu,\dots\}}(q_1,q_2,m_0,m_1,m_2) &=& 
\frac{(2\pi\mu)^{2\eps}}{\ri\pi^2} \int\rd^D q\, 
\frac{\{1,q_\mu,q_\mu q_\nu,\dots\}}{d_0d_1d_2},
\nn\\
D_{\{0,\mu,\mu\nu,\dots\}}(q_1,q_2,q_3,m_0,m_1,m_2,m_3) &=&
\frac{(2\pi\mu)^{2\eps}}{\ri\pi^2} \int\rd^D q\, 
\frac{\{1,q_\mu,q_\mu q_\nu,\dots\}}{d_0d_1d_2d_3},
\nn\\[.5em]
E_{\{0,\mu,\mu\nu,\dots\}}(q_1,q_2,q_3,q_4,m_0,m_1,m_2,m_3,m_4) &=&
\frac{(2\pi\mu)^{2\eps}}{\ri\pi^2} \int\rd^D q\, 
\frac{\{1,q_\mu,q_\mu q_\nu,\dots\}}{d_0d_1d_2d_3d_4},
\nn\\[.5em]
\lefteqn{ \hspace{-15em}
\mbox{with} \quad d_0=q^2-m_0^2+\ri 0, \quad d_j=(q+q_j)^2-m_j^2+\ri 0, 
\quad j=1,\dots,4.}
\label{eq:onelint}
\eeqar
Here $+\ri 0$ is the infinitesimal imaginary part as required by causality, 
$D=4-2\eps$ is the space-time dimension, and $\mu$ is the arbitrary
reference scale of dimensional regularization, which we may identify 
with the renormalization scale. The UV and IR divergences
appear as poles in $\eps$; it is convenient to use the abbreviations
\beqar
\Delta_1(\mu) &=& \frac{\Gamma(1+\eps)}{\eps}
\left(\frac{4\pi\mu^2}{\Mt^2}\right)^\eps,
\qquad
\Delta_2(\mu) = \frac{\Gamma(1+\eps)}{\eps^2}
\left(\frac{4\pi\mu^2}{\Mt^2}\right)^\eps
\eeqar
for these divergences. [The top-quark mass has been used to define 
a characteristic
energy scale.] If we distinguish between UV and IR divergences,
we explicitly write $\Delta^{\UV,\IR}_n(\mu)$.

\subsubsection{Renormalization}
\label{se:ren}

The renormalization is performed in the $\overline{\mathrm{MS}}$ scheme
with the top-quark mass defined on shell. The top quark is decoupled from
the running of the strong coupling $\alpha_{\mathrm{s}}(\mu)$.
Denoting the bare top-quark mass and the bare strong coupling as
$\Mt^0$ and $\alpha_{\mathrm{s}}^0$, respectively, these conditions 
fix the renormalization parameters in the transformations
$\Mt^0=\Mt+\de\Mt$ and 
$\alpha_{\mathrm{s}}^0=\alpha_{\mathrm{s}}+\de\alpha_{\mathrm{s}}$
as follows,
\beqar
\frac{\de\Mt}{\Mt} &=& {}-\frac{\alpha_{\mathrm{s}}}{3\pi}
        \left[3\Delta^{\UV}_1(\mu)+4\right],
\nn\\
\frac{\de\alpha_{\mathrm{s}}}{\alpha_{\mathrm{s}}} &=&
\frac{\alpha_{\mathrm{s}}}{2\pi} 
\left( \frac{N_f}{3} - \frac{11}{2} \right) \Delta^{\UV}_1(\Mt)
+ \frac{\alpha_{\mathrm{s}}}{6\pi} \Delta^{\UV}_1(\mu),
\eeqar
where $N_f=5$ is the number of light flavours.  The divergence
$\Delta^{\UV}_1(\Mt) = {\Gamma(1+\eps)} {(4\pi)^\eps}/{\eps}$ is
independent of $\mu$ and originates from light-quark and gluon
loops. The term proportional to $\Delta^{\UV}_1(\mu)$, on the other
hand, originates from the top-quark loop in the gluon self-energy that
is subtracted at zero-momentum transfer.  In this scheme the running
of the coupling $\alpha_{\mathrm{s}}(\mu)$ is generated solely by the
finite contributions of the light-quark and gluon loops, while the
top-quark contribution is absorbed completely in the renormalization
condition and thus decouples effectively.  The transformation for
$\Mt^0$ also fixes the renormalization of the Yukawa coupling,
$g^0_{ttH} = \Mt^0/v = (\Mt+\de\Mt)/v$, since $v$ does not receive NLO
QCD corrections.

We renormalize the fields of the gluons, $G_{a,\mu}$, of the light
quarks, $\psi_q$, and of the top quark, $\psi_t$, all in the on-shell
scheme, \ie the wave-function renormalization constants $\de
Z_{G,q,t}$, defined by the transformations
\beq
G_{a,\mu}^0 = \left( 1+\de Z_G/2\right) G_{a,\mu}, \qquad
\psi_q^0 = \left( 1+\de Z_q/2\right) \psi_q, \qquad
\psi_t^0 = \left( 1+\de Z_t/2\right) \psi_t, 
\eeq
are adjusted to cancel the external self-energy corrections exactly.
Distinguishing between divergences of UV and IR origin, these
constants can be written as
\beqar
\de Z_G &=& {}-\frac{\alpha_{\mathrm{s}}}{2\pi} 
\left( \frac{N_f}{3}-\frac{5}{2} \right) 
                \left[ \Delta^{\UV}_1(\mu) - \Delta^{\IR}_1(\mu) \right]
-\frac{\alpha_{\mathrm{s}}}{6\pi} \Delta^{\UV}_1(\mu),
\nn\\
\de Z_q &=& {}-\frac{\alpha_{\mathrm{s}}}{3\pi}
                \left[ \Delta^{\UV}_1(\mu) - \Delta^{\IR}_1(\mu) \right],
\nn\\
\de Z_t &=& {}-\frac{\alpha_{\mathrm{s}}}{3\pi}
                \left[ \Delta^{\UV}_1(\mu) + 2\Delta^{\IR}_1(\mu) +4 \right].
\eeqar

The contributions of the counterterms and the external self-energies
to the one-loop matrix elements $\M^{q\bar q,gg}_{\onel}$ can be
derived easily. For the $q\bar q$ channel they explicitly read
\beqar
\M_{\ren}^{q \bar q} &=& 
\left( \frac{\de\alpha_{\mathrm{s}}}{\alpha_{\mathrm{s}}}
        + \frac{\de\Mt}{\Mt} + \de Z_q + \de Z_t \right) 
\M_{\LO}^{q \bar q}
\nn\\ && {}
-\C^{q\bar q} \, 
\frac{\pi\alpha_{\mathrm{s}} g_{ttH}}{\hat{s}}\,\de\Mt
\left[ \phantom{+} \frac{2\Mt}{(s_{35}-\Mt^2)^2}
        \left(\M^{q\bar q}_2+\M^{q\bar q}_3+2\M^{q\bar q}_{13}\right)
\right.
\nn\\
&& \left. \phantom{-\C^{q\bar q} \, 
          \frac{\pi\alpha_{\mathrm{s}} g_{ttH}}{\hat{s}}\,\de\Mt\Biggl[} {}
      +\frac{2\Mt}{(s_{45}-\Mt^2)^2}
        \left(\M^{q\bar q}_2+\M^{q\bar q}_3-2\M^{q\bar q}_{9}\right)
\right.
\nn\\
&& \left. \phantom{-\C^{q\bar q} \, 
          \frac{\pi\alpha_{\mathrm{s}} g_{ttH}}{\hat{s}}\,\de\Mt\Biggl[} {}
- \left( \frac{1}{s_{35}-\Mt^2}
        +\frac{1}{s_{45}-\Mt^2} \right) \M^{q\bar q}_1
\ \right].
\eeqar

The first term accounts for the external self-energy corrections as well as
the counterterms of the strong and Yukawa coupling renormalization,
while the remaining terms proportional to $\de\Mt$ come from the
renormalization of the top-quark mass.  For the $gg$ channel the
former corrections have the same structure (proportional to the LO
amplitude), but the mass renormalization involves cumbersome
expressions.

\subsubsection{Pentagon diagrams}
\label{se:pentagons}

\paragraph{General strategy}

Four different types of pentagon diagrams occur in the calculation; a
representative of each type is shown in the last row of
\reffi{fig:NLOdiags}. Since the incoming quark $q$ is massless, three
different types of 5-point functions are generated which involve one,
two, or three massless propagators in the loop.  Each of the three
types is IR divergent, so that we need explicit results for scalar and
tensor 5-point functions in $D\ne 4$ dimensions.  As known for a long
time \cite{Melrose:1965kb}, in $D=4$ dimensions all 5-point functions
can be expressed in terms of 4-point functions, simplifying the
calculation considerably. In the following we describe an elegant way
how this reduction can be generalized for singular 5-point functions
to $D\ne 4$ up to terms of order $\O(D-4)$ in dimensional regularization.

In order to make use of the reduction of 5-point functions to related
4-point functions, which works in four space-time dimensions, we first
translate the dimensionally regularized integral $E^{(D)}$ into
another regularization scheme that is defined in $D=4$ dimensions.
To this end, we first endow all massless propagators in the loop with
an infinitesimal mass $\la$ and call the new integral
$E^{(\mathrm{mass},D)}$, which is identical to $E^{(D)}$ if $\la=0$.
Next we determine a related {\it well-defined} integral, denoted
$E^{(\mathrm{mass},D)}_{\mathrm{sing}}$, with the same IR (soft and
collinear) singularity structure as $E^{(\mathrm{mass},D)}$. This
integral is obtained by decomposing the {\it integrand} of the 5-point
function in the collinear/soft limit in terms of 3-point integrands
with regularization-scheme-independent kinematical prefactors [explicit 
examples are given below]. The difference of the two integrals,
$E^{(\mathrm{mass},D)}-E^{(\mathrm{mass},D)}_{\mathrm{sing}}$, has a
uniquely-determined finite integrand and is therefore
regularization-scheme independent, \ie the limits $D\to 4$ and $\la\to 0$ 
commute in this quantity. In total, we have generated in this way
the relation
\beq
E^{(D)}-E^{(D)}_{\mathrm{sing}} + \O(D-4) =
E^{(\mathrm{mass},D=4)}-E^{(\mathrm{mass},D=4)}_{\mathrm{sing}} +\O(\la).
\label{eq:Etrans}
\eeq
If artificial UV divergences are introduced via
$E^{(\mathrm{mass},D)}_{\mathrm{sing}}$ these divergences can be
controlled by distinguishing the space-time dimensions $D_\IR$ and
$D_\UV$ for the IR and UV domains explicitly.  As shown below, the
singular parts are given in terms of 3-point functions, and
$E^{(\mathrm{mass},D=4)}$ can be expressed in terms of 4-point
functions. Consequently, solving Eq.~\refeq{eq:Etrans} for $E^{(D)}$,
\beq
E^{(D)} = E^{(D)}_{\mathrm{sing}} + \left[
E^{(\mathrm{mass},D=4)}-E^{(\mathrm{mass},D=4)}_{\mathrm{sing}} 
\right]+\dots,
\label{eq:Edim}
\eeq
this integral is expressed (up to irrelevant terms indicated by the
dots) in terms of 3-point and 4-point functions.

\paragraph{Determination of singular subintegrals in 5-point functions}
\label{par:Esing}

Having described the general strategy, we now determine the singular
terms in the relevant 5-point functions. The following derivation is
valid for the integrals $E^{(\mathrm{mass},D)}$, which are defined for
arbitrary $D$ in the vicinity of $D=4$ and which include infinitesimal
mass regulators $\la$ for internal massless particles.  The final
formulae given for $E_{\mathrm{sing}}$ below are thus valid for both
$E^{(D)}_{\mathrm{sing}}$ and
$E^{(\mathrm{mass},D=4)}_{\mathrm{sing}}$.  We do not write the
regulator $\la$ explicitly, since the relations
for $E_{\mathrm{sing}}$ also hold for other choices of mass regularizations.%
\footnote{Alternatively, we have introduced an 
  infinitesimal quark mass $m_q$ obeying $\Mt\gg m_q\gg\la>0$, and
  obtained the same final result for $E^{(D)}$.}

We first consider the case of one vanishing internal mass, as shown in
the second diagram of the last row in \reffi{fig:NLOdiags}, for
example.  There is only a purely soft (logarithmic) divergence which
is located at $q\to 0$ in the integrand if the vanishing mass is
$m_0=0$. In this soft limit the numerator of the integral and the
denominators $d_2$ and $d_3$ are regular. As a result, the
singularity structure of the integral remains unchanged if we set
$q=0$ in the numerator and the denominators $d_2$ and $d_3$, since the
divergence is logarithmic.  This, in particular, shows that all such
tensor integrals $E_{\mu\dots}$ are finite, and only the scalar
integral $E_0$ develops a soft singularity, which is already contained
in a scalar 3-point function.  There are two different types of 5-point
integrals with one internal massless line, corresponding to the second
diagram of the last row in \reffi{fig:NLOdiags} and the additional
diagram generated by shifting the Higgs line along the top-quark line.
Explicitly the singular parts of the corresponding integrals are given
by
\beqar 
\lefteqn{
  E_0(-p_3,p_1-p_3,p_4-p_2,p_4,0,\Mt,\Mt,\Mt,\Mt)_{\mathrm{sing}} }
\qquad
\nn\\*
&=& \frac{1}{(t_{13}-\Mt^2)(t_{24}-\Mt^2)}\, C_0(-p_3,p_4,0,\Mt,\Mt),
\nn\\[.5em]
\lefteqn{
  E_0(-p_3,p_1-p_3,p_4+p_5,p_4,0,\Mt,\Mt,\Mt,\Mt)_{\mathrm{sing}} }
\qquad
\nn\\*
&=& \frac{1}{(t_{13}-\Mt^2)(s_{45}-\Mt^2)}\, C_0(-p_3,p_4,0,\Mt,\Mt).
\eeqar

Next we consider the case with two internal massless lines, as
generated, for example, in the third diagram of the last row in
\reffi{fig:NLOdiags}. In this case, the denominators $d_i$ are given
by
\beqar
d_0 &=& q^2, \qquad
d_1 = (q+p_1)^2, \qquad
d_2 = (q+p_1-p_4)^2-\Mt^2,
\nn\\
d_3 &=& (q+p_3-p_2)^2-\Mt^2, \qquad
d_4 = (q+p_3)^2-\Mt^2, 
\eeqar
where the imaginary $i0$ terms are suppressed in the notation.  The
integral becomes singular in the collinear limit $q\to x p_1$ with
arbitrary $x$, and in the soft limits $q\to 0$ and $q\to -p_1$, which are
included in the collinear case for the limits $x\to 0$ and $x\to -1$, 
respectively. The non-singular
denominators $d_{2,3,4}$ behave in the collinear limit as
\beqar
d_2 &\sim& (xp_1+p_1-p_4)^2-\Mt^2 = (1+x)(t_{14}-\Mt^2),
\nn\\
d_3 &\sim& (xp_1+p_3-p_2)^2-\Mt^2 = {}-x(t_{13}-\Mt^2)-x\hat{s}+t_{23}-\Mt^2,
\nn\\
d_4 &\sim& (xp_1+p_3)^2-\Mt^2 = {}-x(t_{13}-\Mt^2),
\eeqar
and thus their inverse products as
\beqar
\frac{1}{d_2 d_3 d_4} &\sim&
\frac{1}{(s_{45}-\Mt^2)(t_{13}-\Mt^2)}\, \frac{1}{d_2}
- \frac{(s_{45}-t_{23})^2}
{(s_{45}-\Mt^2)(t_{13}-\Mt^2)(t_{14}-\Mt^2)(t_{23}-\Mt^2)}\,
\frac{1}{d_3}
\nn\\ && {}
+ \frac{1}{(t_{14}-\Mt^2)(t_{23}-\Mt^2)}\, \frac{1}{d_4}.
\eeqar
This leads to the following decomposition of the singular part of 
the 5-point integral,
\beqar
\lefteqn{
E_{\dots}(p_1,p_1-p_4,p_3-p_2,p_3,0,0,\Mt,\Mt,\Mt)_{\mathrm{sing}} } \qquad
\nn\\*
&=& 
\frac{1}{(s_{45}-\Mt^2)(t_{13}-\Mt^2)}\,
C_{\dots}(p_1,p_1-p_4,0,0,\Mt)
\nn\\ && {}
- \frac{(s_{45}-t_{23})^2}
{(s_{45}-\Mt^2)(t_{13}-\Mt^2)(t_{14}-\Mt^2)(t_{23}-\Mt^2)}\,
C_{\dots}(p_1,p_3-p_2,0,0,\Mt)
\nn\\ && {}
+ \frac{1}{(t_{14}-\Mt^2)(t_{23}-\Mt^2)}\,
C_{\dots}(p_1,p_3,0,0,\Mt). 
\eeqar

This relation is valid for all tensors; in contrast to the case with
only one vanishing mass, in this case also tensor integrals become
singular. Note that $E_{\mathrm{sing}}$ does not only include the
singularities but also regular terms in a {\it well-defined} way,
which is important when switching from one regularization scheme to
the other.

Finally, we consider the case with three internal masses zero, as encountered,
e.g., in the first and last diagram of the last row in
\reffi{fig:NLOdiags}. Here three soft and two collinear limits give rise
to divergences. Identifying $m_0=m_1=m_4=0$, the collinear limits
are realized by setting $q\to xp_1\,$ or $\,q\to yp_2$, and the three soft 
limits are reached for particular values of $x$ and/or $y$ 
(\ie $x=y=0$, $x=-1$, or $y=1$). The decomposition for the two
collinear limits can be worked out in analogy to the previous case.
When combining the two limits, however, double-counting of the soft
singularity at $\,x=y=0\,$ has to be avoided. The decomposition of the 
singular structure of the 5-point integral finally reads
\beqar
\lefteqn{
E_{\dots}(p_1,p_1-p_3,p_4-p_2,-p_2,0,0,\Mt,\Mt,0)_{\mathrm{sing}} } \qquad
\nn\\*
&=& 
\frac{1}{(t_{13}-\Mt^2)(t_{24}-\Mt^2)}\,
C_{\dots}(p_1,-p_2,0,0,0)
\nn\\ && {}
+ \frac{1}{\hat{s}(s_{35}-\Mt^2)}\,
C_{\dots}(p_1,p_1-p_3,0,0,\Mt)
\nn\\ && {}
- \frac{(t_{24}-s_{35})^2}{\hat{s}(s_{35}-\Mt^2)(t_{13}-\Mt^2)(t_{24}-\Mt^2)}\,
C_{\dots}(p_1,p_4-p_2,0,0,\Mt)
\nn\\ && {}
+ \frac{1}{\hat{s}(s_{45}-\Mt^2)}\,
C_{\dots}(p_4-p_2,-p_2,0,\Mt,0)
\nn\\ && {}
- \frac{(t_{13}-s_{45})^2}{\hat{s}(s_{45}-\Mt^2)(t_{13}-\Mt^2)(t_{24}-\Mt^2)}\,
C_{\dots}(p_1-p_3,-p_2,0,\Mt,0), 
\label{eq:E0_00mmm}
\eeqar
which is again valid for all tensors.

\paragraph{The scalar 5-point function}

In $D=4$ space-time dimensions the scalar 5-point function $E_0$ can
be expressed \cite{Melrose:1965kb} in terms of the five related
4-point functions $D_0(i)$, obtained from $E_0$ by omitting the
denominator $d_i$, as defined in Eq.~\refeq{eq:onelint}.  
Using the shorthand notations
\beq\label{eq:defY}
Y_{00} = 2 m_0^2, \quad Y_{i0} = Y_{0i} = m_0^2 + m_i^2-q_i^2, \quad
Y_{ij} = m_i^2 + m_j^2 - (q_i-q_j)^2, \quad i,j=1,2,3,4,
\eeq
the relation reads
\beq \label{eq:E0red}
0 = \left\vert \barr{cccccc}
{}-E^{(\mathrm{mass},D=4)}_{0} &\:D^{(\dots)}_{0}(0)&\:
D^{(\dots)}_{0}(1)&\:D^{(\dots)}_{0}(2)
&\:D^{(\dots)}_{0}(3)&\:D^{(\dots)}_{0}(4)\\
  1   &  Y_{00}   &  Y_{01}   &  Y_{02}   &  Y_{03}   &  Y_{04}   \\
  1   &  Y_{10}   &  Y_{11}   &  Y_{12}   &  Y_{13}   &  Y_{14}   \\
  1   &  Y_{20}   &  Y_{21}   &  Y_{22}   &  Y_{23}   &  Y_{24}   \\
  1   &  Y_{30}   &  Y_{31}   &  Y_{32}   &  Y_{33}   &  Y_{34}   \\
  1   &  Y_{40}   &  Y_{41}   &  Y_{42}   &  Y_{43}   &  Y_{44}
\earr \right\vert,
\eeq 
where the scalar integrals are defined for $D=4$. Solving for 
$E^{(\mathrm{mass},D=4)}_0$ we get
\beq\label{E0red2}
E^{(\mathrm{mass},D=4)}_0 = {}-\frac{1}{\det(Y)} \, \sum_{i=0}^4 
\,\det(Y_i)\,D^{(\mathrm{mass},D=4)}_0(i),
\eeq 
where $Y=(Y_{ij})$, and $Y_i$ is obtained from $Y$ by replacing all
entries in the $i$th column by~1.

Although Eq.~\refeq{eq:E0red} was derived for $D=4$, it is valid 
(in all cases considered here)
even for arbitrary $D$ up to terms of order $\O(D-4)$:
\beq\label{E0red3}
E^{(D)}_0 = {}-\frac{1}{\det(Y)} \, \sum_{i=0}^4 \,\det(Y_i)\,D^{(D)}_0(i)
+ \O(D-4).
\eeq
This generalized relation simply follows from the result derived in
\refpar{par:Esing} that the same linear relation between
$E_{\mathrm{sing}}$ and 3-point subintegrals holds in all mass/dimensional
regularization schemes.

\paragraph{Tensor 5-point functions}

We evaluate all tensor integrals by calculating the coefficients to
the Lorentz covariants that span the tensor. For the pentagons, tensor
integrals $E_{\mu\nu\dots}$ appear up to rank 4. In practice, however,
these integrals are needed only up to rank 3, since one momentum
factor $q$ in the numerator of a top-quark propagator can always be
cancelled against a propagator denominator. The covariant
decomposition of the relevant 5-point tensor functions reads
\beqar
E_\mu &=& \sum_{i=1}^4 q_{i,\mu} E_i,
\nn\\
E_{\mu\nu} &=& \sum_{i,j=1}^4 q_{i,\mu} q_{j,\nu} E_{ij}
+ g_{\mu\nu}E_{00},
\nn\\
E_{\mu\nu\rho} &=& \sum_{i,j,k=1}^4 q_{i,\mu} q_{j,\nu}  q_{k,\rho} E_{ijk}
+\sum_{i=1}^4 \left( q_{i,\mu} g_{\nu\rho} 
+q_{i,\nu} g_{\mu\rho} +q_{i,\rho}g_{\mu\nu} \right)
E_{00i}.
\label{eq:Etensors}
\eeqar
Since the tensors
$E_{\mu\nu\dots}$ are symmetric, the tensor coefficients $E_{ij}$ and
$E_{ijk}$ are symmetric in the indices.  
The coefficients $E_i$, $E_{ij}$, etc., can be recursively reduced
to the scalar integrals $E_0$, $D_0$, etc., by an algorithm known
as Passarino--Veltman reduction, as described in
\citeres{Passarino:1979jh,Denner:1991qq}, or by special techniques
\cite{Bern:1992em,Etensor}
developed for higher $n$-point tensor integrals.
The resulting expressions are cumbersome and do not
reveal any new insight; therefore we do not list the expressions
explicitly.

Note that the metric tensor appearing in Eq.~\refeq{eq:Etensors}
is $D$-dimensional.
In contrast to the $D$-dimensional metric tensor $g_{\al\be}$,
its four-dimensional part, $g^{(D=4)}_{\al\be}$, is spanned by the 
linearly independent base vectors $q_i$:
\beq
g^{(D=4)}_{\al\be} = \sum_{i,j=1}^4 \left(Z^{-1}\right)_{ij} 
q_{i,\al} q_{j,\be}, \qquad Z_{ij} = (q_i q_j),
\label{eq:g4decomp}
\eeq
where $Z$ is the Gram matrix of the momenta $q_i$.
The difference $\Delta g_{\al\be}=g_{\al\be}-g^{(D=4)}_{\al\be}$
is, thus, of $\O(D-4)$.
It is interesting to notice that the terms proportional to $\Delta
g_{\al\be}$ do not contribute in our case. 
Since $\Delta g_{\al\be}$ is an object of $\O(D-4)$ it could
only contribute if an accompanying tensor coefficient, such as
$E_{00}$, is divergent. The explicit tensor reduction to scalar
integrals, however, reveals that all such coefficients are finite.
This fact can also be derived from the following argument.  In tensor
integrals the only covariants that come with divergent coefficients
are built up by the singular regions in momentum space. Therefore, a
collinear divergence occurring for $q\to xp$ can only show up in
tensors built up by the momentum $p$ alone.  For purely soft
(logarithmic) divergences occurring at $q\to 0$ power-counting shows
that only the scalar integral is divergent; shifting the integration
momentum by $p$, \ie $q\to q+p$, only leads to divergent coefficients 
related to tensors of $p$ alone.  UV divergences could occur in covariants
containing $\Delta g_{\al\be}$; however, all relevant 5-point
integrals are UV-finite, which is easily proven by power-counting.

\paragraph{Alternatives in practice}
\label{par:Epract}

The results on 5-point functions described above admit two conceptually
different ways to evaluate the pentagon diagrams. We followed both
ways, as described below, and we found perfect numerical agreement.

In the first approach the scalar integral $E_0$ is evaluated in $D$
dimensions using Eq.~\refeq{E0red3} and also the reduction of the
tensor coefficients defined in Eq.~\refeq{eq:Etensors} is performed
consistently in $D\ne 4$. In this approach all divergent scalar 3- and
4-point integrals $C_0$ and $D_0$ are needed in dimensional
regularization. We have listed them in \refapp{app:C0D0}.

Alternatively one can make consistent use of Eq.~\refeq{eq:Edim} and
its analogous version for 4-point functions, in order to replace all
$D$-dimensional 4- and 5-point integrals by four-dimensional integrals
and $D$-dimensional 3-point functions that contain the IR (soft and
collinear) singularities. The reduction of the 4- and 5-point tensor
coefficients then works in four IR dimensions\footnote{UV divergences,
which artificially occur in subexpressions during the tensor
reduction, still have to be treated in $D$ dimensions.}.  An advantage
of this procedure is the easy analytical control of all IR
singularities which are contained in 3-point functions only. In this
approach all divergent $D_0$ functions are needed in the
mass-regularization scheme, and all divergent $C_0$ functions in both
the dimensional and the mass-regularization schemes.  The
mass-regularized functions can be derived from the results of
\refapp{app:C0D0}. Another benefit of evaluating pentagon diagrams in
four dimensions is the direct application of the tensor decomposition
of~\citere{Etensor}. This procedure consistently avoids leading Gram
determinants (i.e.\ those of four momenta) that are potential sources
of numerical instabilities, as described in the next section.

\subsubsection{Numerical stabilization}

As explained above, all tensor integrals occurring in the calculation
can be algebraically reduced to scalar integrals following the
well-known Passarino--Veltman algorithm \cite{Passarino:1979jh}. This
procedure is based on a decomposition of the tensor integral into a
complete set of covariants that are spanned by the metric tensor and
all independent momenta involved in the integral. The coefficients of
the covariants are obtained by inverting a set of linear equations,
which is non-singular as long as the momenta $q_i$ spanning the
covariants are linearly independent, i.e.\ as long as their Gram
determinant $G=\det(q_i q_j)$ is non-zero. Each tensor rank adds a
factor $G^{-1}$ to the expression of the tensor integral.  Near the
boundary of phase space, some determinants $G$ become arbitrarily
small, since some momenta become linearly dependent at the
boundary. The huge factors $G^{-n}$ are balanced by cancellations
between different tensor coefficients that are calculated in terms of
scalar integrals; the whole virtual correction remains well behaved at
the phase-space boundary. These cancellations lead to severe numerical
instabilities in the evaluation of the correction.  The instabilities
are most pronounced in the pentagon diagrams, which do not only
possess the highest tensor ranks, but also involve additional sources
of cancellations. For instance, the decomposition
\refeq{eq:g4decomp} of the four-dimensional metric tensor also leads to
a Gram determinant in the denominator, and the reduction \refeq{eq:E0red} 
of the scalar five-point function $E_0$ in terms of five four-point functions
involves Caley determinants $\det(Y_{\dots})$ that can become very small.

If the virtual correction is not evaluated with particular care at the
phase-space boundary, a phase-space point will soon be selected  
in the Monte Carlo integration where the result for the
correction is totally wrong. The wrong result will typically be 
a large number owing to incomplete cancellations in the evaluation.
Such unphysically large contributions occur more and more frequently
in the adaptive integration procedure and they thus destroy 
the result of the phase-space integration completely .

When using the Passarino--Veltman procedure, we avoid the numerical
instabilities by extrapolating the virtual correction from the
numerically safe inner region of phase space to the numerically
delicate boundary. To this end, we first divide the full phase space
into two regions, one where the correction is evaluated directly and
another where the correction is obtained from an extrapolation out of
the first region.

The extrapolation procedure is applied to the full one-loop
correction, since any subset of diagrams (such as the pentagons alone)
in general shows stronger variations over the phase space than the
full correction, owing to cancellations between different diagram
types. Therefore, we have to inspect the whole boundary of phase
space, which is characterized by the zero of the Gram determinant $G$
of the four momenta $p_1,\dots,p_4$,
\beq
G = \frac{\hat s^2}{16} \, 
\kappa^2({\bf p}_3^2,{\bf p}_4^2,{\bf p}_5^2) \,
\sin^2\theta_1 \, \sin^2\chi,
\eeq
where $\kappa$ is defined in Eq.~\refeq{eq:kappa} and $p_5$ is fixed by
momentum conservation, $p_1+p_2=p_3+p_4+p_5$. The three-momenta
${\bf p}_i$ and the angles $\theta_1$, $\chi$ are defined in the
partonic centre-of-mass frame. The angle $\theta_1$ is the angle between
${\bf p}_3$ and the beam axis, and $\chi$ denotes the (azimuthal)
angle between the ${\bf p}_4$--${\bf p}_5$ plane and the plane spanned by
${\bf p}_3$ and the beam axis. We identify the ``safe'' region by demanding 
$G/\hat s^4$ as well as $\sin\chi$ to be larger than certain numerical
cut values. The extrapolation into the ``problematic'' region is done
as follows:
\begin{enumerate}
\item
Regarding the integrand as a function of the variable $\chi$, thereby
keeping ${\bf p}_3$ and $|{\bf p}_4|$ fixed, we expand it in terms of
a Fourier series. As a set of basis functions we took trigonometric
functions or Legendre polynomials. The first few coefficients of the
series are determined by numerical integration, for which we took
Simpson or Gaussian quadrature. If the extrapolation result is close
to the directly calculated result for the correction, the direct
result is taken over.
\item
If the agreement between direct and extrapolated result is too bad,
the first step is repeated again with twice as many coefficients
in the Fourier series and twice as many points in the related integration
for the coefficients. The first step is repeated at most twice.
If the directly calculated result is still not in good agreement with
the extrapolation, but instead the various extrapolations look convergent,
the most precise extrapolated result is taken over.
\item
If the procedure still has not stopped yet, step one is modified by
changing the extrapolation direction by varying also ${\bf p}_3$ and 
$|{\bf p}_4|$ during the change in $\chi$. Step two is carried out as 
above.
\item
The modified extrapolation of step three is repeated if still no
success has been reached. Finally, if none of the extrapolations looks
convincing, the integration point is discarded, but this does not
happen very often.
\end{enumerate}
The extrapolation procedure solves the problem of numerical
instabilities at the expense of computing time.  Of course, the
described procedure has to be optimized by selecting appropriate cut
values on $G/\hat s^4$ and $\sin\chi$ and by choosing appropriate
techniques for the calculation of the Fourier
coefficients. Nevertheless the final phase-space integration has to be
stable against any change in the ex\-tra\-po\-la\-tion procedure. We
have checked this stability by changing the cut parameters and by
choosing different systems of orthogonal functions as well as
different integration techniques, as mentioned above.

In addition to adopting the usual Passarino--Veltman reduction
of 5-point tensor coefficients, we have applied the method of
\citere{Etensor} where the strategy 
\cite{Melrose:1965kb}
for scalar 5-point functions is generalized and 5-point tensor
coefficients are reduced to 4-point integrals directly. In this
approach inverse Gram determinants of four momenta are avoided
completely. Applying this alternative renders the virtual correction
near the phase-space boundary much more stable than in the usual
Passarino--Veltman approach. The results obtained by the two methods
mutually agree with each other.

\subsubsection{One-loop soft and collinear divergences}
\label{se:UVIRdivs}

There are two sources of IR (soft and collinear) singularities in the
one-loop diagrams: the self-energy corrections of the external fields
and various kinds of 3-point subintegrals whenever a (nearly on-shell)
gluon is exchanged between external lines. The former singularities
can be read off from the results of \refse{se:ren}, and the strategy
to determine the latter singularities has already been explained in
\refse{se:pentagons}~\refpar{par:Epract}.  Here we list the explicit
results:
\newcommand{\fqqdiv}{f^{q\bar q}}
\newcommand{\fggdiv}{f^{gg}}
\beqar
\M^{q\bar q}_{\onel}\Big|_{\IR} &=& 
\frac{\alpha_{\mathrm{s}}}{24\pi} 
\left( \fqqdiv_{12}-7\fqqdiv_{13}-2\fqqdiv_{23}-2\fqqdiv_{14}
-7\fqqdiv_{24}+\fqqdiv_{34} \right)
\M^{q\bar q}_{\LO},
\label{eq:Mdivqq}
\\[.5em]
\M^{gg}_{\onel}\Big|_{\IR} &=& 
{}-\frac{\alpha_{\mathrm{s}}}{24\pi} \Biggl[
\left(9\fggdiv_{12}+9\fggdiv_{13}+9\fggdiv_{24}-\fggdiv_{34}\right) 
\left(\frac{2}{3}\C^{gg}_1+\C^{gg}_2+\C^{gg}_3\right) 
\left(M_{\mathrm{direct}}+\frac{1}{2}M_{\mathrm{fusion}}\right)
\nn\\ && \quad {}
+\left( 
9\fggdiv_{12}+9\fggdiv_{23}+9\fggdiv_{14}-\fggdiv_{34} \right) 
\left(\frac{2}{3}\C^{gg}_1-\C^{gg}_2+\C^{gg}_3\right) 
\left(M_{\mathrm{crossed}}-\frac{1}{2}M_{\mathrm{fusion}}\right)
\nn\\ && \quad {}
+6(\fggdiv_{12}-\fggdiv_{14}-\fggdiv_{23}+\fggdiv_{34})\,\C^{gg}_1 
\left(M_{\mathrm{direct}}+\frac{1}{2}M_{\mathrm{fusion}}\right)
\nn\\ && \quad {}
+6(\fggdiv_{12}-\fggdiv_{13}-\fggdiv_{24}+\fggdiv_{34})\,\C^{gg}_1 
\left(M_{\mathrm{crossed}}-\frac{1}{2}M_{\mathrm{fusion}}\right)
\Biggr].
\label{eq:Mdivgg}
\eeqar
The functions $M_{\dots}$ are given in Eq.~\refeq{eq:Mggauc},
and the divergent factors $\fqqdiv_{ij}$ and $\fggdiv_{ij}$ read
\beqar
\fqqdiv_{12} &=& 2\Delta^{\IR}_2(\mu)
      -2\Delta^{\IR}_1(\mu)\left[\ln\left({}-\frac{\hat{s}}{\Mt^2}-\ri 0\right)
              -\frac{3}{2} \right],
\nn\\
\fggdiv_{12} &=& 2\Delta^{\IR}_2(\mu)
      -2\Delta^{\IR}_1(\mu)\left[\ln\left({}-\frac{\hat{s}}{\Mt^2}-\ri 0\right)
              -\frac{11}{6}+\frac{N_f}{9} \right],
\nn\\
\fqqdiv_{ij} &=& \Delta^{\IR}_2(\mu)
      -\Delta^{\IR}_1(\mu)\left[2\ln\left(1-\frac{t_{ij}}{\Mt^2}\right)
              -\frac{5}{2} \right],
\nn\\
\fggdiv_{ij} &=& \Delta^{\IR}_2(\mu)
      -\Delta^{\IR}_1(\mu)\left[2\ln\left(1-\frac{t_{ij}}{\Mt^2}\right)
              -\frac{17}{6}+\frac{N_f}{9} \right],
\qquad i=1,2, \quad j=3,4,
\nn\\
\fqqdiv_{34} &=& \fggdiv_{34} = 
2\Delta^{\IR}_1(\mu)\left[ \frac{s_{34}-2\Mt^2}{s_{34}\beta_{s_{34}}}
        \ln\left(\frac{\beta_{s_{34}}-1}{\beta_{s_{34}}+1}\right)+1 \right],
\qquad
\beta_{s_{34}} = \sqrt{1-\frac{4\Mt^2}{s_{34}}+\ri 0}\,.
\hspace{2em}
\label{eq:fdiv}
\eeqar
These results are in agreement with \citere{Catani:2001ef}, where
the general singularity structure of QCD amplitudes has been presented.
The agreement becomes apparent after rewriting Eqs.~\refeq{eq:Mdivqq}
and \refeq{eq:Mdivgg}
in the following generic form which is valid for both the $q\bar q$
and $gg$ channels,
\beq
\M_{\onel}\Big|_{\IR} = \M_{\LO} \otimes
\frac{\alpha_{\mathrm{s}}}{4\pi} 
\sum_{i,j=1 \atop i<j}^4 ({\bf T}_i\cdot {\bf T}_j)
\, f_{ij}.
\label{eq:Mdiv}
\eeq
Here ${\bf T}_i$ are the colour operators defined in 
\citeres{Catani:1996jh,Catani:2002hc,Catani:2001ef},
and the symbol 
$\otimes$ denotes the colour correlations. Since we will make
repeated use of this notation in the remainder of our report, a short
description can be found in \refapp{app:colour}.

\subsection{Real corrections}
\label{se:realcorr}

The real corrections at relative order ${\cal O}(\alpha_{\rm s})$
include processes with real-gluon radiation as well as $t\bar{t}H$
production in gluon-(anti)quark scattering. After a brief description
of the parton reactions and their evaluation, we will discuss in some
detail the isolation of IR divergences for soft and collinear parton
configurations as well as the treatment of collinear initial-state
singularities by means of mass factorization.

\subsubsection{Parton processes and their evaluation}

\begin{figure}
\SetScale{0.5}
{\unitlength 0.5pt 
\begin{picture}(170,120)(-50,0)
\ArrowLine(20, 80)(50,50)
\ArrowLine(50, 50)(20,20)
\Gluon(50,50)(100,50){4}{5}
\Vertex(50,50){3}
\Vertex(100,50){3}
\Vertex(125,25){3}
\Vertex(125,75){3}
\DashLine(125, 25)(150,25){5}
\ArrowLine(125,75)(150, 75)
\ArrowLine(100,50)(125, 75)
\ArrowLine(125,25)(100, 50)
\ArrowLine(150, 0)(125, 25)
\Gluon(125,75)(150,100){4}{3}
\put(158,100){$g$}
\put( 8,80){$q$}
\put( 8,20){$\bar q$}
\put(158, 70){$t$}
\put(158,20){$H$}
\put(158,-10){$\bar{t}$}
\end{picture}
\hspace*{1em}
\begin{picture}(170,120)(-50,0)
\Gluon(50,100)(100,100){4}{5}
\Gluon(100,100)(150,100){4}{5}
\Gluon(100,100)(100,70){4}{3}
\Gluon(50,  0)(100,  0){4}{5}
\Vertex(100,100){3}
\Vertex(100, 70){3}
\Vertex(100,  0){3}
\Vertex(100, 35){3}
\DashLine(100, 35)(150,35){5}
\ArrowLine(100, 70)(150, 70)
\ArrowLine(100, 35)(100, 70)
\ArrowLine(100,  0)(100, 35)
\ArrowLine(150,  0)(100,  0)
\put(158,100){$g$}
\put(158, 65){$t$}
\put(158,30){$H$}
\put(158,-5){$\bar{t}$}
\put(35,100){$g$}
\put(35,  0){$g$}
\end{picture}
\hspace*{1em}
\begin{picture}(170,120)(-50,0)
\ArrowLine(50,100)(100,100)
\ArrowLine(100,100)(150,100)
\Gluon(100,100)(100,70){4}{3}
\Gluon(50, 25)(100, 25){4}{5}
\Vertex(100,100){3}
\Vertex(100, 70){3}
\Vertex(100, 25){3}
\Vertex(125, 25){3}
\DashLine(125, 25)(150, 25){5}
\ArrowLine(100, 70)(150, 70)
\ArrowLine(100, 25)(100, 70)
\ArrowLine(125, 25)(100, 25)
\ArrowLine(150, 0)(125, 25)
\put(158, 95){$q$}
\put(158, 65){$t$}
\put(158,20){$H$}
\put(158,-10){$\bar{t}$}
\put(35, 95){$q$}
\put(35, 25){$g$}
\end{picture}
\hspace*{1em}
\begin{picture}(170,120)(-50,0)
\ArrowLine(100,100)(50,100)
\ArrowLine(150,100)(100,100)
\Gluon(100,100)(100,70){4}{3}
\Gluon(50,  0)(100,  0){4}{5}
\Vertex(100,100){3}
\Vertex(100, 70){3}
\Vertex(100,  0){3}
\Vertex(100, 35){3}
\DashLine(100, 35)(150,35){5}
\ArrowLine(100, 70)(150, 70)
\ArrowLine(100, 35)(100, 70)
\ArrowLine(100,  0)(100, 35)
\ArrowLine(150,  0)(100,  0)
\put(158, 65){$t$}
\put(158,30){$H$}
\put(158,-5){$\bar{t}$}
\put(35,  0){$g$}
\put(158, 95){$\bar q$}
\put(35, 95){$\bar q$}
\end{picture}
}
\caption[]{Generic set of Feynman diagrams contributing to real gluon emission 
and gluon-(anti)quark scattering.}
\label{fig:NLO-real-diags}
\end{figure}
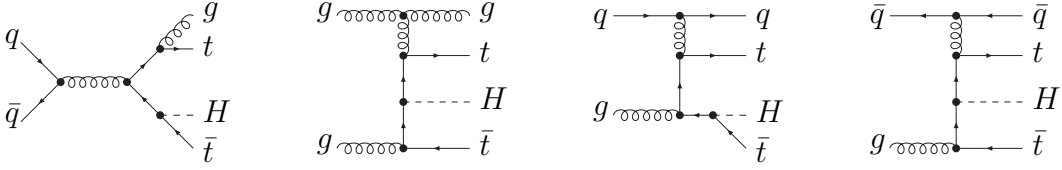

The evaluation of the ${\cal O}(\alpha_{\rm s}^3\,g_{ttH}^2)$ cross
section requires the calculation of the gluon bremsstrahlung processes
\beqar
q(p_1)+\bar q(p_2) &\to& t(p_3)+\bar t(p_4)+H(p_5) + g(p_6), \nonumber \\
g(p_1)+g(p_2)      &\to& t(p_3)+\bar t(p_4)+H(p_5) + g(p_6),
\label{eq:real-gluon}
\eeqar
and the gluon-(anti)quark scattering reactions
\beqar
q(p_1)+g(p_2) &\to& t(p_3)+\bar t(p_4)+H(p_5) + q(p_6), \nonumber \\
g(p_1)+\bar{q}(p_2) &\to& t(p_3)+\bar t(p_4)+H(p_5) + \bar{q}(p_6).
\label{eq:real-quark}
\eeqar
A representative set of Feynman diagrams, contributing to the real 
corrections, is given in \reffi{fig:NLO-real-diags}. 

Helicity amplitudes for the processes (\ref{eq:real-gluon},
\ref{eq:real-quark}) have been generated and evaluated using the
program packages {\sl MadGraph}~\cite{Stelzer:tk} and {\sl
HELAS}~\cite{Murayama:1992gi}. The result has been verified by an
independent calculation based on standard trace techniques.

\subsubsection{Soft and collinear divergences}

The $2\to 4$ cross sections contain singularities in the limits where
the splitting $g\to gg$, $g\to q\bar{q}$, $q\to qg$, or $\bar{q}\to
\bar{q}g$ becomes soft or collinear. We have used the dipole subtraction
method~\cite{Catani:1996jh}, as formulated in \citere{Catani:2002hc}
for massive QCD partons, to extract the singular part of the real
cross section in $D$ dimensions.  The dipole subtraction method is
based on the fact that soft and collinear limits of a real-emission
matrix element can be expressed by convoluting process-independent
splitting functions with the LO matrix element. By subtracting these
splitting-function expressions as counterterms from the squared real
matrix element, the IR (soft and collinear) singularities cancel and
the phase-space integration can be performed numerically in four
dimensions. The counterterms are constructed such that they can be
integrated analytically in $D$ dimensions over the phase space of the
extra emitted parton, leading to poles in $\epsilon = (4-D)/2$. When
the integrated subtraction counterterms and the virtual cross sections
are combined, the soft singularities cancel.  The remaining
initial-state collinear divergences have to be absorbed into a
redefinition of the parton distribution functions at NLO. Below we
will describe the subtraction formalism in more detail and we will
give explicit results for the soft and collinear contributions.

The NLO contribution of the 
cross section can symbolically be written as 
\cite{Catani:1996jh,Catani:2002hc}
\begin{eqnarray}\label{sNLO4}
\sigma^{ab}_{\rm NLO}(p_1,p_2,\mu_F^2) &=& 
\sigma_{\rm NLO}^{ab\,\{4\}}(p_1,p_2) + \sigma_{\rm NLO}^{ab\,\{3\}}(p_1,p_2) 
\nn\\[2mm]
& & {}+\int_0^1 \rd x\;\biggl[ 
\hat{\sigma}_{\rm NLO}^{ab\,\{3\}}(x;xp_1,p_2,\mu_F^2) 
+ \hat{\sigma}_{\rm NLO}^{ab\,\{3\}}(x;p_1,xp_2,\mu_F^2) \biggr]
\nn\\[2mm]
&=&
\int_4 \biggl\{ 
\Bigl[ \rd\sigma^{ab}_{\rm R}(p_1,\dots,p_6) \Bigr]_{\epsilon=0}\! 
- \biggl[\,\sum_{\mathrm{dipoles}} \!\!\Bigl( \rd\sigma_{\rm LO}^{ab}\otimes\;
           \rd V_{\mathrm{dipole}}\Bigr)(\tilde{p}_1,\dots,\tilde{p}_5,p_6)
  \biggr]_{\epsilon=0} \,\biggr\} 
\nn\\[1mm]
&& {}+\int_3 \left[ \rd\sigma^{ab}_{\rm V}(p_1,\dots,p_5) 
+ \rd\sigma^{ab}_{\rm LO}(p_1,\dots,p_5) \otimes {\cal I}^{ab}
\right]_{\epsilon=0} 
\nn \\[2mm]
&& {}+\sum_{a'=g,q,\bar q}\, \int_0^1 \rd x \int_3 
\left[ \rd\sigma_{\rm LO}^{a'b}(xp_1,p_2,\tilde{p}_3,\tilde{p}_4,\tilde{p}_5)
       \otimes \Gamma^{a,a'}_{t\bar tH + b}(x,\mu_F^2) 
\right]_{\epsilon=0}
\nn\\[1mm]
&& {}+\sum_{b'=g,q,\bar q}\, \int_0^1 \rd x \int_3 
\left[ \rd\sigma_{\rm LO}^{ab'}(p_1,xp_2,\tilde{p}_3,\tilde{p}_4,\tilde{p}_5)
       \otimes \Gamma^{b,b'}_{t\bar tH + a}(x,\mu_F^2) 
\right]_{\epsilon=0}\;.
\end{eqnarray}
This symbolic expression is valid under the assumption that no
massless particles are identified in the final state, \ie the
additional gluon or light (anti)quark is integrated out. The indices
$a$ and $b$ denote the flavours of the incoming partons, \ie $ab=q\bar
q,\,gg,\,qg,\,g\bar q$ in the processes
(\ref{eq:real-gluon},\ref{eq:real-quark}), and $\mu_F$ is the
factorization scale at which the redefinition of the parton
distribution functions is performed at NLO. The notation for the
integrals indicates that the real-emission contribution (labelled by
$\rm R$) involves four final-state particles [one gluon or light
(anti)quark more than in LO], while the virtual contribution (labelled
by $\rm V$) has LO three-particle kinematics.

The dipole subtraction counterterm is given in terms of universal
splitting functions, $\rd V_{\mathrm{dipole}}$ (see
\citere{Catani:2002hc} for explicit expressions), and an appropriate
colour, spin, and flavour projection of the LO cross section,
$\rd\sigma_{\rm LO}$. The symbol $\otimes$ denotes such possible
colour, spin, and flavour correlations.  In particular, in the sum
over ``dipoles'' the flavour correlations can change the incoming
partons $a$ or $b$ in $\rd\sigma_{\rm LO}^{ab}$.  The final state of
$\rd\sigma_{\rm LO}$ is fixed to $t\bar tH$, since no light
(anti)quark can be radiated from the top quarks. As far as the
four-particle phase space is concerned, each individual dipole
counterterm is evaluated for a specific set of auxiliary momenta
$\tilde{p}_1,\dots,\tilde{p}_5$. This involves combining the emitting
particle (with momentum $p_j$) and the emitted light particle (with
momentum $p_6$) into a single parent particle called ``emitter'' (with
momentum $\tilde{p}_j$). In order to balance the total momentum, one
of the other external particles (with momentum $p_k$) is replaced by a
``spectator'' particle (with the same quantum numbers, but modified
momentum $\tilde{p}_k$). Apart from the situation where both emitter
and spectator are incoming partons, the momenta of the remaining
external particles are not altered%
\footnote{In the case of both an initial-state emitter and an initial-state 
          spectator, the momentum of the spectator is left unaltered.
          Consequently, the momenta of all final-state particles
          undergo a Lorentz transformation.}. This phase-space mapping
          is constructed in such a way that \vspace*{-1ex}
\begin{itemize}
\itemsep -2pt
  \item[-] the soft and (quasi-)collinear momentum configurations are 
           reproduced point-wise;
  \item[-] both the emitter and spectator are on their respective mass shells;
  \item[-] momentum conservation is preserved;
  \item[-] the phase space can be factorized into a hard-scattering phase 
           space and a single-particle phase-space factor corresponding to 
           $p_6$.
\end{itemize}
Many of these phase-space mappings either have an initial-state emitter or 
an initial-state spectator. In those cases the corresponding initial-state 
momentum is rescaled by a longitudinal momentum fraction $x$, resulting in a 
boosted reference frame for the final-state momenta and a phase-space
factorization that involves a convolution over the boost parameter $x$. 
For explicit phase-space mappings we refer to the literature 
\cite{Catani:1996jh,Catani:2002hc}. 

The readded subtraction counterterm, integrated over the phase space
of the emitted extra parton, can be split into two parts. The first
part, involving the universal insertion operator ${\cal I}^{ab}$, has
LO kinematics and contains all the poles in $\eps$ that are necessary
to cancel the soft singularities in $\rd\sigma^{ab}_{\rm V}$. These
so-called endpoint contributions are obtained by integrating the
dipoles over the complete phase space of the emitted extra parton and
by using the $(\dots)_+$ prescription to extract the singular terms
related to the soft endpoint ($x\to 1$) of the $x$-integration of the
dipoles. The second part is a finite remainder that is left after
factorization of initial-state collinear singularities into the parton
distribution functions at NLO. It involves LO cross sections with
boosted three-particle kinematics, $x$-dependent structure functions
$\Gamma(x,\mu_F^2)$, and an additional one-dimensional integration
with respect to the longitudinal momentum fraction $x$ of an incoming
parton. The structure functions $\Gamma(x,\mu_F^2)$ depend on the
factorization scheme; only the terms proportional to $\ln(\mu_F^2)$
are factorization-scheme independent. The summations over the flavours
$a'$ and $b'$ represent the afore-mentioned flavour
correlations. Non-zero contributions are obtained if both
$\rd\sigma_{\rm LO}^{a'b}$ and $\Gamma^{a,a'}_{t\bar tH + b}$ are
non-zero, or if both $\rd\sigma_{\rm LO}^{ab'}$ and
$\Gamma^{b,b'}_{t\bar tH + a}$ are non-zero.

In the remainder of this subsection we give the explicit expressions
for the readded subtraction counterterms, which are based on the
results of \citere{Catani:2002hc}. The explicit construction of the
dipole terms, which is not described here, can also be found there.
We start with the insertion operator ${\cal I}^{ab}$, which depends on
the colour charges, momenta, and masses of the final-state particles
in the LO matrix element:
\beq
{\cal I}^{qg} = {\cal I}^{g\bar q} = 0, \qquad
  {\cal I}^{ab} = -\,\frac{\alpha_{\mathrm{s}}}{2\pi} \sum_{i,j=1 \atop i<j}^4 
                  ({\bf T}_i\cdot {\bf T}_j) \, {\cal J}_{ij}^{ab},
\qquad ab = q\bar q,gg,
\eeq
with
\beqar
  {\cal J}_{12}^{ab} &=& 2\Delta^{\IR}_2(\mu)
    - 2\Delta^{\IR}_1(\mu)\left[ \ln\left(\frac{\hat{s}}{\Mt^2}\right)
                                 - \frac{\gamma_a}{{\bf T}_a^2} \right] 
    + \ln^2\left(\frac{\hat{s}}{\Mt^2}\right)
    - 2\,\frac{\gamma_a}{{\bf T}_a^2}\ln\left(\frac{\hat{s}}{\Mt^2}\right)
    - \frac{2}{3}\,\pi^2,
\nn\\[1mm]
  {\cal J}_{34}^{ab} &=& 
      2\,\frac{s_{34}-2\Mt^2}{s_{34}\beta_{s_{34}}} \Biggl\{ \,
      \left[ \Delta^{\IR}_1(\mu) - \ln\left(\frac{s_{34}}{\Mt^2}\right) \right]
      \ln(\rho) - 4\ln(\rho)\ln(1-\rho) - \frac{1}{2}\ln^2(\rho)
\nn\\[1mm]
                          & & 
      {}- 2\Li(1-\rho^2) - 2\Li(1-\rho) \,\Biggr\}
\nn\\[1mm]
                          & & 
    {} + 2\Delta^{\IR}_1(\mu) - 2\ln\left(\frac{s_{34}}{\Mt^2}\right)
    + 6 + \frac{2s_{34}}{s_{34}-2\Mt^2}
          \ln\left(\frac{\sqrt{s_{34}}-\Mt}{\Mt}\right)
\nn\\[1mm]
                          & & 
    {} - 4\ln\left(\frac{\sqrt{s_{34}}-2\Mt}{\Mt}\right) 
    - \frac{2\Mt}{\sqrt{s_{34}}-\Mt} 
    - \frac{4\Mt (\sqrt{s_{34}}-2\Mt)}{s_{34}-2\Mt^2},
\nn\\[1mm]
  {\cal J}_{ij}^{ab} &=& \Delta^{\IR}_2(\mu)
    - \Delta^{\IR}_1(\mu)\left[2\ln\left(1-\frac{t_{ij}}{\Mt^2}\right)
                               - \frac{\gamma_a}{{\bf T}_a^2}-1 \right]
    + \frac{1}{2}\ln\left(2-\frac{t_{ij}}{\Mt^2}\right)
    + \ln^2\left(1-\frac{t_{ij}}{\Mt^2}\right)
\nn\\
      &&              
    - \left(\frac{\gamma_a}{{\bf T}_a^2}+2\right)
      \ln\left(1-\frac{t_{ij}}{\Mt^2}\right) 
    + \frac{3}{2} - \frac{2}{3}\,\pi^2 
    - 2\Li\left(\frac{-\,1}{1-t_{ij}/\Mt^2}\right)
    + \frac{1}{4-2\,t_{ij}/\Mt^2},
\nn\\[2mm]
      && \rm{for} \quad i=1,2 \quad \rm{and} \quad j=3,4.  
\label{eq:Jij} 
\eeqar
The variable $\beta_{s_{34}}$ is defined in Eq.~\refeq{eq:fdiv} and
$\rho$ is given by $\rho = (1-\beta_{s_{34}})/(1+\beta_{s_{34}})$. The
dilogarithm $\Li(x)$ is defined in the usual way as
\beq
  \Li(x)= {}-\int_0^1 \rd t\,\frac{\ln(1-xt)}{t}.
\eeq 
The Casimir operators ${\bf T}_a^2$ and anomalous dimensions $\gamma_a$
read explicitly
\beqar
  {\bf T}_q^2 = {\bf T}_{\bar q}^2 
  \equiv C_{\mathrm{F}} = \frac{4}{3},        \quad && \quad
  \frac{\gamma_q}{{\bf T}_q^2} 
  = \frac{\gamma_{\bar q}}{{\bf T}_{\bar q}^2} = \frac{3}{2},
\nn\\[1mm]
  {\bf T}_g^2 \equiv C_{\mathrm{A}} = 3, \quad && \quad
  \frac{\gamma_g}{{\bf T}_g^2} = \frac{11}{6} - \frac{N_f}{9}.
\label{eq:Casimir}
\eeqar
Inserting all this in Eq.~\refeq{eq:Jij}, we observe indeed that the
singular $\Delta^{\IR}_{2}(\mu)$ and $\Delta^{\IR}_{1}(\mu)$ terms in
the functions ${\cal J}_{ij}^{ab}$ cancel against the singular terms
in the coefficients $f_{ij}^{ab}$ given in Eq.~\refeq{eq:fdiv}.

Finally, we give a general expression for the structure function 
$\Gamma^{a,a'}_{t\bar tH + b}(x,\mu_F^2)$ in the $\MSbar$ factorization scheme:
\beqar
  \Gamma^{a,a'}_{t\bar tH + b} &=& \frac{\alpha_{\mathrm{s}}}{2\pi} \Biggl\{
        \ln\left(\frac{\hat{s}}{\mu_F^2}\right) P^{aa'}(x)
      + 2P_{\mathrm{reg}}^{aa'}(x)\ln(1-x) + \hat{P}'_{aa'}(x)
      + 4\,\delta^{aa'}{\bf T}_a^2\left[\frac{\ln(1-x)}{1-x}\right]_+ 
                                                                    \Biggr\}
\nn\\[1mm]
                               &+& \frac{\alpha_{\mathrm{s}}}{2\pi}\sum_{j=3}^4
        \frac{({\bf T}_{a'}\cdot {\bf T}_j)}{{\bf T}_{a'}^2} \Biggl\{ 
        \ln\left(\frac{\hat{s}\,\mu_{j1}^2}{\Mt^2}\right) P^{aa'}(x)
      + P_{\mathrm{reg}}^{aa'}(x)\ln(1-x+\mu_{j1}^2)
\nn\\
                               & & 
\hphantom{\frac{\alpha_{\mathrm{s}}}{2\pi}\sum_{j=3}^4
          \frac{({\bf T}_{a'}\cdot {\bf T}_j)}{{\bf T}_{a'}^2}a} {}
      + 2\,\delta^{aa'}{\bf T}_a^2\left[ \frac{1}{1-x}\,\Bigl( 1 +
        \ln[1-x+\mu_{j1}^2] \Bigr) - \frac{1-x}{4(1-x+\mu_{j1}^2)^2}\right]_+
\nn\\
                               & & 
\hphantom{\frac{\alpha_{\mathrm{s}}}{2\pi}\sum_{j=3}^4
          \frac{({\bf T}_{a'}\cdot {\bf T}_j)}{{\bf T}_{a'}^2}a} {}
      + 2\,\delta^{ga'}{\bf T}_a^2\,\frac{\mu_{j1}^2}{x}
        \ln\left(\frac{1-x+\mu_{j1}^2}{\mu_{j1}^2}\right)           \Biggr\}.
\label{eq:Gamma}
\eeqar
The various Altarelli--Parisi splitting functions $P^{aa'}(x)$, 
$P_{\mathrm{reg}}^{aa'}(x)$, and $\hat{P}'_{aa'}(x)$ can be found in 
\refapp{app:splitting}, and the kinematical variable $\mu_{j1}^2$ is given by 
\beq
  \mu_{j1}^2 = \frac{\Mt^2}{2p_1\tilde{p}_j}\,,
\label{eq:muj1}
\eeq
where $\tilde{p}_j$ is the auxiliary momentum of particle $j$ in the
boosted reference frame, while $p_1$ is the original initial-state
momentum of the incoming parton $a$.  In our convention the
$(\dots)_+$ prescription is defined by
\beqar
\lefteqn{\int_0^1\rd x \int\rd \Phi(x)\,
         \Bigl[ f(x,\mu^{(x)}_{j1}) \Bigr]_+ g(\Phi(x))} \qquad
  \nn\\[2mm]
&& =\ \int_0^1\rd x \int\rd \Phi(x) \int_{-\infty}^{\infty} \rd \bar{\mu}\,
      \,\delta(\bar{\mu}-\mu^{(x)}_{j1})\,
      \Bigl[ f(x,\bar{\mu}) \Bigr]_+ g(\Phi(x))
   \nn\\[2mm]
&& =\ \int_0^1\rd x \int_{-\infty}^{\infty} \rd \bar{\mu}\,
      f(x,\bar{\mu})\,\biggl\{ 
      \,\int\rd \Phi(x)\,g(\Phi(x))\,\delta(\bar{\mu}-\mu^{(x)}_{j1})
   \nn\\
&& \hphantom{=\ \int_0^1\rd x \int_{-\infty}^{\infty} \rd \bar{\mu}\,
             f(x,\bar{\mu})a} {}
      - \int\rd \Phi(1)\,g(\Phi(1))\,\delta(\bar{\mu}-\mu^{(1)}_{j1}) \Biggr\}
   \nn\\
&& =\ \int_0^1 \rd x \,\biggl\{
      \,\int\rd \Phi(x)\,f(x,\mu^{(x)}_{j1})\,g(\Phi(x))
      - \int\rd \Phi(1)\,f(x,\mu^{(1)}_{j1})\,g(\Phi(1)) \biggr\}\;,
\eeqar
where the label $(x)$ of $\,\mu^{(x)}_{j1}$ indicates that $\tilde{p}_j$ 
belongs to the phase space $\Phi(x)$ of the $x$-boosted reference frame and 
therefore $\mu^{(1)}_{j1}=\Mt/\sqrt{2p_1p_j}$. In fact, this is the same as 
saying that $\mu_{j1}$ should be kept fixed during the $x$-integration, 
\eg by using it as a phase-space integration variable in $\rd \Phi(x)$. 
The expression for the structure function $\Gamma^{b,b'}_{t\bar tH +
a}(x,\mu_F^2)$ can be obtained from Eqs.~\refeq{eq:Gamma} and
\refeq{eq:muj1} by interchanging $(a,a',\mu_{j1},p_1)$ and
$(b,b',\mu_{j2},p_2)$.

The formulas presented above are spelled out for total cross sections,
but they can also be used to evaluate any kind of IR-safe differential
distributions by filling histograms during the Monte Carlo
integration.  In this context, the only subtlety in the dipole
subtraction approach is due to the fact that contributions of
different dipole terms in general contribute to different histogram
bins.  The IR safety of an observable guarantees that singular terms
of the original integrand and their corresponding counterterms
contribute to the same bin in the soft and collinear limits.%
\footnote{If a direct calculation of a differential cross section is 
aimed at, i.e.\ without an integration over the full phase space,
further effort is needed, since the kinematics in some dipole terms 
would have to be changed for this purpose.}

\subsubsection{Slicing method for the $q\bar q$ channel}

In the phase-space slicing approach (see \citere{Harris:2001sx} for a
review) the soft and collinear regions are excluded from phase space
by appropriate phase-space cuts, generically denoted by $\De_\cut$.
The singular integration over these regions is then carried out
analytically using factorization properties.  The full real correction
is obtained by the sum of these two contributions in the limit
$\De_\cut\to 0$. In practice, a plateau is required for small but
finite values of $\De_\cut$.

In the following paragraphs we describe the so-called two-cutoff
method for the $q\bar q$ channel, where the soft region for gluon
bremsstrahlung is defined by a cut $\De E$ on the gluon energy and the
collinear region for gluon radiation off the initial state by a cut
$\De\theta$ on the emission angle.

\paragraph{Soft-gluon region}

Denoting the momentum of the radiated gluon by $k$, the soft
region is defined by 
\beq
0<k_0<\Delta E\ll \Mt,
\eeq
where $k_0$ is the gluon energy in the partonic centre-of-mass
frame. In the soft region, gluon radiation is described by an eikonal
current (see \eg \citere{Catani:1996jh}), and the contribution to the
bremsstrahlung cross section is given by
\beq
\rd\sigma_{\soft}^{q \bar q} = {}-\rd\sigma_{\LO}^{q \bar q} \otimes
\frac{\alpha_{\mathrm{s}}}{2\pi} \sum_{i,j=1 \atop i<j}^4 
({\bf T}_i\cdot {\bf T}_j) \, g_{ij}(p_i,p_j),
\eeq
where 
$g_{ij}$ denote the soft integrals
\beq
g_{ij}(p_i,p_j) =
\frac{(2\pi\mu)^{2\eps}}{2\pi} \int_{k_0<\Delta E}
\frac{\rd^{D-1}{\bf k}}{k_0} \,
\left[ \frac{2(p_i p_j)}{(p_i k)(p_j k)}
-\frac{p_i^2}{(p_i k)^2}-\frac{p_j^2}{(p_j k)^2} \right].
\eeq

The integrals are obtained as follows: $g_{12}$ is calculated easily;
$g_{34}$ can be taken over from \citere{Denner:1993kt}, where this
integral is given for an infinitesimal gluon mass $m_{\Pg}$ that
translates into $1/\eps$ via the substitution $\ln(m_{\Pg}^2) \;\to\;
(4\pi\mu^2)^\epsilon\,\Gamma(1+\epsilon)/\eps$; the remaining $g_{ij}$
can be derived using the auxiliary integrals (C.20) and (C.25) of
\citere{Beenakker:1989bq}.  The results are
\beqar
g_{12}(p_1,p_2) &=&
\left(\frac{4\pi\mu^2}{4\Delta E^2}\right)^\eps \Gamma(1+\eps)
\left\{\frac{2}{\eps^2}-\frac{2\pi^2}{3}\right\}
\nn\\[.5em]
&=& 
2\Delta^{\IR}_2(\mu) 
- 2\ln\left(\frac{4\Delta E^2}{\Mt^2}\right)\Delta^{\IR}_1(\mu)
+\ln^2\left(\frac{4\Delta E^2}{\Mt^2}\right)-\frac{2\pi^2}{3},
\\[1em]
g_{ij}(p_i,p_j) &=&
\left(\frac{4\pi\mu^2}{4\Delta E^2}\right)^\eps \Gamma(1+\eps)
\Biggl\{ \frac{1}{\eps^2}
-\frac{2}{\eps}\ln\left(\frac{p_i p_j}{p_{i,0}\Mt}\right)
-\frac{\pi^2}{3}
-\frac{1}{2}\ln^2\left(\frac{1+\beta_j}{1-\beta_j}\right)
\nn\\ && \quad {}
-2\Li\left(1-\frac{p_{i,0}p_{j,0}(1-\beta_j)}{p_i p_j}\right)
-2\Li\left(1-\frac{p_{i,0}p_{j,0}(1+\beta_j)}{p_i p_j}\right)
\nn\\ && \quad {}
+\frac{1}{\eps}
-\frac{1}{\beta_j}\ln\left(\frac{1-\beta_j}{1+\beta_j}\right)
\Biggr\}
\nn\\[.5em]
&=& \Delta^{\IR}_2(\mu) 
-\ln\left(\frac{4\Delta E^2}{\Mt^2}\right)\Delta^{\IR}_1(\mu)
+\frac{1}{2}\ln^2\left(\frac{4\Delta E^2}{\Mt^2}\right)
-2\ln\left(\frac{p_i p_j}{p_{i,0}\Mt}\right)\Delta^{\IR}_1(\mu)
\nn\\ && {}
+2\ln\left(\frac{4\Delta E^2}{\Mt^2}\right)
\ln\left(\frac{p_i p_j}{p_{i,0}\Mt}\right)-\frac{\pi^2}{3}
-\frac{1}{2}\ln^2\left(\frac{1+\beta_j}{1-\beta_j}\right)
\nn\\ && {}
-2\Li\left(1-\frac{p_{i,0}p_{j,0}(1-\beta_j)}{p_i p_j}\right)
-2\Li\left(1-\frac{p_{i,0}p_{j,0}(1+\beta_j)}{p_i p_j}\right)
\nn\\ && {}
+\Delta^{\IR}_1(\mu)-\ln\left(\frac{4\Delta E^2}{\Mt^2}\right)
-\frac{1}{\beta_j}\ln\left(\frac{1-\beta_j}{1+\beta_j}\right),
\nn\\[.5em]
&& 
\mbox{for} \quad i=1,2, \;\; j=3,4, \quad
\mbox{with} \quad \beta_j=\sqrt{1-\frac{\Mt^2}{p_{j,0}^2}},
\\[1em]
g_{34}(p_3,p_4) &=&
\left(\frac{4\pi\mu^2}{4\Delta E^2}\right)^\eps \Gamma(1+\eps)\,
\frac{s_{34}-2\Mt^2}{s_{34}\beta_{s_{34}}}\, \Biggl\{
{}-\frac{2}{\eps}\ln(\alpha)
\nn\\ && {}
+\left[\frac{1}{2}\ln^2\left(\frac{u_0-|{\bf u}|}{u_0+|{\bf u}|}\right)
+2\Li\left(1-\frac{u_0+|{\bf u}|}{v}\right)
+2\Li\left(1-\frac{u_0-|{\bf u}|}{v}\right) 
\right]^{u=\alpha p_3}_{u=p_4} \Biggr\}
\nn\\ && {}
+\left(\frac{4\pi\mu^2}{4\Delta E^2}\right)^\eps \Gamma(1+\eps)
\Biggl\{ \frac{2}{\eps}
-\frac{1}{\beta_3}\ln\left(\frac{1-\beta_3}{1+\beta_3}\right)
-\frac{1}{\beta_4}\ln\left(\frac{1-\beta_4}{1+\beta_4}\right)
\Biggr\}
\nn\\[.5em] 
&=&
\frac{s_{34}-2\Mt^2}{s_{34}\beta_{s_{34}}}\, \Biggl\{
{}-2\ln(\alpha)\Delta^{\IR}_1(\mu)
+2\ln\left(\frac{4\Delta E^2}{\Mt^2}\right)\ln(\alpha)
\nn\\ && {}
+\left[\frac{1}{2}\ln^2\left(\frac{u_0-|{\bf u}|}{u_0+|{\bf u}|}\right)
+2\Li\left(1-\frac{u_0+|{\bf u}|}{v}\right)
+2\Li\left(1-\frac{u_0-|{\bf u}|}{v}\right) 
\right]^{u=\alpha p_3}_{u=p_4} \Biggr\}
\nn\\ && {}
+2\Delta^{\IR}_1(\mu)-2\ln\left(\frac{4\Delta E^2}{\Mt^2}\right)
-\frac{1}{\beta_3}\ln\left(\frac{1-\beta_3}{1+\beta_3}\right)
-\frac{1}{\beta_4}\ln\left(\frac{1-\beta_4}{1+\beta_4}\right),
\nn\\[.5em] && 
\mbox{with} \quad 
\alpha=\frac{1+\beta_{s_{34}}}{1-\beta_{s_{34}}}=\frac{1}{\rho}, \quad
v=\frac{\Mt^2(\alpha^2-1)}{2(\alpha p_{3,0}-p_{4,0})}.
\label{eq:gij}
\eeqar
The $1/\eps^2$ poles, \ie the $\Delta^{\IR}_2(\mu)$ terms, of the functions
$g_{ij}$ are the same as in the coefficients $\fqqdiv_{ij}$ given in
Eq.~\refeq{eq:fdiv}. This implies that these terms cancel in the sum
of virtual and soft corrections, as it should be.

\paragraph{Collinear gluon emission from the initial state}

The region of collinear gluon emission from the initial state
is defined by
\beq
k_0>\Delta E, \qquad
0<\theta({\bf p}_i,{\bf k})<\Delta\theta\ll 1, \qquad i=1,2,
\eeq
in the partonic centre-of-mass frame. According to the QCD
factorization formula (see \eg \citere{Catani:1996jh}), the
contribution of collinear gluon radiation off parton $i$ with $i=1,2$
reads
\beqar
\rd\sigma_{\coll,i}^{q \bar q}(p_i) &=& 
\frac{\alpha_{\mathrm{s}}}{2\pi}\, C_{\mathrm{F}}
\left(\frac{4\pi\mu^2}{\hat{s}}\right)^\eps \Gamma(1+\eps)
\int_0^{1-2\Delta E/\sqrt{\hat{s}}}\rd x\,\rd\sigma_{\LO}^{q \bar q}(xp_i)
\nn\\ && {} \times
\left\{\frac{1+x^2}{1-x}\left[ -\frac{1}{\eps} 
+\ln\left(\frac{\Delta\theta^2}{4}\right)+2\ln(1-x) \right]
+1-x \right\} 
\nn\\[.5em]
&=&
\frac{\alpha_{\mathrm{s}}}{2\pi} 
\int_0^1\rd x\,\rd\sigma_{\LO}^{q \bar q}(xp_i)
\left\{ P^{qq}(x) \left[ {}-\Delta^{\IR}_1(\mu)
+\ln\left(\frac{\hat{s}\Delta\theta^2}{4\Mt^2}\right) \right] \right.
\nn\\ && \qquad \left. {} 
+C_{\mathrm{F}}\left[2\ln(1-x)\frac{1+x^2}{1-x} +1-x \right]_+
\right\}
\nn\\ &+&  
\frac{\alpha_{\mathrm{s}}}{2\pi}\, C_{\mathrm{F}}\,
\rd\sigma_{\LO}^{q \bar q}(p_i)
\left\{ \left[\Delta^{\IR}_1(\mu)
-\ln\left(\frac{\hat{s}\Delta\theta^2}{4\Mt^2}\right)\right]
\left[\ln\left(\frac{4\Delta E^2}{\hat{s}}\right)+\frac{3}{2}\right] \right.
\nn\\ && \qquad \left. {} 
-\frac{1}{2}\ln^2\left(\frac{4\Delta E^2}{\hat{s}}\right) +4\right\},
\eeqar
where the Casimir operator $C_{\mathrm{F}}=4/3$ has been defined in 
Eq.~\refeq{eq:Casimir} and the Altarelli--Parisi splitting function $P^{qq}(x)$
can be found in \refapp{app:splitting}. These results can, for instance, 
be easily derived by evaluating the
$x$ and $v$ integrals in Section 5.5 of \citere{Catani:1996jh} with
the $\Delta E$ and $\Delta\theta$ cuts.

{}From the results given above it is easy to verify that all
divergences $\Delta^{\IR}_1(\mu)$ cancel in the sum of the virtual
corrections (see \refse{se:UVIRdivs}), the soft corrections
$\rd\sigma_{\soft}^{q \bar q}$, and the part of the collinear
corrections $\rd\sigma_{\coll}^{q \bar q}$ with the LO kinematics,
\ie the part proportional to $\rd\sigma_{\LO}^{q \bar q}(p_i)$.
The remaining $P^{qq} (x)\,\Delta^{\IR}_1(\mu)$ term is absorbed 
into the redefined parton densities.

\paragraph{Independent check within the mass-regularization scheme}

The sum of $\rd\sigma_{\soft}^{q \bar q}+\rd\sigma_{\coll}^{q \bar q}$
has been checked by deriving all formulae for infinitesimal quark
masses. Apart from $g_{34}$ this calculation is completely independent
to the procedure in paragraphs {\it (i)} and {\it (ii)}. 
These are the steps to be followed:
\begin{enumerate}
\item Derive $\rd\sigma_{\soft}^{q \bar q}$ and $\rd\sigma_{\coll}^{q
    \bar q}$ for infinitesimal quark masses for $q$ and $\bar q$; see
  \citere{Denner:1993kt} for $\rd\sigma_{\soft}^{q \bar q}$, and
  \citere{Dittmaier:1994bj}, in particular Eq.~(26), for
  $\rd\sigma_{\coll}^{q \bar q}$.
\item The translation of the continuum part of $\rd\sigma_{\coll}^{q
    \bar q}$ [the part with 
  $P^{qq}(x)$ and the $(\dots)_+$ prescription] to the
  dimensional regularization scheme can be performed by comparing the
  renormalization of the parton distribution functions within the two 
  regularization schemes.
  Specifically, Eq.~(6.6) of \citere{Catani:1996jh} has to be compared
  with Eq.~(23) of \citere{Baur:1999kt} (see also
  \citere{Dittmaier:2001ay}).
\item The translation of the endpoint part of $\rd\sigma_{\soft}^{q
    \bar q}+\rd\sigma_{\coll}^{q \bar q}$ [this is the part
  proportional to $\rd\sigma_{\LO}^{q \bar q}(p_i)$] is described in
  Eqs.~(23) and (41) of \citere{Catani:2001ef}.
\end{enumerate}

\subsection{Leading threshold corrections} \label{subsec:threshold}

{}In view of the complexity of the NLO QCD corrections, as detailed in
the previous subsections, any additional cross-check is welcome.  The
partonic energy regime that is most suited for performing such
cross-checks is the threshold region, $\,\sqrt{\hat{s}} \gsim
2\Mt+\MH$, where the QCD corrections are dominated by soft-gluon
effects. The leading soft-gluon effects fall into two categories:
(virtual) Coulombic gluon exchange between the massive final-state
particles and IR-enhanced edge-of-phase-space effects from (real)
gluon radiation. Both effects can be calculated analytically.

Apart from providing explicit cross-checks, the leading threshold
corrections can also serve as a simple, intuitive tool for
understanding the differences between the hadronic NLO corrections at
the Tevatron and the LHC. This should not be overrated, though. In
$t\bar tH$ production, with its massive three-particle phase space,
the influence of the threshold region is substantially less than for
processes with massive two-particle final states, like $t\bar t$
production (see Section~\ref{sec:parton}).
  
\subsubsection{The Sommerfeld rescattering correction}

\begin{figure}
\SetScale{0.5}
\centerline{\unitlength 0.5pt 
\begin{picture}(200,120)
\Line(50, 95)( 90, 50)
\Line(50,  5)( 90, 50)
\ArrowLine(190,  0)(150, 20)
\ArrowLine(150, 20)( 90, 50)
\ArrowLine(100, 50)(155, 50)
\ArrowLine(155, 50)(200, 50)
\Vertex(155, 50){3}
\Vertex(150, 20){3}
\DashLine(90,50)(190,95){5}
\Gluon(155, 50)(150, 20){4}{4}
\GCirc( 90, 50){20}{.5}
\put(200, 90){$H$}
\put(210,44){$t$}
\put(200,-8){$\bar{t}$}
\put(10, 95){$q,g$}
\put(10,  2){$\bar q,g$}
\end{picture}
}
\caption[]{Generic diagram leading to the Coulomb singularity}
\label{fig:Couldiag}
\end{figure}
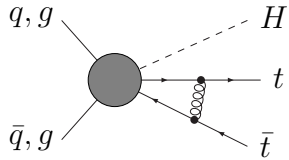

For energies near the threshold
the virtual QCD corrections are enhanced by Coulombic gluon
exchange between the top and antitop quark in the final state.  
The corresponding generic diagram is shown in \reffi{fig:Couldiag}.
When the gluon momentum tends to zero a so-called Coulomb singularity is 
generated.
In this limit the contribution of the singularity to the matrix
element, denoted by $\M_{\Coul}$, is given by
\beq
\M_{\Coul} = {}-\M_{\LO} \otimes
\frac{\pi\alpha_{\mathrm{s}}}{4\beta_{s_{34}}} 
({\bf T}_3\cdot {\bf T}_4),
\eeq
and it increases inversely proportional to the $t/\bar t$
velocity $\beta_{s_{34}}$ in the $t\bar t$ centre-of-mass frame, 
as defined in Eq.~\refeq{eq:fdiv}.
This is the well-known Sommerfeld rescattering correction \cite{Sommerfeld}.
Upon averaging $1/\beta_{s_{34}}$ over the phase space
near threshold, we obtain the following singular contribution to
the partonic cross section
\beq
\label{eq:beta}
\si_{\Coul} = \frac{8\alpha_{\mathrm{s}}}{3}\,
\sqrt{\frac{2\Mt}{2\Mt+\MH}}\,\frac{C_{\Coul}}{\beta}\, \si_{\LO},
\eeq
where
\beq
\beta=\sqrt{1-\frac{(2\Mt+\MH)^2}{\hat{s}}}.
\eeq
The coefficient $\beta$ is proportional to the maximal relative velocity
of the $t\bar t$ system. The rescattering coefficient $C_{\Coul}$ depends 
on the production mechanism of the $t\bar t$ pair,
\beq
C_{\Coul}^{e^+e^-} = {}+\frac{4}{3}, 
\qquad
C_{\Coul}^{q\bar q} = {}-\frac{1}{6}, 
\qquad
C_{\Coul}^{gg} = \frac{11}{42}\,
\frac{(4\Mt^2-\MH^2)^2-{\text\frac{9}{11}}\MH^4}
{(4\Mt^2-\MH^2)^2+{\text\frac{9}{7}}\MH^4}.
\eeq

In $e^+e^-$ annihilation the $t\bar t$ pair is generated in a
colour-singlet state in which the quark and antiquark attract each
other, and with $C_{\Coul}^{e^+e^-}={}+4/3$ the correction is
positive. This leads to a strong enhancement of the $e^+e^-\to t\bar t
H$ annihilation cross section near threshold \cite{Dittmaier:1998dz}.
By contrast, in $q\bar q$ annihilation the $t\bar t$ pair is generated
in a colour-octet state in which the force is repulsive, and with
$C_{\Coul}^{q\bar q}={}-1/6$ the correction is negative and relatively
small. This configuration is realized in the dominant channel at the
Tevatron, $q\bar q\to t\bar tH$. Finally, in $gg$ fusion the various
colour channels contribute differently, and the relative weighting
depends on the ratio $\MH^2/4 \Mt^2$.  For not too large $\MH$ the
rescattering coefficient $C_{\Coul}^{gg}$ is moderately positive, and
for $\MH^2/4 \Mt^2 \ll 1$ the rescattering coefficient approaches
$C_{\Coul}^{gg}(\MH=0)=11/42$ for $gg\to t\bar t$ (see
\citeres{Beenakker:1989bq,Nason:1988xz}).

In the close vicinity of threshold, our numerical results for the NLO
partonic cross sections indeed reproduce these Coulomb singularities
(see Section~\ref{sec:parton}).

\subsubsection{Threshold logarithms from soft-gluon radiation}

Owing to the lack of available phase space near threshold, enhanced
real-radiation effects from the initial state (ISR) and the final
state (FSR) are only noticeable in the vicinity of soft IR poles. In
that case the phase-space suppression is compensated by the inverse
dependence on the energy of the emitted massless particle, resulting
in an integrated correction that is logarithmic in the phase-space
volume -- in contrast to the Rutherford pole in the Sommerfeld
correction. The calculation of these so-called threshold logarithms
simply amounts to a convolution of the LO cross section with the
soft-pole parts [$\propto 1/(1-x)$] of the dipole functions for
real-gluon emission.

The LO matrix elements of the reactions $\,q\bar q,gg \to t\bar tH$
are basically constant near threshold. This leads to the following
threshold behaviour of the LO cross section [using Eqs.~\refeq{eq:LOxsection}
and \refeq{eq:beta}]:
\beq
\label{eq:lo_thres}
  \sigma_{\LO}^{ab}(\hat{s}) 
  \ \approx \  
  \Theta\biggl( \sqrt{\hat{s}}-2\Mt-\MH \biggr)\,
  \beta^4 \,
  \overline{\sigma}^{\,ab}_{\LO},
\eeq
where $\overline{\sigma}^{\,ab}_{\LO}$ represents the part of the LO
cross section that remains constant near threshold. The strong
suppression of the LO cross section near threshold $\propto \beta^4$
is a consequence of the massive three-particle phase space.

The threshold logarithms at NLO are given by
\beq
\sigma_{\log}^{ab} = 
  \int\limits_0^1 \rd x\,\sigma_{\LO}^{ab}(x\hat{s}) \otimes \Gamma_{\log}(x)
  \ \approx\ 
  \overline{\sigma}^{\,ab}_{\LO} \otimes 
  \Biggl[\,\,\int\limits_0^1 \rd x \,
  \Gamma_{\log}(x)\,
  \biggl( 1-\frac{1-\beta^2}{x} \biggr)^2 \,\Theta(x-1+\beta^2) \Biggr]\,.
\eeq
The function $\Gamma_{\log}(x)$ comprises all IR-sensitive parts of
the structure functions $\,\Gamma^{a,a'}_{t\bar tH + b}\,$ and 
$\,\Gamma^{b,b'}_{t\bar tH + a}\,$ in the limit $\beta\to 0$:
\beq
  \Gamma_{\log}(x) \ \approx\ 
      \frac{\alpha_{\mathrm{s}}}{\pi}\,
      \sum_{i=1}^2 \Biggl\{\, 
\left(\frac{1}{1-x}\right)_+\, \Biggl[
2{\bf T}_i^2\, \ln\biggl(\frac{2\Mt+\MH}{\mu_F}\biggr)
      + \sum_{j=3}^4 {\bf T}_i\cdot {\bf T}_j \Biggr]
+2{\bf T}_i^2\,\left(\frac{\ln(1-x)}{1-x}\right)_+\,
      \Biggr\}, 
\eeq
where $\mu_F$ denotes the factorization scale.
Exploiting charge conservation and keeping only the terms involving 
$\ln\beta$, one obtains finally
\beqar \label{eq:thrlog}
\sigma_{\log}^{ab} &=& \sigma_{\LO}^{ab} \otimes
\frac{2\alpha_{\mathrm{s}}}{\pi} 
\biggl\{ \left( {\bf T}_1^2+{\bf T}_2^2 \right)
        \left[ 2\ln^2\beta-3\ln\beta
               -2\ln\beta\ln\biggl(\frac{\mu_F}{2\Mt+\MH}\biggr) \right]
\nn\\[1mm]    && \qquad \qquad \qquad {}
                 - \left( {\bf T}_3+{\bf T}_4 \right)^2\ln\beta
\biggr\}
\nn\\[1mm] 
&=& \sigma_{\LO}^{ab} \,
\frac{2\alpha_{\mathrm{s}}}{\pi} 
\left\{ C^{ab}_{\ISR}
        \left[ 2\ln^2\beta-3\ln\beta
               -2\ln\beta\ln\biggl(\frac{\mu_F}{2\Mt+\MH}\biggr) \right]
       -C^{ab}_{\FSR} \ln\beta
\right\}, \qquad
\eeqar
with
\beq
C_{\ISR}^{q\bar q} = 2\,C_{\mathrm{F}} = \frac{8}{3}, 
\qquad
C_{\ISR}^{gg} = 2\,C_{\mathrm{A}} = 6,
\eeq
and
\beq
C_{\FSR}^{q\bar q} = C_{\mathrm{A}} = 3, 
\qquad
C_{\FSR}^{gg} = \frac{15}{7}\,
\frac{(4\Mt^2-\MH^2)^2+{\text\frac{9}{5}}\MH^4}
{(4\Mt^2-\MH^2)^2+{\text\frac{9}{7}}\MH^4}.
\eeq

The contributions from final-state gluon radiation exhibit a behaviour
in colour space that is parallel to the Coulomb singularities. In
$q\bar q$ annihilation the $t\bar t$ pair is generated in a
colour-octet state, resulting in an overall colour charge
$C_{\FSR}^{q\bar q} = 3$. In the case of $gg$ fusion both colour-octet
and colour-singlet states are superimposed. However, soft gluons
cannot be emitted from colour-singlet $t\bar t$ final states near
threshold owing to the overall vanishing colour charge and the long
wavelength of the emitted IR gluons -- as evident from $e^+e^-$
annihilation \cite{Jersak:sp}. Therefore only the colour-octet state
receives a logarithmic correction, causing the overall coefficient
$C_{\FSR}^{gg}$ to depend on the ratio $\MH^2/4 \Mt^2$.

In total, the threshold logarithms give rise to a positive NLO
correction near threshold, as long as the factorization scale is not
chosen too small. Furthermore we observe that the corrections in the
$gg$ channel are larger than the corrections in the $q \bar q$ channel
by roughly a factor of two.  A threshold analysis of our numerical
results for the partonic NLO cross sections indeed confirms the
above-given threshold logarithms (see Section~\ref{sec:parton}).

\section{Numerical Results}

In this section we present the NLO QCD results for the total partonic
cross sections and for the final hadronic cross sections at the
Tevatron and at the LHC, including the differential distributions in
transverse momentum and rapidity of the Higgs boson and of the top
quarks. As discussed previously, renormalization and factorization
have been performed in the $\overline{\mathrm{MS}}$ scheme with the
top-quark mass defined on-shell. The top quark is decoupled from the
running of the strong coupling $\alpha_{\mathrm{s}}(\mu)$ and from the
evolution of the parton densities.  For the evaluation of the hadronic
cross sections we have adopted the MRST~\cite{Martin:2001es} parton
densities at LO and NLO, corresponding to the QCD parameters
$\Lambda_5^{\mathrm{LO}}=167\MeV$ and
$\Lambda_5^{\overline{\mathrm{MS}}}=239\MeV$ at the one- and two-loop
level of $\alpha_{\mathrm{s}}(\mu)$, respectively. The strength of the
SM Yukawa coupling is fixed by $g_{ttH}=m_t/v$, where $v=
(\sqrt{2}G_F)^{-1/2}$ is the vacuum-expectation value of the Higgs
field and $G_F = 1.16639\!\times\!{}10^{-5}$~GeV${}^{-2}$. The
top-quark mass is set to $m_t=174\GeV$.

\subsection{Partonic cross sections}\label{sec:parton}

To explore the NLO QCD results at the parton level for the
production of $t\bar{t}H$ in quark and gluon collisions, we introduce
the scaling functions $f_{ab}(\eta)$, 
\beq\label{eq:scaling_fcts}
\sigma^{ab} = \frac{\alpha_{\rm s}^{2}(\mu^2)}{\mu_0^2}
\left\{f_{ab}^{(0)}(\eta) + 4\pi\alpha_{\rm s}(\mu^2)
\left[f_{ab}^{(1)}(\eta)+\bar{f}_{ab}^{\,(1)}(\eta)
\ln\left(\frac{\mu^2}{\mu_0^2}\right)
\right]
\right\}.
\eeq
The centre-of-mass energy of the partonic reaction, $\sqrt{\hat{s}}$,
is expressed in terms of the quantity $\eta = \hat{s}/4\mu_0^2 - 1$,
which is better suited for analyzing the scaling functions in the
various regions of interest. The reference scale $\mu_0 = \Mt + \MH/2$
has been set to half the partonic threshold energy. The index pair
$a,b$ indicates the partonic $q,\bar{q},g$ initial state of the
reaction $ab \to t\bar{t}H+X$. We have identified the renormalization
scale with the factorization scale: $\mu_R=\mu_F=\mu$. The scaling
functions are divided into the LO term $f^{(0)}$, the
scale-independent NLO term $f^{(1)}$, and the scale-dependent NLO
contribution $\bar{f}^{(1)}\ln(\mu^2/\mu_0^2)$.

\begin{figure}

\hspace*{5mm}
\epsfig{file=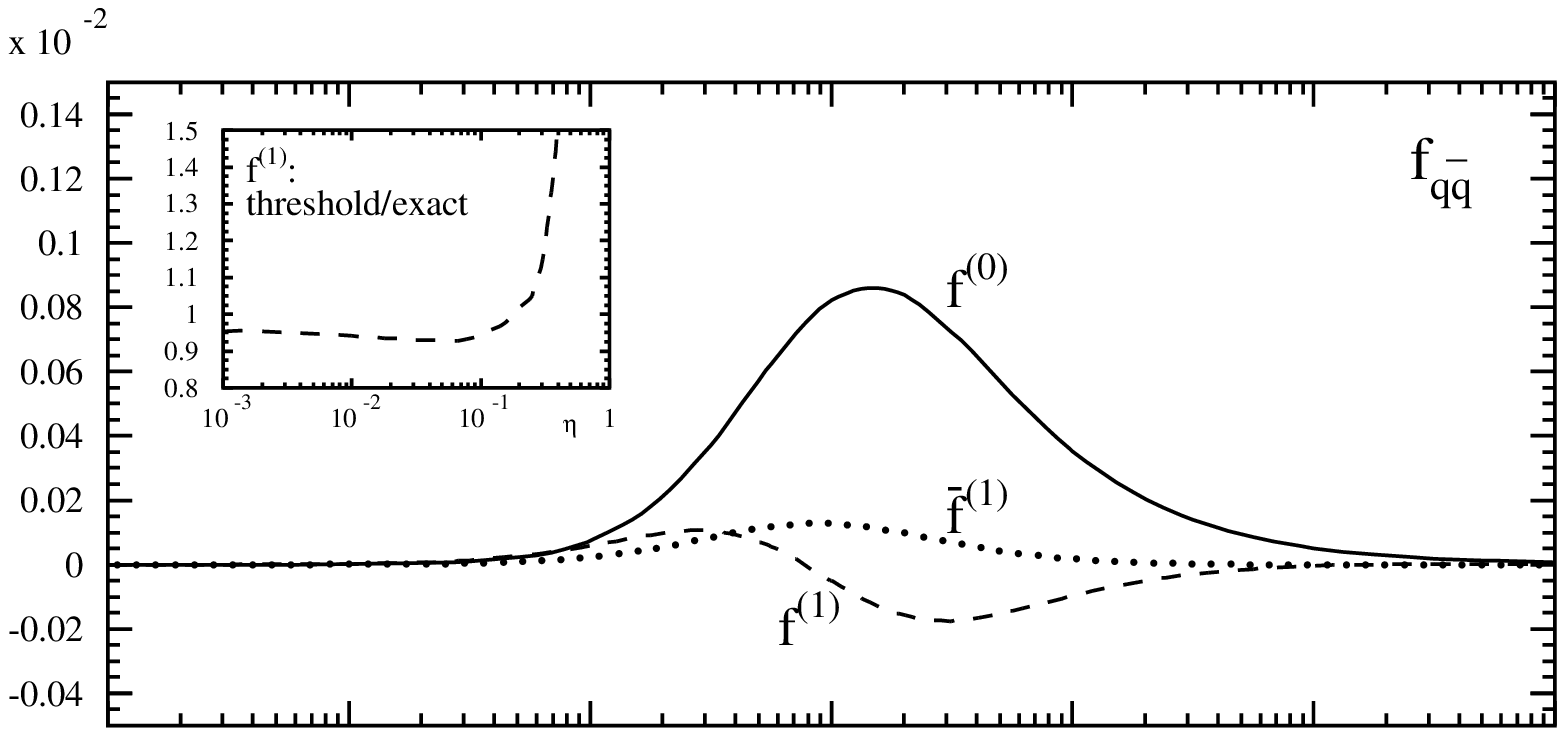,%
        bbllx=45pt,bblly=335pt,bburx=530pt,bbury=560pt,%
        width=14.25cm,height=7cm,angle=0,clip=}

\vspace*{-0.3cm}

\hspace*{5mm}
\epsfig{file=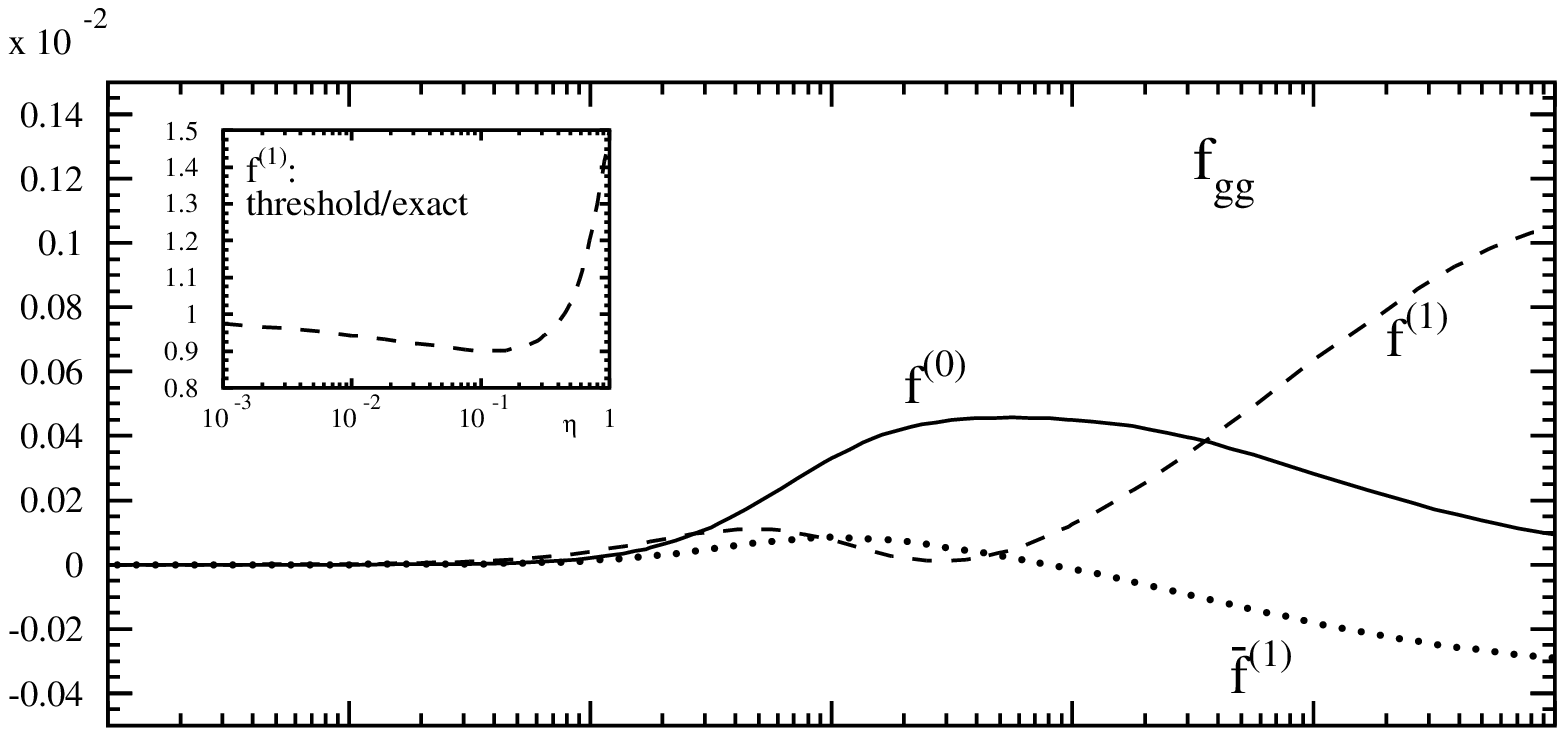,%
        bbllx=45pt,bblly=335pt,bburx=530pt,bbury=560pt,%
        width=14.25cm,height=7cm,angle=0,clip=}

\vspace*{-0.3cm}

\hspace*{5mm}
\epsfig{file=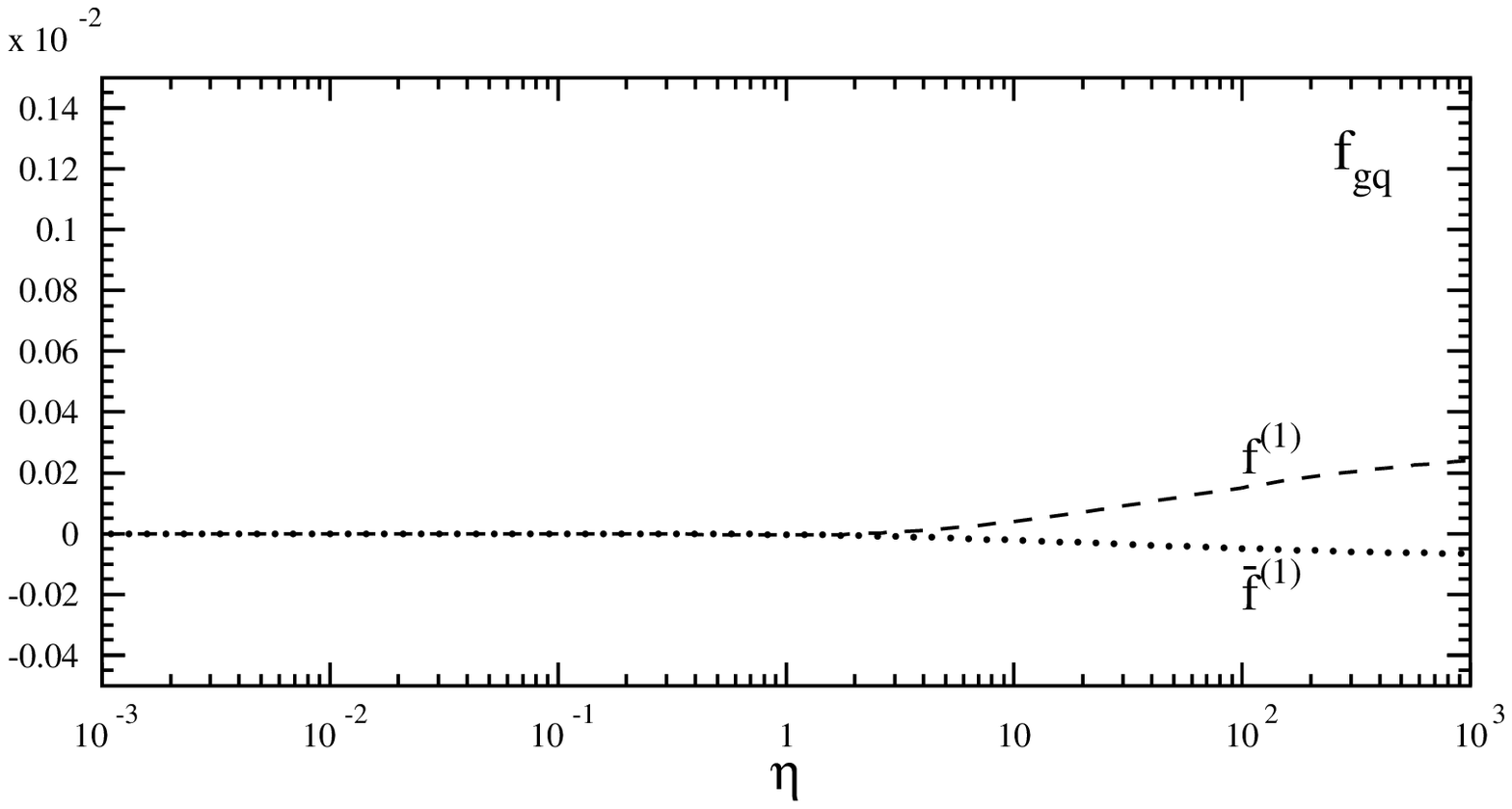,%
        bbllx=45pt,bblly=335pt,bburx=530pt,bbury=560pt,%
        width=14.25cm,height=7cm,angle=0,clip=}

\vspace*{8mm}

\caption[]{The scaling functions $f_{ab}$ for the partonic processes 
         $ab \to t\bar{t}H$ as a function of $\eta = \hat{s}/4\mu_0^2
         - 1$. The notation follows Eq.~\refeq{eq:scaling_fcts}. The
         inserts show the ratio of the analytic threshold
         approximation and the complete numerical calculation. The
         Higgs-boson mass has been set to \MH = 120~GeV.}

\label{fig:scaling}

\end{figure}

The scaling functions are displayed in \reffi{fig:scaling} for the
quark--antiquark, gluon--gluon, and (anti)quark--gluon channels. 
The Higgs-boson mass is set to $\MH = 120\GeV$. As described in
Section~\ref{subsec:threshold}, the parton cross section near the
production threshold ($\eta\approx\beta^2\ll 1$) is dominated by soft-gluon
effects which can be calculated analytically. The inserts in
\reffi{fig:scaling} show the ratio of the threshold approximation, 
$\si_{\rm thres} = \si_{\Coul} + \si_{\log}$, Eqs.~\refeqs{eq:beta}
and \refeq{eq:thrlog}, and the numerically integrated, complete NLO
scaling functions.  It is evident that the complete NLO calculation
approaches the threshold approximation as $\eta\to 0$. However,
because of the $\beta^4$ suppression of the massive three-particle
phase space, see Eq.~\refeq{eq:lo_thres}, the impact of the Coulomb
singularity and the threshold logarithms from soft-gluon radiation is
suppressed. This behaviour is significantly different from
processes with massive two-particle final states, where the phase
space near threshold scales $\propto \beta$ and where soft-gluon
radiation provides a source of large NLO corrections.

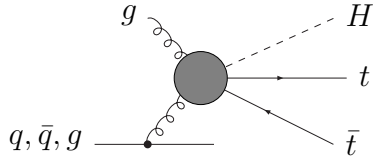
\begin{figure}
\SetScale{0.5}
\centerline{\unitlength 0.5pt 
\begin{picture}(200,120)
\Gluon(50, 95)( 90, 50){4}{4}
\Gluon(50,  0)( 90, 50){4}{4}
\Line(10,  0)(100,  0)
\Vertex(50, 0){3}
\ArrowLine(190,  0)( 90, 50)
\ArrowLine(100, 50)(200, 50)
\DashLine(90,50)(190,95){5}
\GCirc( 90, 50){20}{.5}
\put(200, 90){$H$}
\put(210,44){$t$}
\put(200,-8){$\bar{t}$}
\put(30, 95){$g$}
\put(-55,  0){$q,\bar q,g$}
\end{picture}
}
\caption[]{Generic diagram leading to the high-energy plateau in 
the scaling functions $f^{(1)}_{gg},\bar{f}^{(1)}_{gg}$ and
$f^{(1)}_{gq},\bar{f}^{(1)}_{gq}$.}
\label{fig:plateau}
\end{figure}

At high energies the NLO partonic cross sections in the gluon--gluon
and (anti)quark--gluon channels approach non-zero limits (plateaus)
asymptotically, resulting in a ratio $f_{gg}^{(1)}/f_{gq}^{(1)} \to
2C_{\mathrm{A}}/C_{\mathrm{F}}=9/2$. This has to be contrasted with
the effective scaling behaviour $\propto 1/\hat{s}$ of the LO cross
section in the invariant energy range of the partons far above
threshold, $\sqrt{\hat{s}}\gg \mu_0$ (but at the same time
$\,\ln\sqrt{\hat{s}}\, \dsl{\gg}\, \ln\Mt\,$, so that $t$-channel
top-exchange diagrams are not yet logarithmically enhanced).  The
enhancement of the NLO diagrams is caused by the exchange of soft,
nearly on-shell, space-like gluons, associated with the radiation of a
hard parton from the initial state (see \reffi{fig:plateau}).
This gives rise to a radiative-return phenomenon, causing the
$gg$-fusion subprocess induced by the virtual gluon to effectively
take place at much lower energies, \ie
$\sqrt{\hat{s}_{\mathrm{eff}}}={\cal O}(\mu_0)$. In analogy with the
Drell--Yan process, the cross section for the $q\bar{q}$ channel
vanishes asymptotically in leading and next-to-leading
order. Therefore we expect the hadronic cross sections at the Tevatron
to be affected substantially less by these high-energy plateaus than
the corresponding cross sections at the LHC.

The average CM energy $\langle \sqrt{\hat{s}}\rangle$ for the parton
processes at the Tevatron and LHC colliders is approximately $650\GeV$
and $850\GeV$, respectively, corresponding to $\eta \approx
1-2$. From~\reffi{fig:scaling} we can conclude that in this range of
$\eta$ the NLO corrections at the central scale $\mu_0$ are moderately
negative for the quark--antiquark channel, which dominates the cross
section at the Tevatron, and moderately positive for the gluon--gluon
channel, which is the prominent production mechanism at the LHC. The
(anti)quark--gluon channel is strongly suppressed at both the Tevatron
and the LHC.

\subsection{Hadronic cross sections}

The numerical analyses of the hadronic cross sections have been
performed for the Fermilab Tevatron $p\bar{p}$ collider with a
centre-of-mass energy of $\sqrt{s} = 2$~TeV, and for the CERN LHC with
a $pp$ centre-of-mass energy of $\sqrt{s} = 14$~TeV. The hadronic
cross sections are obtained by convoluting the partonic cross sections
with the parton distribution functions of the initial-state hadrons as
specified in Eq.~\refeq{eq:hadxs}.

\subsubsection{Total cross section and scale dependence}

\begin{figure}
\begin{center}
\epsfig{file=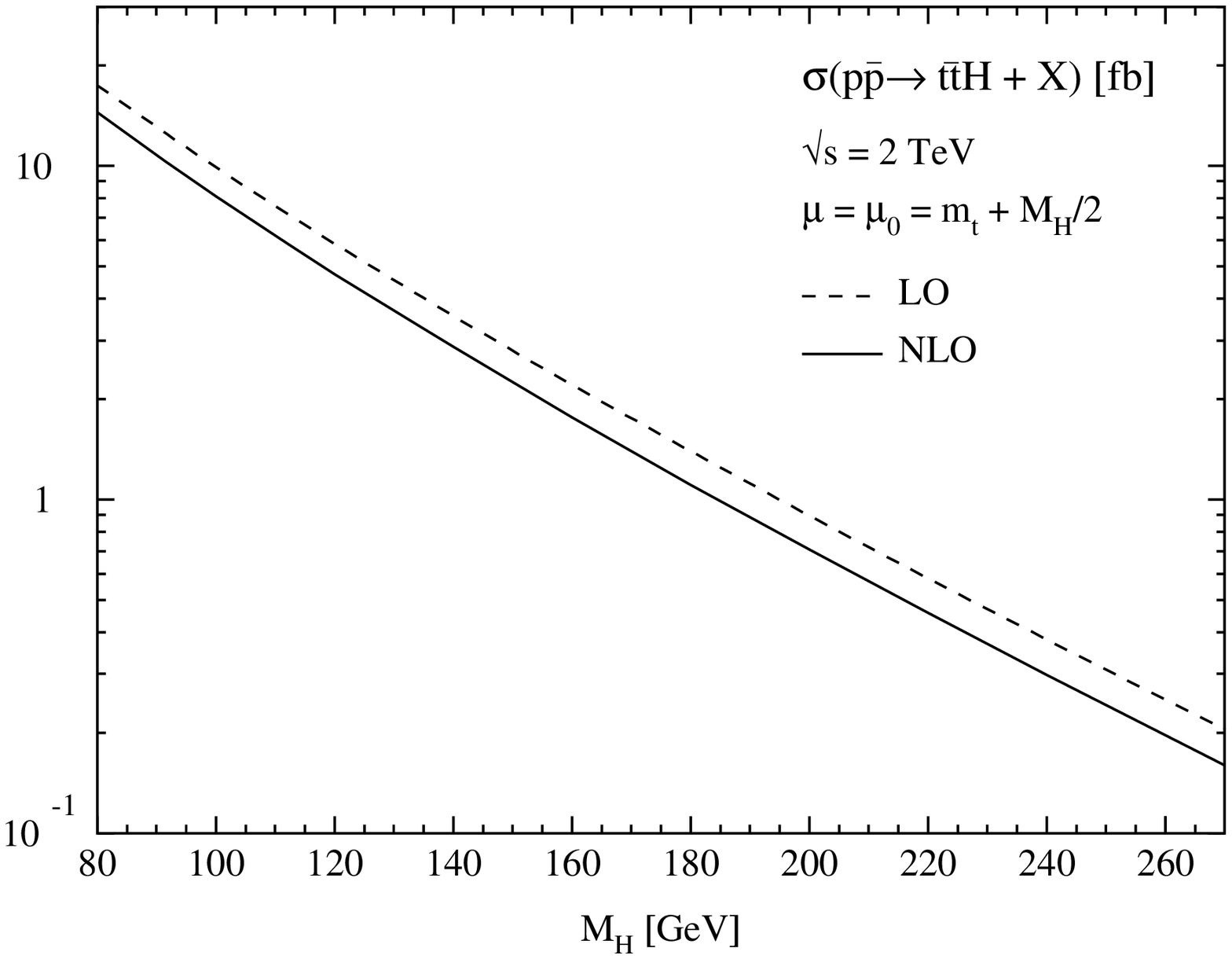,%
        bbllx=30pt,bblly=230pt,bburx=580pt,bbury=625pt,scale=0.7}
\caption[]{Total cross section for $\,p\bar{p}\to t\bar{t}H + X$
 at the Tevatron in LO and NLO approximation, with the renormalization
 and factorization scales set to $\mu_0 = (2m_t+M_H)/2$.}
\label{fig:cs_mh_tev}
\end{center}
\begin{center}
\epsfig{file=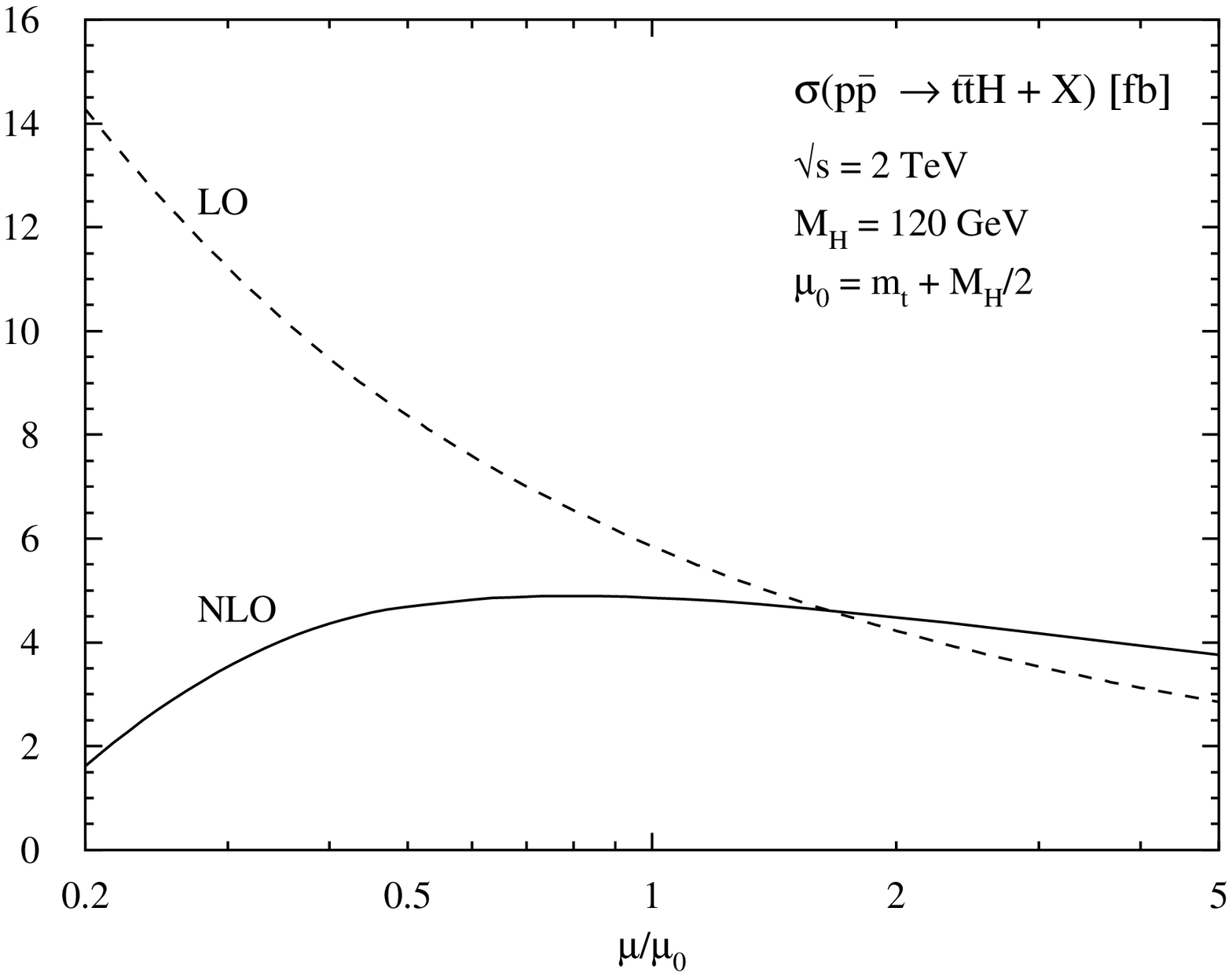,%
        bbllx=30pt,bblly=230pt,bburx=580pt,bbury=625pt,scale=0.7}
\caption[]{Variation of the LO and NLO cross sections with the renormalization 
         and factorization scales for $\,p\bar p\to t\bar tH$ at the
         Tevatron.}
\label{fig:cs_mu_tev}
\end{center}
\end{figure}

The total cross section at the Tevatron is displayed in
\reffis{fig:cs_mh_tev} and \ref{fig:cs_mu_tev}. Representative results
are listed in \refta{tab:totalxs}. For a Higgs mass between $100$ and
$150~\GeV$, the cross section varies between about $10\fb$ and $1\fb$,
if the central value $\mu\to\mu_0=(2m_t+M_H)/2$ is chosen for the
renormalization and factorization scales. In NLO the theoretical
prediction is remarkably stable with very little variation for $\mu$
between $\sim\mu_0/3$ and $\sim 3\mu_0$, in contrast with the Born
approximation, for which the production cross section changes by more
than a factor of 2 within the same interval. It has been checked that
no accidental compensation of scale dependences is introduced by
identifying renormalization and factorization scales. Although the
cross section at the Tevatron is strongly dominated by the $q\bar q$
annihilation channel for scales $\mu\sim\mu_0$, the proper study of
the scale dependence requires the inclusion of the $gg$, $qg$, and
$g\bar q$ channels. If $\mu$ is chosen too low, large logarithmic
corrections spoil the convergence of perturbation theory, and the NLO
cross section would even turn negative for $\mu \lsim \mu_0/5$.

As apparent from \reffi{fig:cs_mu_tev}, the $K$ factor, $K =
\sigma_\NLO/ \sigma_\LO$ with all quantities calculated consistently
in lowest and next-to-leading order, varies from $\sim 0.8$ at the
central scale $\mu = \mu_0$ to $\sim 1.0$ at the threshold scale
$\mu=2\mu_0$. The $K$ factor is nearly independent of $M_H$ in the
relevant Higgs-mass range (see \refta{tab:totalxs}). As argued in
Ref.~\cite{Beenakker:2001rj}, the $K$ factor below unity is in
qualitative agreement with results obtained in the fragmentation
picture proposed in Ref.~\cite{Dawson:1998im}.

\begin{figure}
\begin{center}
\epsfig{file=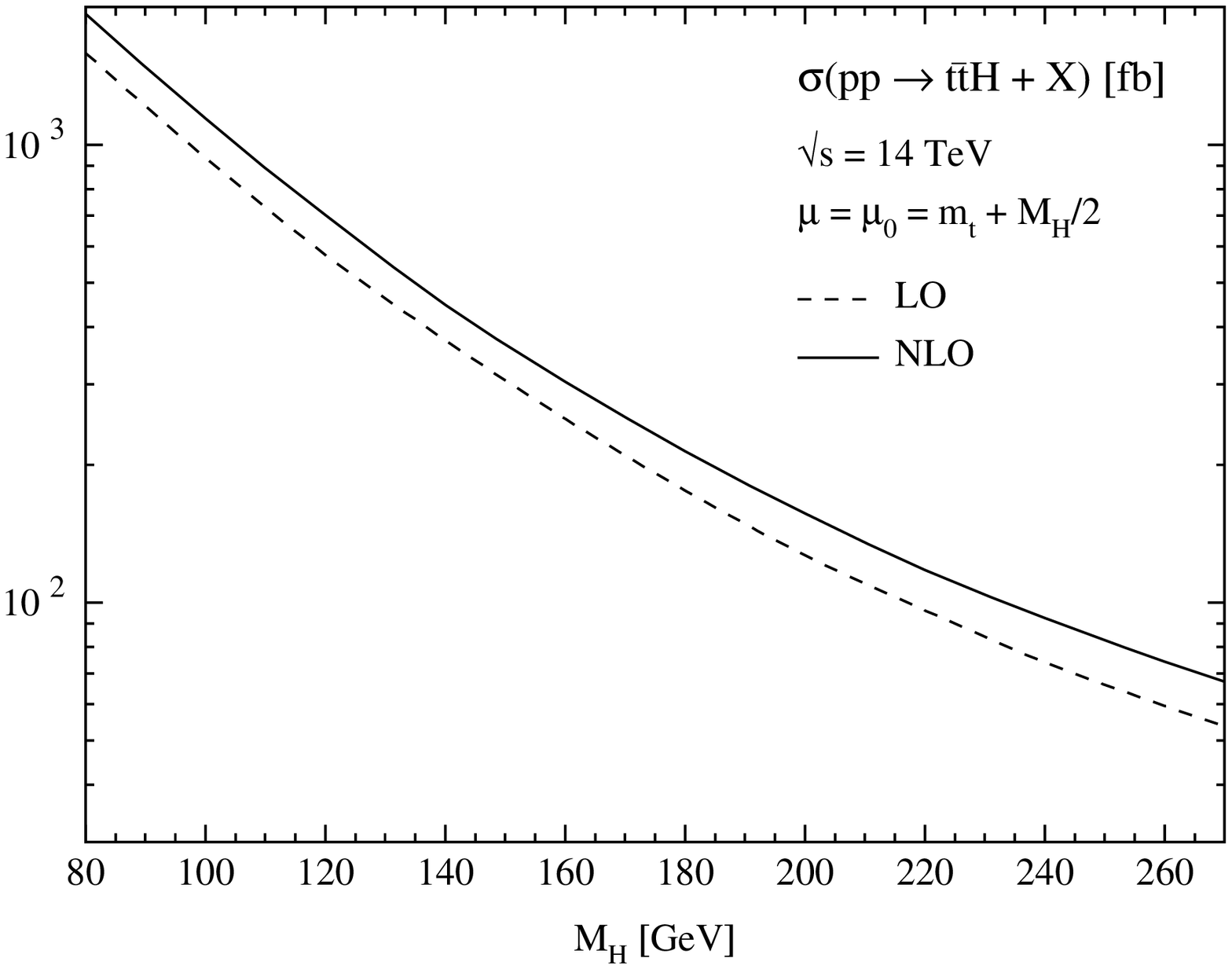,%
        bbllx=30pt,bblly=230pt,bburx=580pt,bbury=625pt,scale=0.7}
\caption[]{Total cross section for $\,pp\to t\bar{t}H + X$
 at the LHC in LO and NLO approximation, with the renormalization
 and factorization scales set to $\mu_0 = (2m_t+M_H)/2$.}
\label{fig:cs_mh_lhc}
\end{center}
\begin{center}
\epsfig{file=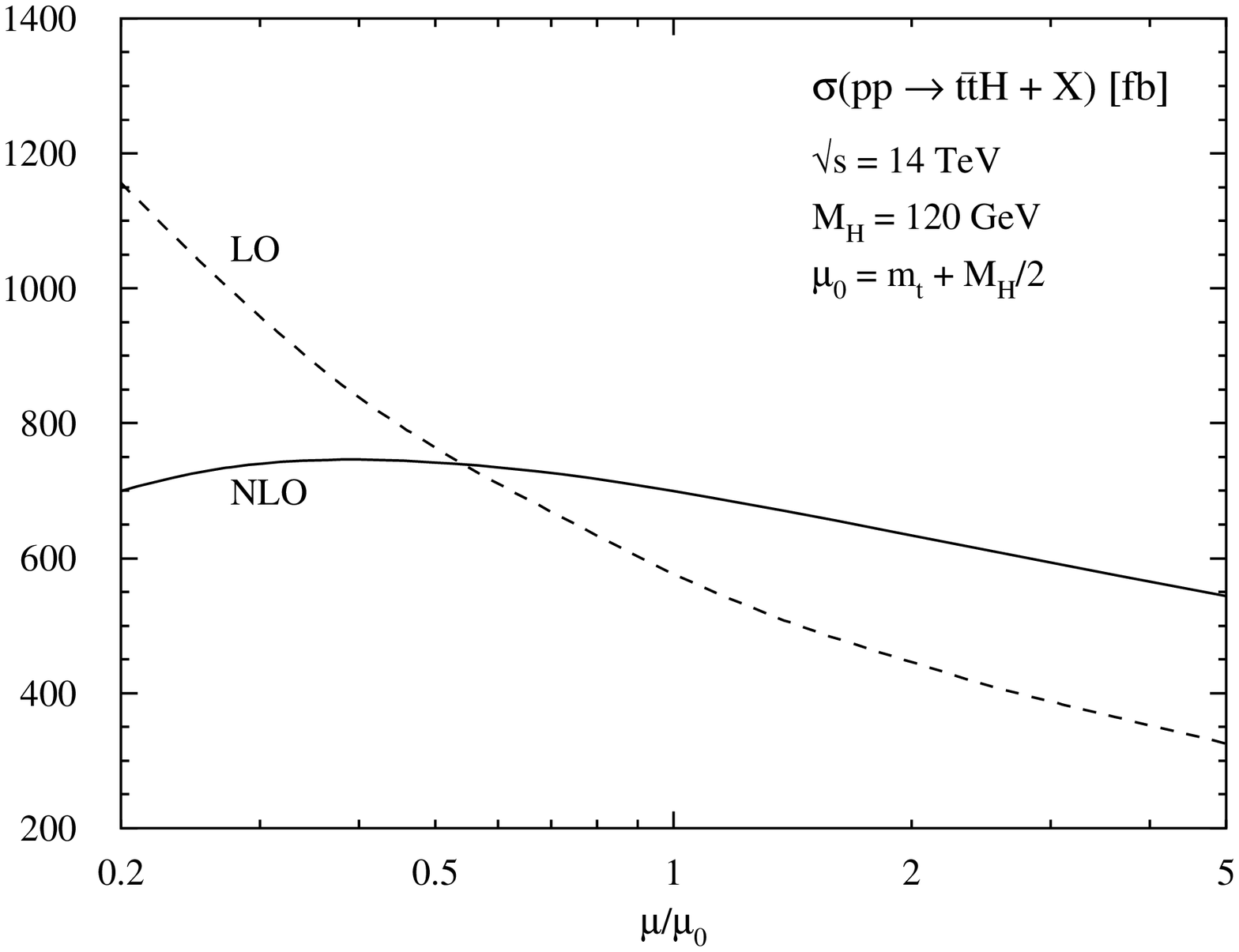,%
        bbllx=30pt,bblly=230pt,bburx=580pt,bbury=625pt,scale=0.7}
\caption[]{Variation of the LO and NLO cross sections with the renormalization
         and factorization scales for $\,pp\to t\bar tH$ at the LHC.}
\label{fig:cs_mu_lhc}
\end{center}
\end{figure}

The cross section at the LHC is dominated by the gluon--gluon process,
which receives positive NLO corrections in the vicinity of the central
renormalization/factorization scale $\mu_0$. At $\mu_0$ we obtain
$K\sim 1.2$, increasing to $\sim 1.4$ at the threshold value
$\mu=2\mu_0$, see \reffis{fig:cs_mh_lhc} and
\ref{fig:cs_mu_lhc}. Again these values are nearly independent of $M_H$ in
the relevant Higgs-mass range and compatible at the qualitative level
with estimates obtained in the fragmentation
picture~\cite{Dawson:1998im}. The NLO corrections significantly reduce
the renormalization and factorization scale dependence and stabilize
the theoretical predictions for the cross section at the LHC.

\begin{table}
\begin{center}
\begin{tabular}{|l||c|c|c|c|c||c|}
\hline
 & & \multicolumn{4}{c||}{\small MRST(LO/NLO)} &
 \rule[-3mm]{0mm}{8mm} {\small CTEQ6} \\ \cline{3-6}
 & \raisebox{1.9ex}[-1.9ex]{ $M_H$~[GeV]} & 
 $\sigma_\LO$~[fb] & $\sigma_\NLO$~[fb] & $K$ &
 $q\bar{q} : gg$~(LO) & \rule[-3mm]{0mm}{8mm} $\sigma_\NLO$~[fb] \\ 
 \hline\hline
         & 120 & 5.846(2) & 4.857(8) & 0.83 & 0.99 : 0.01 & 4.871(8) \\ 
         & 140 & 3.551(1) & 2.925(4) & 0.82 & 0.99 : 0.01 & 2.931(5) \\
\raisebox{1.5ex}[-1.5ex]{Tevatron} 
         & 160 & 2.205(1) & 1.806(2) & 0.82 & 0.99 : 0.01 & 1.808(3)  \\
         & 180 & 1.393(1) & 1.132(1) & 0.81 & 0.996 : 0.004 & 1.133(2) \\\hline
         & 120 & 577.3(4) & 701.5(18) & 1.22 & 0.30 : 0.70 & 665.0(19) \\ 
         & 140 & 373.4(3) & 452.3(12) & 1.21 & 0.33 : 0.67 & 431.8(13) \\
\raisebox{1.5ex}[-1.5ex]{LHC} 
         & 160 & 251.6(2) & 305.6(8) & 1.21 & 0.35 : 0.65 & 290.6(8)\\
         & 180 & 176.0(1) & 214.0(6) & 1.22 & 0.36 : 0.64 & 203.4(5) \\\hline
\end{tabular}
\caption[]{\label{tab:totalxs} Total cross sections and $K$ factors 
         for $\,p\bar p\to t\bar tH$ at the Tevatron and for $\,pp \to
         t\bar tH$ at the LHC. The renormalization and factorization
         scales have been set to $\mu = \mu_0=(2m_t+M_H)/2$. The
         MRST~\cite{Martin:2001es} parton densities have been adopted
         as default, whereas in the rightmost column a comparison is
         performed with the CTEQ6~\cite{Pumplin:2002vw} parton
         densities. The relative weight of $q\bar q$ and $gg$ initial
         states at LO is given in the second-to-last column.}
\end{center}
\end{table}

As argued in Section~\ref{sec:parton}, the impact of the Coulomb
singularity and the threshold logarithms from soft-gluon radiation is
suppressed because of the $\beta^4$ behaviour of the massive
three-particle phase space near threshold. Nevertheless the threshold
region gives us insight into the different colour factors that
determine the NLO corrections at the Tevatron and the LHC.  Just like
the partonic high-energy region, where the fragmentation picture
applies and the high-energy plateaus dominate the NLO corrections,
also the low-energy threshold region hints at $K$ factors at the LHC
that are substantially larger than the ones at the Tevatron (see
Section~\ref{subsec:threshold}).

We have also studied the uncertainty in the cross-section prediction
due to the error in the parametrization of the parton densities. To
this end we have compared the NLO cross section evaluated using the
default MRST~\cite{Martin:2001es} parametrization with the cross
section evaluated using the CTEQ6~\cite{Pumplin:2002vw}
parametrization. The results are collected in the rightmost column of
\refta{tab:totalxs}. Both MRST and CTEQ include a systematic treatment of
parton-distribution-function uncertainties and provide the means for a
quantitative estimate of the corresponding uncertainties in the cross
sections. Using the parton-distribution-error
packages~\cite{Martin:2001es,Pumplin:2002vw}, we find that the
uncertainty due to the parametrization of the parton densities in NLO
is less than $\approx 5\%$ at the Tevatron, where the quark--antiquark
initial states dominate, and less than $\approx 10\%$ at the LHC,
where gluon--gluon initial states are more prominent.

\subsubsection{Differential distributions}

\begin{figure}[H]
\begin{center}
\epsfig{file=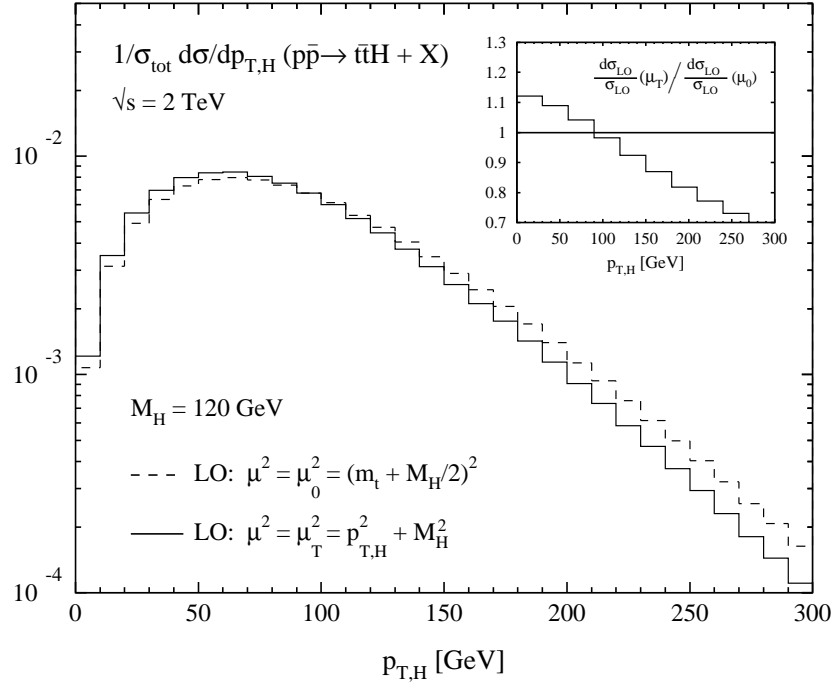,%
        bbllx=30pt,bblly=200pt,bburx=580pt,bbury=635pt,scale=0.6}

\vspace*{10mm}

\epsfig{file=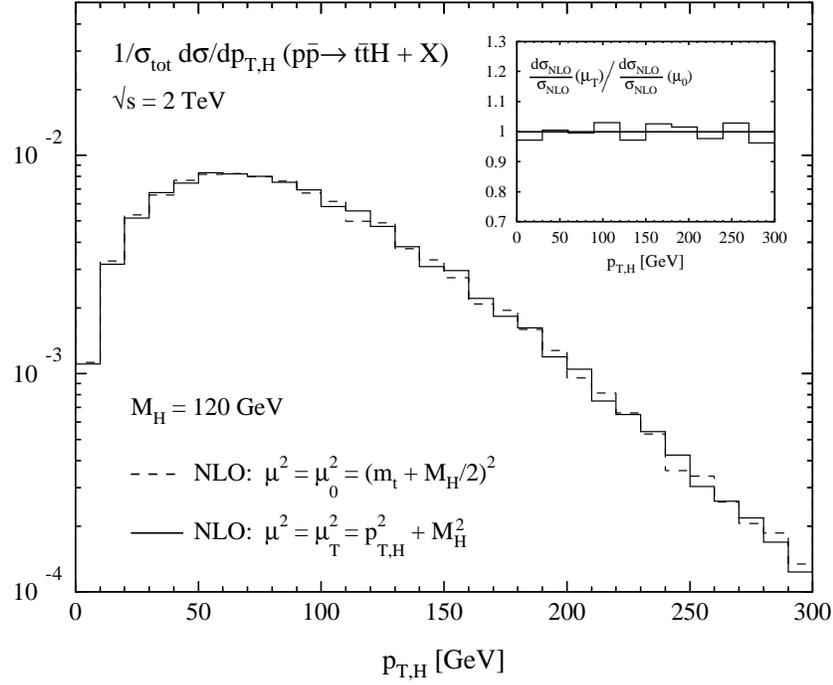,%
        bbllx=30pt,bblly=200pt,bburx=580pt,bbury=635pt,scale=0.6}
\caption[]{Normalized transverse-momentum distribution [$\,GeV^{-1}$] 
 of the Higgs boson for $\,p\bar p\to t\bar{t}H + X$ at the Tevatron 
 in LO (upper figure) and NLO (lower figure) approximation, for two 
 choices of the 
 renormalization and factorization scales: 
 $\mu^2 = \mu_0^2 = (\Mt+\MH/2)^2$ and $\mu^2 = p_{T,H}^2 + \MH^2$.}
\label{fig:ptH-scale}
\end{center}
\end{figure}

\begin{figure}[H]
\begin{center}
\epsfig{file=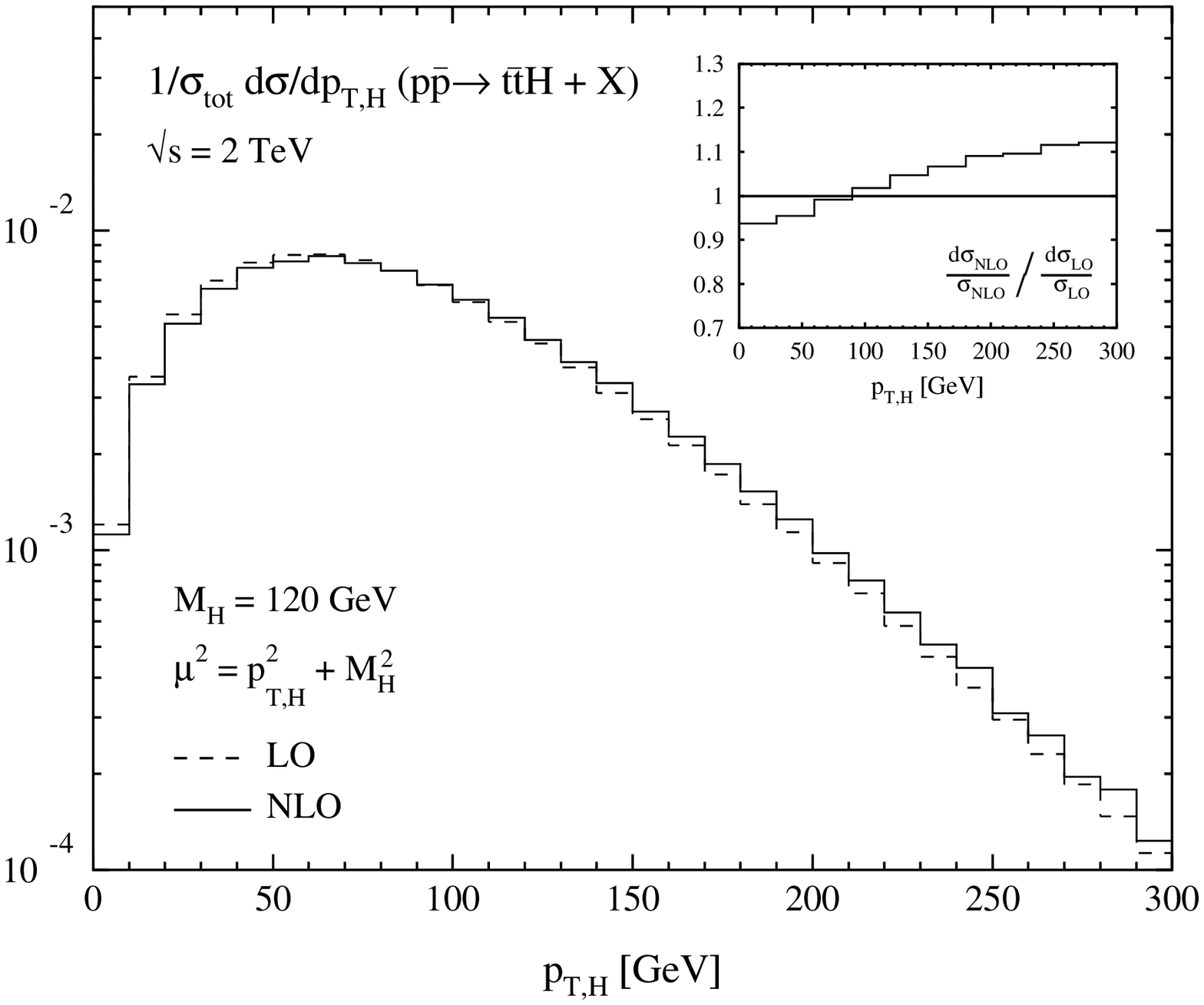,%
        bbllx=30pt,bblly=200pt,bburx=580pt,bbury=635pt,scale=0.6}
\caption[]{Normalized transverse-momentum distribution [$\,GeV^{-1}$] 
 of the Higgs 
 boson for $\,p\bar p\to t\bar{t}H + X$ at the Tevatron in LO and NLO
 approximation, with the renormalization and factorization scales set
 to $\mu^2 = p_{T,H}^2 + \MH^2$.}
\label{fig:ptH-tev}
\end{center}
\begin{center}
\epsfig{file=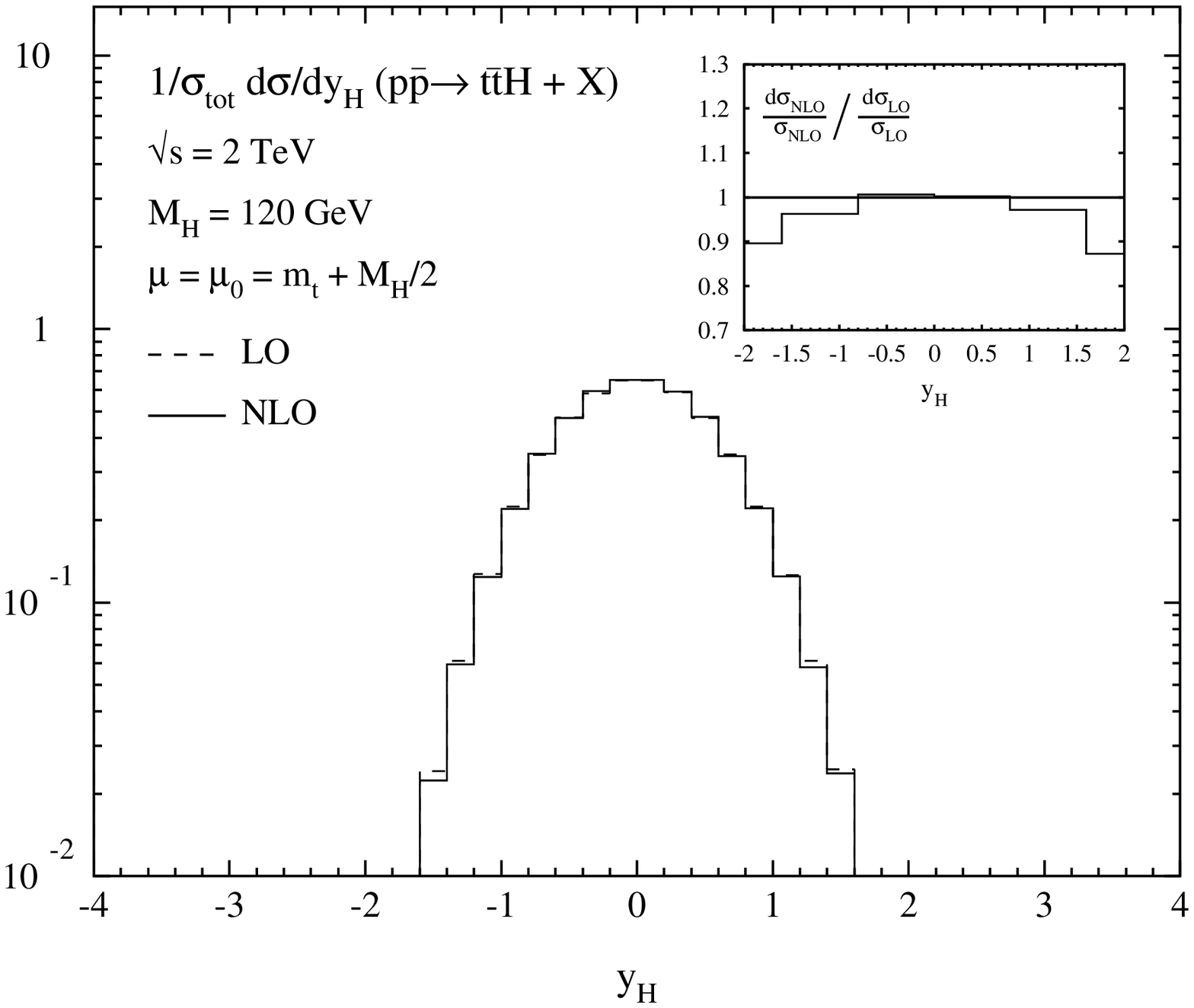,%
        bbllx=30pt,bblly=200pt,bburx=580pt,bbury=635pt,scale=0.6}
\caption[]{Normalized rapidity distribution of the Higgs boson 
 for $\,p\bar p\to t\bar{t}H + X$ at the Tevatron in LO and NLO
 approximation, with the renormalization and factorization scales set
 to $\mu_0 = (2m_t+M_H)/2$.}
\label{fig:yH-tev}
\end{center}
\end{figure}

\begin{figure}[H]
\begin{center}
\epsfig{file=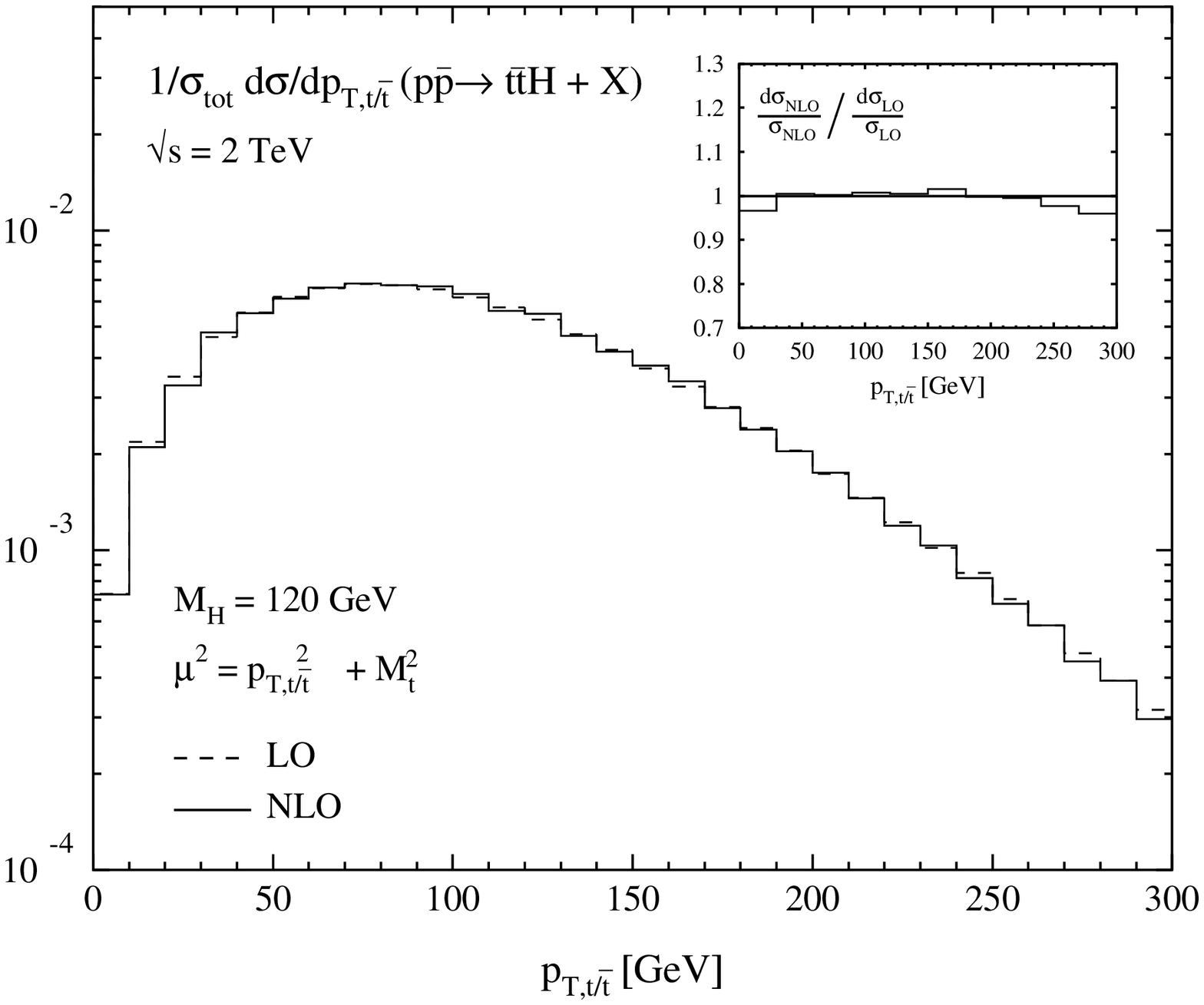,%
        bbllx=30pt,bblly=200pt,bburx=580pt,bbury=635pt,scale=0.6}
\caption[]{Normalized transverse-momentum distribution [$\,GeV^{-1}$] of 
 the top and antitop quarks, $(\mbox{d}\sigma/\mbox{d} p_{T,t} +
 \mbox{d}\sigma/\mbox{d} p_{T,\bar{t}})/2$, for $\,p\bar p\to
 t\bar{t}H + X$ at the Tevatron in LO and NLO approximation, with the
 renormalization and factorization scales set to $\mu^2 = p_{T,t/\bar
 t}^2 + \Mt^2$.}
\label{fig:ptQQb-tev}
\end{center}
\begin{center}
\epsfig{file=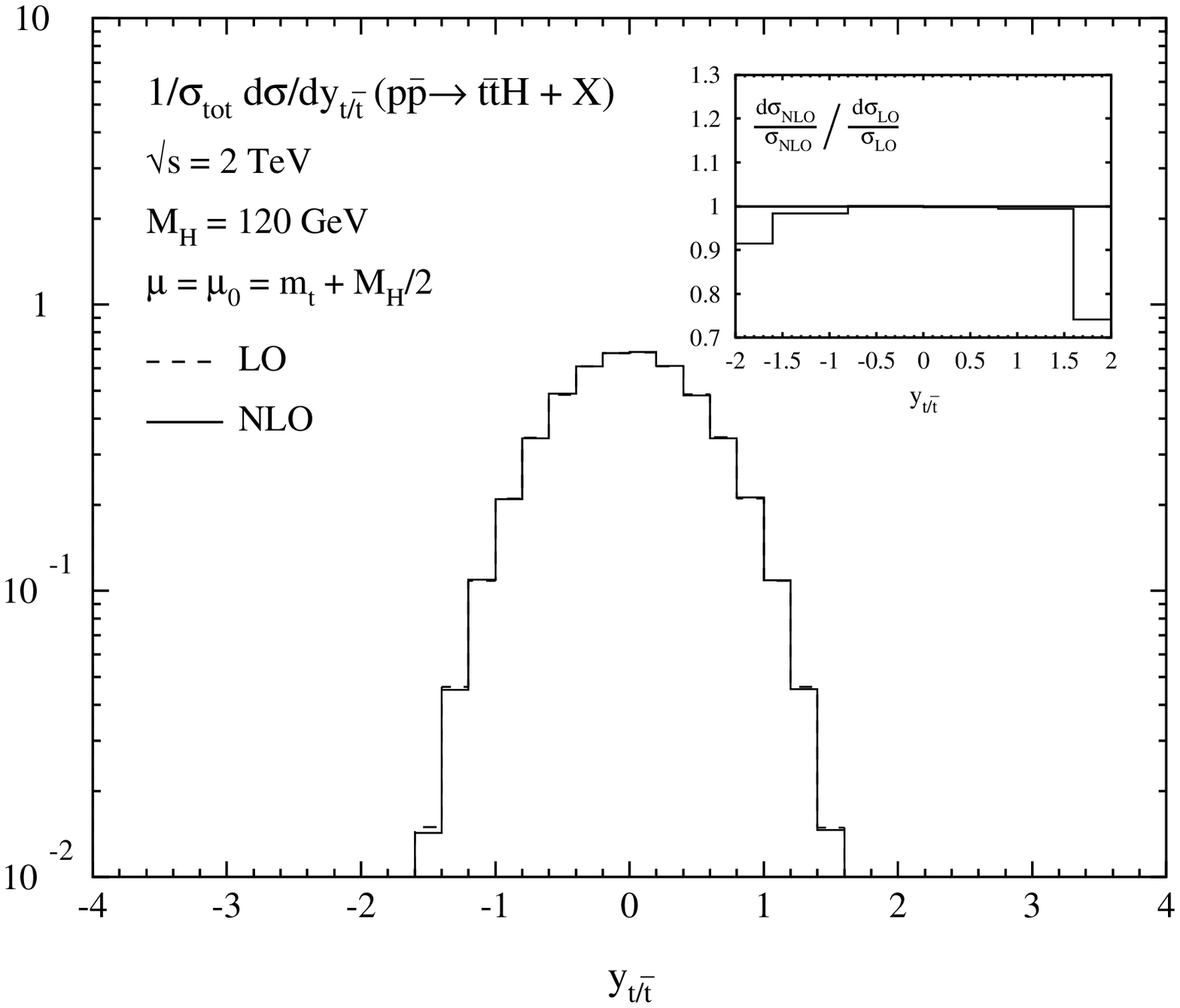,%
        bbllx=30pt,bblly=200pt,bburx=580pt,bbury=635pt,scale=0.6}
\caption[]{Normalized rapidity distribution of the top and antitop quarks, 
 $(\mbox{d}\sigma/\mbox{d} y_{t} + \mbox{d}\sigma/\mbox{d}
 y_{\bar{t}})/2$, for $\,p\bar p\to t\bar{t}H + X$ at the Tevatron in
 LO and NLO approximation, with the renormalization and factorization
 scales set to $\mu_0 = (2m_t+M_H)/2$.}
\label{fig:yQQb-tev}
\end{center}
\end{figure}

In \reffis{fig:ptH-scale}--\ref{fig:yQQb-tev} the normalized
transverse-momentum and rapidity distributions of the Higgs boson and
the top quarks [$t$ and $\bar t$ not discriminated] are shown for
$\,p\bar p\to t\bar tH$ at the Tevatron in LO and NLO
approximation. For the transverse-momentum distributions we choose the
transverse mass as the default renormalization and factorization
scale, i.e.\ $\mu^2 = p_{T,H}^2 + \MH^2$ and $\mu^2 =
p_{T,t/\bar{t}}^2 +
\Mt^2$ for Higgs-boson and top/antitop distributions, respectively,
which provides a more natural choice for large transverse
momenta.\footnote{The integrated cross sections
$\sigma_{\mathrm{tot}}=\int\mbox{d}p_{T}\,\mbox{d}\sigma/
\mbox{d}p_{T}$ reproduce the total cross sections with renormalization and
factorization scales fixed to $\mu_0$ within better than 5\% at NLO.}

The scale dependence of the distributions in rapidity and transverse
momentum of the Higgs boson and top quarks is greatly reduced at NLO.
The improvement is exemplified in \reffi{fig:ptH-scale} which compares
the transverse-momentum distribution of the Higgs boson, once calculated
with the renormalization and factorization scales set to the
transverse mass, and once with both scales fixed to $\mu_0$. We observe
differences of up to $30\%$ at LO, while the shape of the $p_{T,H}$
distribution at NLO is practically identical for the two choices of
scales.

The ratio of the (normalized) NLO and LO transverse-momentum
distributions of the Higgs boson, displayed in the insert in
\reffi{fig:ptH-tev}, reveals that the simple rescaling of the LO
$p_{T,H}$ distribution by a constant $K$ factor reproduces the NLO
distribution only within $\pm 10\%$ in the relevant
transverse-momentum range, if the transverse mass is chosen as the
scale. The increase of the ratio
$\mbox{d}\sigma_{\NLO}/\mbox{d}\sigma_{\LO}$ with increasing $p_{T,H}$
can be understood in terms of the scale dependences of
$\mbox{d}\sigma_{\NLO}$ and $\mbox{d}\sigma_{\LO}$: with $\mu^2 =
p_{T,H}^2 + \MH^2$ rising with increasing $p_{T,H}$,
$\mbox{d}\sigma_{\LO}$ decreases with increasing $\mu$ faster than
$\mbox{d}\sigma_{\NLO}$ (see~\reffi{fig:cs_mu_tev}). We note that
choosing a fixed scale for the LO $p_{T,H}$ distribution provides a
better description of the NLO shape.

For the transverse-momentum distribution of the top quarks, the
increase of the ratio $\mbox{d}\sigma_{\NLO}/\mbox{d}\sigma_{\LO}$
with increasing $p_{T,H}$ due to the scale dependence is balanced by
gluon radiation off top quarks at NLO, which reduces the energies and
momenta of the top particles and thereby depletes the region of large
transverse momenta. As a net result, the shape of the
transverse-momentum distribution of the top quarks is barely affected
by the NLO corrections in the relevant range $p_{T,t/\bar{t}}\,
\lsim\, 300\GeV$, if the transverse top-quark mass is adopted for 
the renormalization and factorization scales
(see~\reffi{fig:ptQQb-tev}).

The reduction in energies and momenta of the massive final-state
particles due to gluon radiation at NLO is also expected to enhance
the central rapidity region. However, as evident from
\reffis{fig:yH-tev} and \ref{fig:yQQb-tev}, the effect on the
(normalized) rapidity distributions of the massive final-state
particles is small and becomes noticeable at the edges of phase space
only.

\begin{figure}[H]
\begin{center}
\epsfig{file=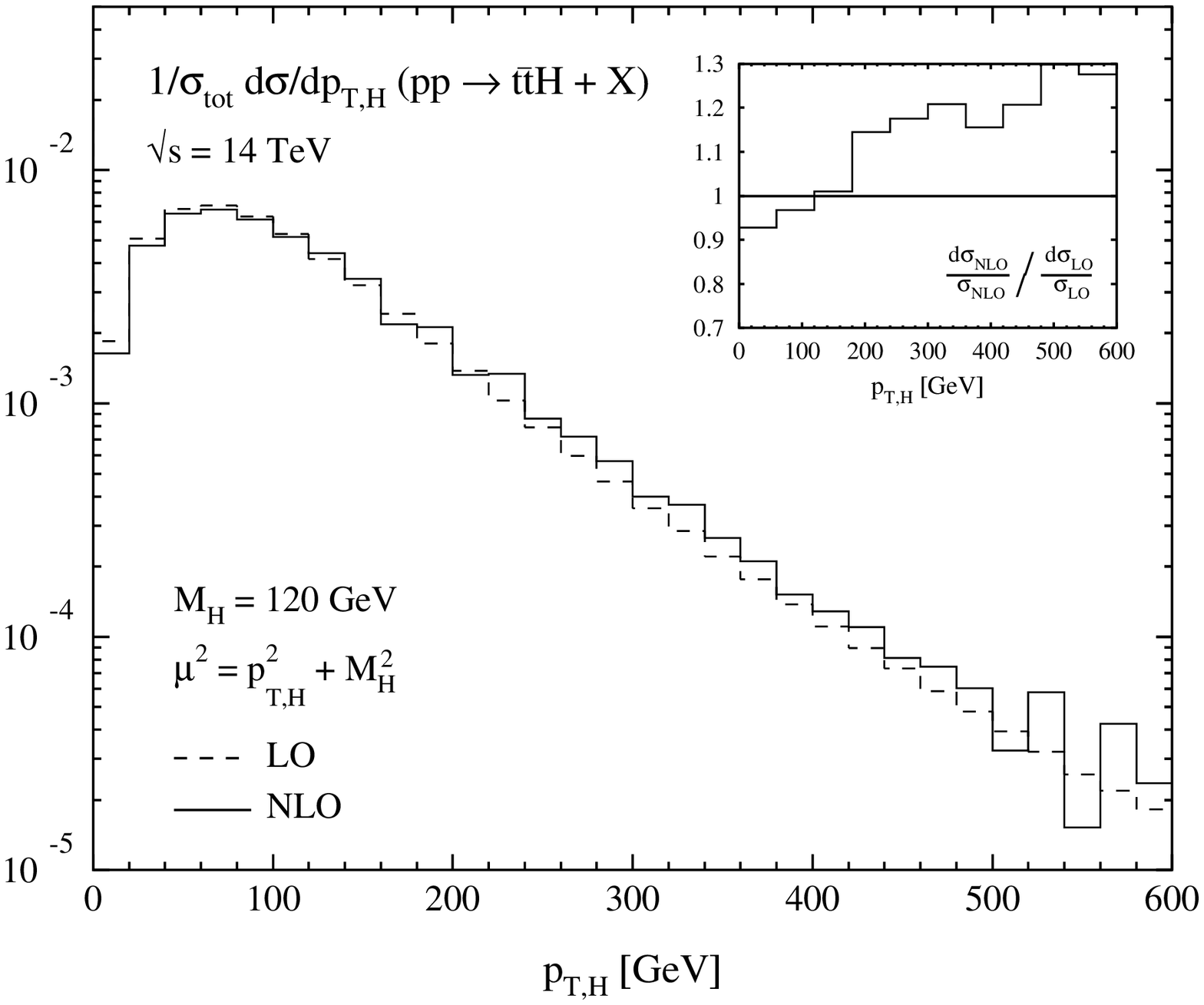,%
        bbllx=30pt,bblly=200pt,bburx=580pt,bbury=635pt,scale=0.6}
\caption[]{Normalized transverse-momentum distribution [$\,GeV^{-1}$] 
 of the Higgs 
 boson for $\,p p\to t\bar{t}H + X$ at the LHC in LO and NLO
 approximation, with the renormalization and factorization scales set
 to $\mu^2 = p_{T,H}^2 + \MH^2$.}
\label{fig:ptH-lhc}
\end{center}
\begin{center}
\epsfig{file=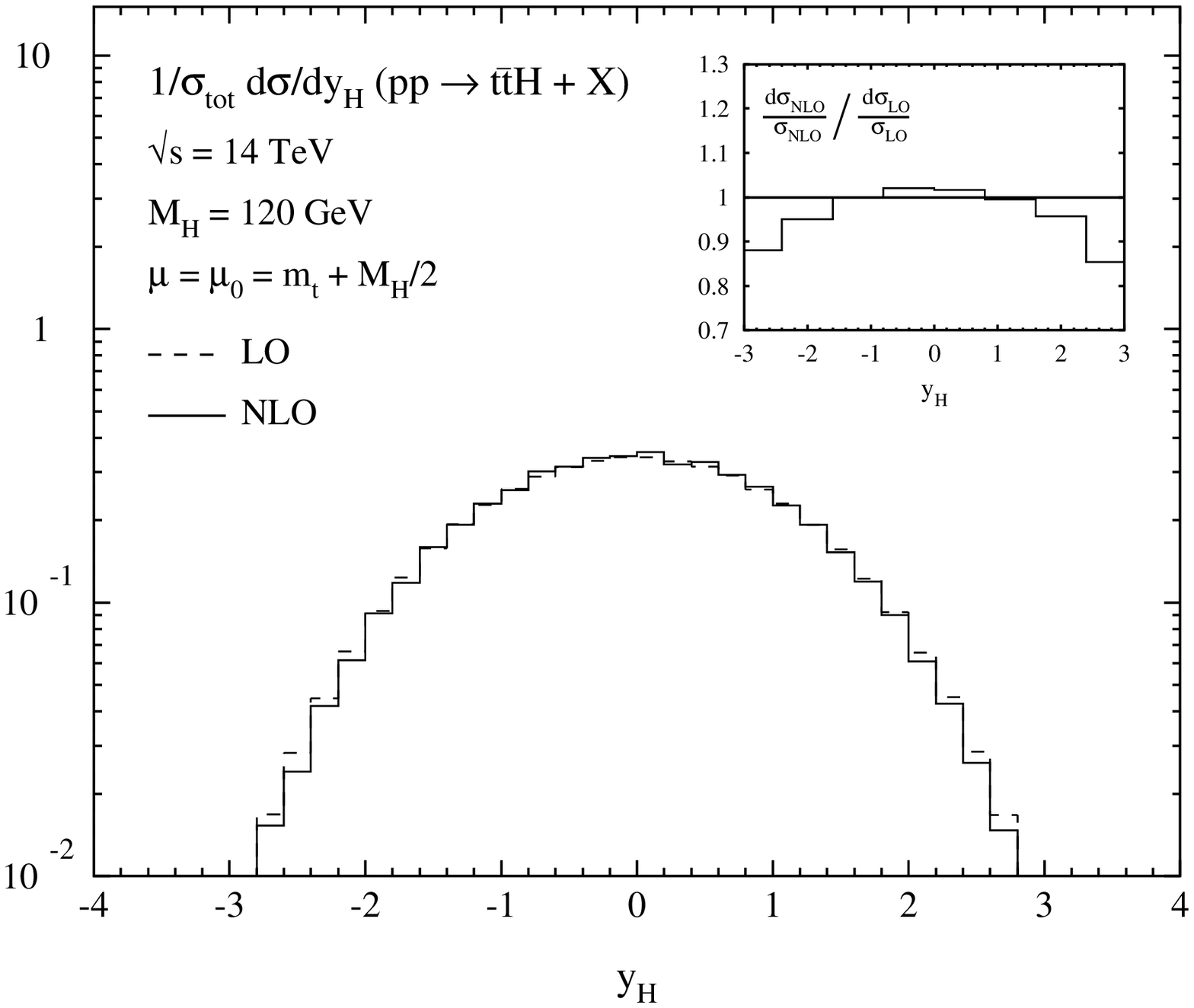,%
        bbllx=30pt,bblly=200pt,bburx=580pt,bbury=635pt,scale=0.6}
\caption[]{Normalized rapidity distribution of the Higgs boson 
 for $\,p p\to t\bar{t}H + X$ at the LHC in LO and NLO approximation,
 with the renormalization and factorization scales set to $\mu_0 =
 (2m_t+M_H)/2$.}
\label{fig:yH-lhc}
\end{center}
\end{figure}

\begin{figure}[H]
\begin{center}
\epsfig{file=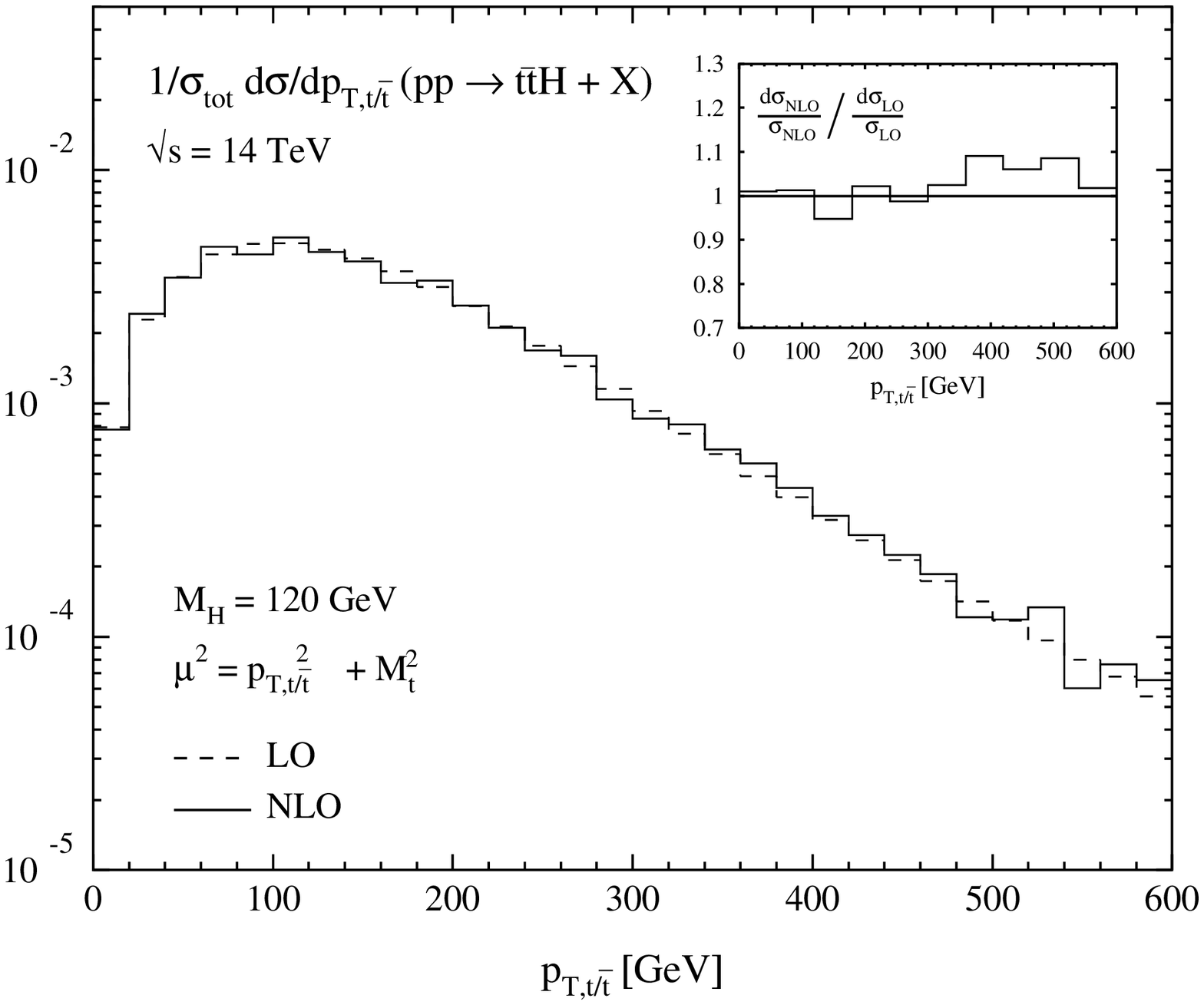,%
        bbllx=30pt,bblly=200pt,bburx=580pt,bbury=635pt,scale=0.6}
\caption[]{Normalized transverse-momentum distribution [$\,GeV^{-1}$] of 
 the top and antitop quarks, $(\mbox{d}\sigma/\mbox{d} p_{T,t} +
 \mbox{d}\sigma/\mbox{d} p_{T,\bar{t}})/2$, for $\,p p\to t\bar{t}H +
 X$ at the LHC in LO and NLO approximation, with the renormalization
 and factorization scales set to $\mu^2 = p_{T,t/\bar t}^2 + \Mt^2$.}
\label{fig:ptQQb-lhc}
\end{center}
\begin{center}
\epsfig{file=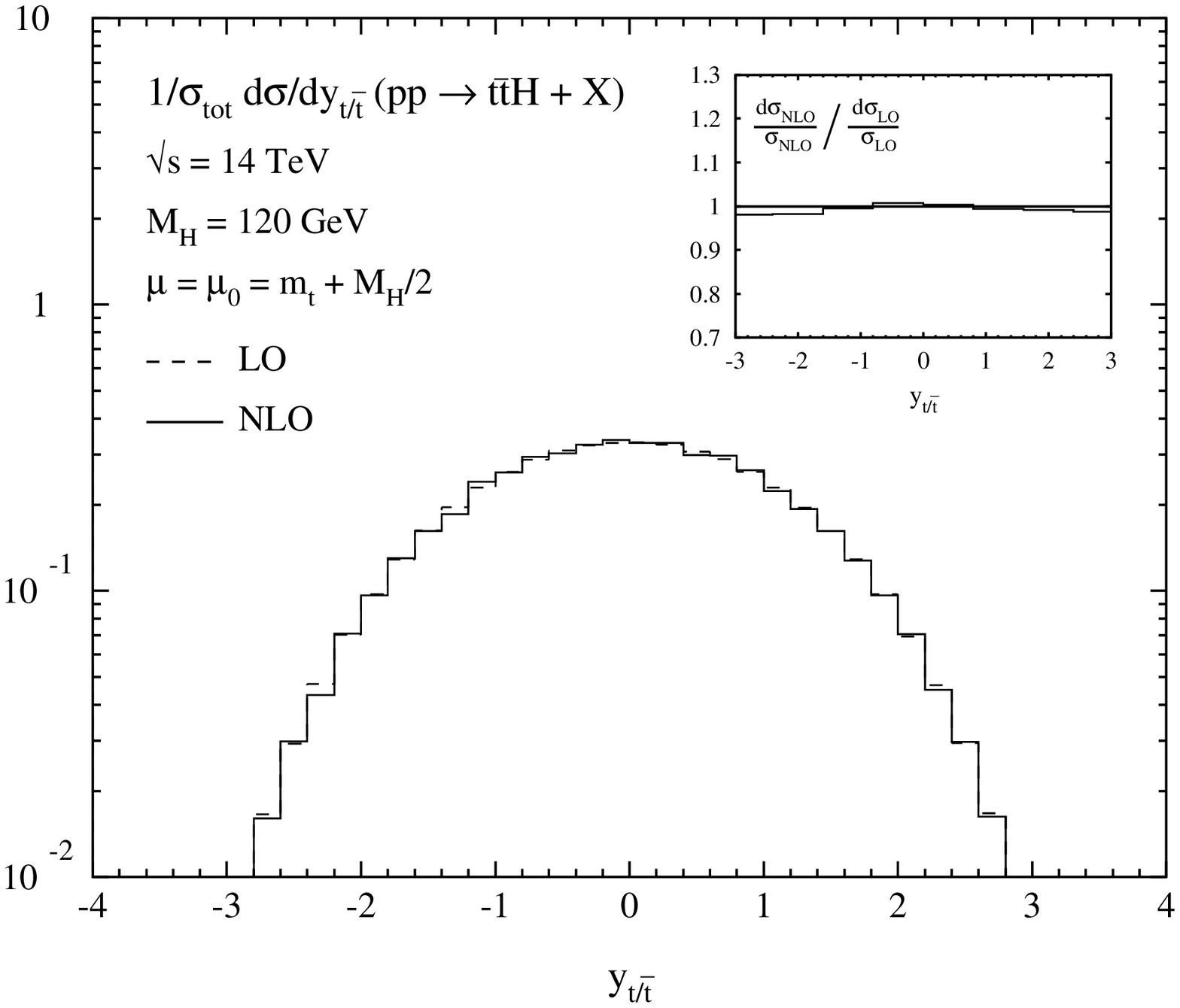,%
        bbllx=30pt,bblly=200pt,bburx=580pt,bbury=635pt,scale=0.6}
\caption[]{Normalized rapidity distribution of the top and antitop quarks, 
 $(\mbox{d}\sigma/\mbox{d} y_{t} + \mbox{d}\sigma/\mbox{d}
 y_{\bar{t}})/2$, for $\,p p\to t\bar{t}H + X$ at the LHC in LO and
 NLO approximation, with the renormalization and factorization scales
 set to $\mu_0 = (2m_t+M_H)/2$.}
\label{fig:yQQb-lhc}
\end{center}
\end{figure}

\textit{Cum grano salis}, the transverse-momentum and rapidity 
distributions of the Higgs boson and the top quarks for $\,p p\to
t\bar tH$ at the LHC, \reffis{fig:ptH-lhc}--\ref{fig:yQQb-lhc}, show
the same qualitative behaviour as observed for the Tevatron. Adopting
the Higgs transverse mass as the scale, the difference between the NLO
and LO Higgs-boson transverse-momentum distributions at the LHC is
naturally more pronounced, with deviations of up to $30\%$ at large
$p_{T,H}$. As for the Tevatron, choosing a fixed scale for the LO
$p_{T,H}$ distribution provides a better description of the NLO shape.

\section{Summary}

In this report we have presented the theoretical analysis 
of the Standard-Model processes $p\bar{p}/pp \to t\bar{t}H+X$ in
next-to-leading order QCD at the Tevatron and the LHC. We have focussed 
on two aspects.

{\bf (i)} The technical basis for the prediction of the cross sections
has been elaborated in detail. Particular emphasis has been put on
evaluating pentagon diagrams, which involves the isolation of the
singularities, the reduction to lower $n$-point functions, and the
stable integration over the phase space. As a second important
technical element we have applied the dipole subtraction formalism to
treat massive final-state particles.  In fact, the processes reported
on here, represent the first complex examples that have been treated
in this formalism in the case of massive quarks.

{\bf (ii)} In expanding on results presented in an earlier
letter, \citere{Beenakker:2001rj}, we have analyzed not only the total
cross sections for Higgs bremsstrahlung off top quarks at the Tevatron
and the LHC, but also final-state rapidity and transverse-momentum
distributions for the Higgs boson and the top quarks. Moreover, we
have studied the uncertainty in the next-to-leading order cross section 
prediction due to different parametrizations of the parton densities.

When including higher-order QCD corrections, the $K$ factor at the
Tevatron has a value slightly below unity. It varies between $\sim
0.8$ and $\sim 1.0$ for renormalization and factorization scales
between $\mu=\mu_0$ and $2\mu_0$, almost independent of the
Higgs-boson mass in the intermediate range. Here $2\mu_0=2\Mt+\MH$
denotes the threshold CM energy of the partonic subprocesses.
Similarly, the $K$ factor varies between $\sim 1.2$ and $\sim 1.4$ for
the same scales at the LHC. Most importantly, the NLO predictions
for the total cross sections and for the distributions in rapidity and
transverse momentum of the Higgs boson and top quarks are stable when
the renormalization and factorization scales are varied, in contrast
to the Born approximation. The improved cross sections can therefore
serve as a solid base for experimental analyses at the Tevatron and
the LHC.

\section*{Acknowledgement}
The NLO analyses presented in this report have been approached in
parallel by S.~Dawson, L.~Orr, L.~Reina and D. Wackeroth. The results
for the total cross sections have been cross-checked by the two
groups, and perfect agreement has been found. We are grateful to these
authors for the cooperation.

\appendix
\section*{Appendix}

\section{The IR-divergent scalar 3- and 4-point integrals}
\label{app:C0D0}

In this first appendix all relevant IR-divergent scalar
3- and 4-point integrals are listed. They have
been derived in two independent ways, once with the help of Feynman
parametrization, once by means of cutting techniques and dispersion 
integrals. 
In the following the momenta $p_i$ are defined
as in \refse{se:LOconv}, \ie they are on shell:
$p_1^2=p_2^2=0$, $p_3^2=p_4^2=\Mt^2$, $p_5^2=\MH^2$. 
Throughout this appendix we make extensive use of the quantities
\beq
\beta_r = \sqrt{1-\frac{4\Mt^2}{\bar r}},
\qquad x_r = \frac{\beta_r-1}{\beta_r+1},
\eeq
where $r$ represents a generic invariant. Moreover, we adopt 
the convention that invariants with a bar
receive an infinitesimally small, positive imaginary part:
$\bar r=r+\ri 0$.

There are four different IR-divergent 3-point functions that become
relevant. In $D\neq 4$ dimensions they read
\beqar
C^{(D)}_0(p_1,-p_2,0,0,0)      &=& \frac{1}{\hat{s}} \Biggl[ 
  \Delta^{\IR}_2(\mu)
  -\Delta^{\IR}_1(\mu)\ln\biggl(\frac{{}-\bar{\hat{s}}\,}{\Mt^2}\biggr)
  +\frac{1}{2}\ln^2\biggl(\frac{{}-\bar{\hat{s}}\,}{\Mt^2}\biggr)
  -\frac{\pi^2}{6} \Biggr], 
\nn\\[.5em]
C^{(D)}_0(p_1,p_1-p_3,0,0,\Mt)     &=& \frac{1}{t_{13}-\Mt^2} \Biggl[ 
\frac{1}{2}\Delta^{\IR}_2(\mu)
  -\Delta^{\IR}_1(\mu)\ln\biggl(1-\frac{\bar t_{13}}{\Mt^2}\biggr)
  + \Li\biggl( \frac{\bar t_{13}}{\Mt^2} \biggr) 
\nn\\ && \qquad {}
  + \ln^2\biggl(1-\frac{\bar t_{13}}{\Mt^2}\biggr) \Biggr], 
\nn\\[.5em]
C^{(D)}_0(p_1,p_4+p_5,0,0,\Mt) &=& \frac{1}{s_{45}-t_{23}} \Biggl[ 
  \Delta^{\IR}_1(\mu)\ln\biggl(\frac{\bar t_{23}-\Mt^2}
                                    {\bar s_{45}-\Mt^2}\biggr)
  + \Li\biggl( \frac{\bar s_{45}}{\Mt^2} \biggr)
  - \Li\biggl( \frac{\bar t_{23}}{\Mt^2} \biggr) 
\nn\\ && \qquad {}
  + \ln^2\biggl(1-\frac{\bar s_{45}}{\Mt^2}\biggr) 
  - \ln^2\biggl(1-\frac{\bar t_{23}}{\Mt^2}\biggr) \Biggr], 
\nn\\[.5em]
C^{(D)}_0(p_3,-p_4,0,\Mt,\Mt) &=& \frac{1}{s_{34}\beta_{s_{34}}} \Biggl[ 
  \Delta^{\IR}_1(\mu)\ln(x_{s_{34}}) - 2\Li({}-x_{s_{34}})
\nn\\ && \qquad {}
  - 2\ln(x_{s_{34}})\,\ln(1+x_{s_{34}}) + \frac{1}{2}\,\ln^2(x_{s_{34}}) 
  - \frac{\pi^2}{6} \Biggr]. 
\label{eq:C0dim}
\eeqar

We also present these 3-point functions with mass regularization, 
where all internal massless particles are endowed with an infinitesimally
small mass $\la$:
\beqar
C^{(\mathrm{mass},D=4)}_0(p_1,-p_2,\la,\la,\la)      &=& 
\frac{1}{2\hat{s}} \ln^2\biggl(\frac{{}-\bar{\hat{s}}\,}{\la^2}\biggr),
\nn\\[.5em]
C^{(\mathrm{mass},D=4)}_0(p_1,p_1-p_3,\la,\la,\Mt)     &=& 
\frac{1}{t_{13}-\Mt^2} \Biggl[ 
    \ln^2\biggl(\frac{\Mt^2-\bar t_{13}}{\la\Mt}\biggr)
  + \Li\biggl( \frac{\bar t_{13}}{\Mt^2} \biggr) + \frac{\pi^2}{12}
\Biggr], 
\nn\\[.5em]
C^{(\mathrm{mass},D=4)}_0(p_1,p_4+p_5,\la,\la,\Mt) &=& 
\frac{1}{s_{45}-t_{23}} \Biggl[ 
    \ln^2\biggl(\frac{\Mt^2-\bar s_{45}}{\la\Mt}\biggr) 
  - \ln^2\biggl(\frac{\Mt^2-\bar t_{23}}{\la\Mt}\biggr)
\nn\\ && \qquad {}
  + \Li\biggl( \frac{\bar s_{45}}{\Mt^2} \biggr)
  - \Li\biggl( \frac{\bar t_{23}}{\Mt^2} \biggr)  
\Biggr], 
\nn\\[.5em]
C^{(\mathrm{mass},D=4)}_0(p_3,-p_4,\la,\Mt,\Mt) &=& 
\frac{1}{s_{34}\beta_{s_{34}}} \Biggl[ 
    \ln\biggl(\frac{\la^2}{\Mt^2}\biggr)\ln(x_{s_{34}}) - 2\Li({}-x_{s_{34}})
\nn\\ && \qquad {}
  - 2\ln(x_{s_{34}})\,\ln(1+x_{s_{34}}) + \frac{1}{2}\,\ln^2(x_{s_{34}}) 
  - \frac{\pi^2}{6} 
\Biggr]. \nn\\
\label{eq:C0mass}
\eeqar

Taking into account that the decomposition of the 5-point integrals 
leads to additional 4-point integrals, six different IR-divergent 
4-point integrals are needed: 
\beqar
\lefteqn{ D^{(D)}_0(p_1,p_1-p_3,-p_2,0,0,\Mt,0) 
= \frac{1}{\hat{s}(t_{13}-\Mt^2)}
} \quad
\nn\\*
&& {} \times\Biggl[ 
  \frac{3}{2}\Delta^{\IR}_2(\mu) 
  - \Delta^{\IR}_1(\mu)\biggl\{ 2 \ln\biggl(1-\frac{\bar t_{13}}{\Mt^2}\biggr) 
  + \ln\biggl(\frac{{}-\bar{\hat{s}}\,}{\Mt^2}\biggr) 
  - \ln\biggl(1-\frac{\bar s_{45}}{\Mt^2}\biggr) \biggr\} 
\nn\\ && \qquad {}
  - 2\Li\biggl( \frac{\bar t_{13}-\bar s_{45}}{\bar t_{13}-\Mt^2} \biggr)
  + 2 \ln\biggl(\frac{{}-\bar{\hat{s}}\,}{\Mt^2}\biggr) 
      \ln\biggl(1-\frac{\bar t_{13}}{\Mt^2}\biggr) 
  - \ln^2\biggl(1-\frac{\bar s_{45}}{\Mt^2}\biggr) 
  - \frac{2\pi^2}{3} \Biggr], 
\hspace{2em}
\nn\\[.5em]
\lefteqn{ D^{(D)}_0(p_1,p_1-p_3,p_4+p_5,0,0,\Mt,\Mt) 
= \frac{1}{(t_{13}-\Mt^2)(t_{23}-\Mt^2)}
} \quad
\nn\\*
&& {} \times \Biggl[ \frac{1}{2}\Delta^{\IR}_2(\mu) 
  - \Delta^{\IR}_1(\mu)\biggl\{ \ln\biggl(1-\frac{\bar t_{13}}{\Mt^2}\biggr) 
  + \ln\biggl(1-\frac{\bar t_{23}}{\Mt^2}\biggr) 
  - \ln\biggl(1-\frac{\bar s_{45}}{\Mt^2}\biggr) \biggr\} 
\nn\\ && \qquad {}
  - 2\Li\biggl( \frac{\bar t_{23}-\bar s_{45}}{\bar t_{23}-\Mt^2} \biggr)
  - 2\Li\biggl( \frac{\bar t_{13}-\bar s_{45}}{\bar t_{13}-\Mt^2} \biggr)
  + 2\ln\biggl(1-\frac{\bar t_{13}}{\Mt^2}\biggr) 
     \ln\biggl(1-\frac{\bar t_{23}}{\Mt^2}\biggr) 
\nn\\ && \qquad {}
  - \ln^2\biggl(1-\frac{\bar s_{45}}{\Mt^2}\biggr) - \frac{\pi^2}{6} \Biggr], 
\nn\\[.5em]
\lefteqn{ D^{(D)}_0(p_1,p_1-p_3,p_4,0,0,\Mt,\Mt) = 
\frac{1}{(t_{13}-\Mt^2)(t_{14}-\Mt^2)}
} \quad
\nn\\*
&& {} \times \Biggl[ \Delta^{\IR}_2(\mu)
  - \Delta^{\IR}_1(\mu)\biggl\{ \ln\biggl(1-\frac{\bar t_{14}}{\Mt^2}\biggr) 
    + \ln\biggl(1-\frac{\bar t_{13}}{\Mt^2}\biggr) \biggr\} - \ln^2(x_{t_{25}})  
\nn\\ && \qquad {}
  + 2 \ln\biggl(1-\frac{\bar t_{13}}{\Mt^2}\biggr) 
      \ln\biggl(1-\frac{\bar t_{14}}{\Mt^2}\biggr) 
  - \frac{2\pi^2}{3} \Biggr], 
\nn\\[.5em]
\lefteqn{ D^{(D)}_0(p_1,p_1-p_3,p_4-p_2,0,0,\Mt,\Mt)  =
\frac{1}{(t_{13}-\Mt^2)(s_{35}-\Mt^2)}
} \quad
\nn\\*
&& {} \times \Biggl[ \frac{1}{2}\Delta^{\IR}_2(\mu)
  - \Delta^{\IR}_1(\mu)\biggl\{ \ln\biggl(1-\frac{\bar t_{13}}{\Mt^2}\biggr) 
  + \ln\biggl(1-\frac{\bar s_{35}}{\Mt^2}\biggr)
  - \ln\biggl(1-\frac{\bar t_{24}}{\Mt^2}\biggr) \biggr\} 
\nn\\ && \qquad {}
  - \sum_{\rho=\pm 1}\cLi\biggl(\frac{\bar t_{24}-\Mt^2}{\bar t_{13}-\Mt^2},
               x_{\MH^2}^{\,\rho}\biggr)
  - 2\,\Li\biggl(\frac{\bar s_{35}-\bar t_{24}}{\bar s_{35}-\Mt^2} \biggr)
  + 2 \ln\biggl(1-\frac{\bar t_{13}}{\Mt^2}\biggr) 
      \ln\biggl(1-\frac{\bar s_{35}}{\Mt^2}\biggr) 
\nn\\ && \qquad {}
  - \ln^2\biggl(1-\frac{\bar t_{24}}{\Mt^2}\biggr) 
  - \ln^2(x_{\MH^2})
     - \frac{\pi^2}{6} \Biggr], 
\nn\\[.5em]
\lefteqn{ D^{(D)}_0(p_3,p_3-p_1,-p_4,0,\Mt,\Mt,\Mt) =
\frac{1}{(t_{13}-\Mt^2)s_{34}\beta_{s_{34}}}
} \quad
\nn\\*
&& {} \times \Biggl[ \Delta^{\IR}_1(\mu)\ln(x_{s_{34}})
  - 2\sum_{\rho=\pm 1}\cLi(x_{s_{34}},x_{t_{25}}^{\,\rho}) 
  - \Li(x_{s_{34}}^2) - 2\ln(x_{s_{34}}) \ln(1-x_{s_{34}}^2) 
\nn\\ && \qquad {}
  - 2 \ln(x_{s_{34}}) \ln\biggl(1-\frac{\bar t_{13}}{\Mt^2}\biggr) 
  - \ln^2(x_{t_{25}}) + \frac{\pi^2}{6} \Biggr], 
\nn\\[.5em]
\lefteqn{ D^{(D)}_0(p_3,p_3+p_5,-p_4,0,\Mt,\Mt,\Mt) =
\frac{1}{(s_{35}-\Mt^2)s_{34}\beta_{s_{34}}}
} \quad
\nn\\*
&& {} \times \Biggl[ \Delta^{\IR}_1(\mu)\ln(x_{s_{34}})
  - \sum_{\rho,\,\sigma=\pm 1}
          \cLi(x_{s_{34}},x_{\hat{s}}^{\,\rho},x_{\MH^2}^{\sigma})
  - \Li(x_{s_{34}}^2) - 2\ln(x_{s_{34}}) \ln(1-x_{s_{34}}^2) 
\nn\\ && \qquad {}
  - 2 \ln(x_{s_{34}}) \ln\biggl(1-\frac{\bar s_{35}}{\Mt^2}\biggr) 
  - \ln^2(x_{\hat{s}}) - \ln^2(x_{\MH^2}) + \frac{\pi^2}{6} \Biggr].
\label{eq:D0dim}
\eeqar
The following specific combinations of logarithms and dilogarithms
were used,
\begin{eqnarray}
\cLi(x,y)   &=& \Li(1-xy)  + \ln(1-xy) \,\Bigl[ \ln(xy)-\ln(x)-\ln(y)
                                           \Bigr],
\nn\\[.5em]
\cLi(x,y,z) &=& \Li(1-xyz) + \ln(1-xyz)\,\Bigl[ \ln(xyz)-\ln(x)-\ln(y)
                                                 -\ln(z) \Bigr].
\end{eqnarray}
The expressions given for the $C_0$ and $D_0$ functions
are valid for arbitrary real values of the variables
$\hat{s},s_{ij}$, $t_{ij}$, and $\MH^2$, independent of their kinematical
range.

The method described in \refse{se:pentagons}~\refpar{par:Esing}
can also be used to calculate the singularities of the 4-point
functions $D_0$ in terms of 3-point functions $C_0$. 
The corresponding relations between $D_{0,\mathrm{sing}}$ and
the 3-point subintegrals are
\beqar
\lefteqn{ D_0(p_1,p_1-p_3,-p_2,0,0,\Mt,0)_{\mathrm{sing}}
= \frac{1}{t_{13}-\Mt^2}\,C_0(p_1,-p_2,0,0,0) } \quad  
\nn\\
&& {} + \frac{t_{13}-s_{45}}{\hat{s}(t_{13}-\Mt^2)}\,C_0(p_1-p_3,-p_2,0,\Mt,0) 
  + \frac{1}{\hat{s}}\,C_0(p_1,p_1-p_3,0,0,\Mt), 
\hspace{4em}
\nn\\[.5em]
\lefteqn{ D_0(p_1,p_1-p_3,p_4+p_5,0,0,\Mt,\Mt)_{\mathrm{sing}}
= \frac{1}{t_{23}-\Mt^2}\,C_0(p_1,p_1-p_3,0,0,\Mt) } \quad
\nn\\
&& {}  + \frac{t_{23}-s_{45}}{(t_{13}-\Mt^2)(t_{23}-\Mt^2)}\,C_0(p_1,p_4+p_5,0,0,\Mt), 
\nn\\[.5em]
\lefteqn{ D_0(p_1,p_1-p_3,p_4,0,0,\Mt,\Mt)_{\mathrm{sing}} 
= \frac{1}{t_{14}-\Mt^2}\,C_0(p_1,p_1-p_3,0,0,\Mt) } \quad
\nn\\
&& {}  + \frac{1}{t_{13}-\Mt^2}\,C_0(p_1,p_4,0,0,\Mt), 
\nn\\[.5em]
\lefteqn{ D_0(p_1,p_1-p_3,p_4-p_2,0,0,\Mt,\Mt)_{\mathrm{sing}} 
= \frac{1}{s_{35}-\Mt^2}\,C_0(p_1,p_1-p_3,0,0,\Mt) } \quad
\nn\\
&& {}  + \frac{s_{35}-t_{24}}{(t_{13}-\Mt^2)(s_{35}-\Mt^2)}
  \,C_0(p_1,p_4-p_2,0,0,\Mt), 
\nn\\[.5em]
\lefteqn{ D_0(p_3,p_3-p_1,-p_4,0,\Mt,\Mt,\Mt)_{\mathrm{sing}} 
= \frac{1}{t_{13}-\Mt^2}\,C_0(p_3,-p_4,0,\Mt,\Mt), } \quad
\nn\\[.5em]
\lefteqn{ D_0(p_3,p_3+p_5,-p_4,0,\Mt,\Mt,\Mt)_{\mathrm{sing}} 
= \frac{1}{s_{35}-\Mt^2}\,C_0(p_3,-p_4,0,\Mt,\Mt).} 
\eeqar
Obviously the explicit results for $D^{(D)}_0$ and $C^{(D)}_0$ given above
are compatible with these relations. Moreover, the functions
$D_{0,\mathrm{sing}}$ can be used to translate the $D$-dimensional
$D_0$ functions of Eq.~\refeq{eq:D0dim} into their mass-regularized
counterparts $D_0^{(\mathrm{mass},D=4)}$,
\beq
D_0^{(\mathrm{mass},D=4)} = 
D^{(\mathrm{mass},D=4)}_{0,\mathrm{sing}} 
+ \left[ D_0^{(D)} - D^{(D)}_{0,\mathrm{sing}} \right]+\dots,
\eeq
where the $C_0$ functions occurring in 
$D^{(D)}_{0,\mathrm{sing}}$ and $D^{(\mathrm{mass},D=4)}_{0,\mathrm{sing}}$ 
are given in Eqs.~\refeq{eq:C0dim} and \refeq{eq:C0mass}, respectively,
or result from those by mere substitutions. The dots represent infinitesimally
small terms proportional to the regulators $(\,D\!-\!4\,)$ or $\la$.

A simple example:
\beqar
\lefteqn{ D_0^{(\mathrm{mass},D=4)}(p_3,p_3-p_1,-p_4,\la,\Mt,\Mt,\Mt)
= \frac{1}{t_{13}-\Mt^2}\,
C_0^{(\mathrm{mass},D=4)}(p_3,-p_4,\la,\Mt,\Mt)} \quad
\nn\\[.5em]
&& \quad {} + D_0^{(D)}(p_3,p_3-p_1,-p_4,0,\Mt,\Mt,\Mt) 
- \frac{1}{t_{13}-\Mt^2}\,C_0^{(D)}(p_3,-p_4,0,\Mt,\Mt) +\dots
\nn\\[.5em]
&=& \frac{1}{(t_{13}-\Mt^2)s_{34}\beta_{s_{34}}}
\Biggl[ \ln\biggl(\frac{\la^2}{\Mt^2}\biggr)\ln(x_{s_{34}})
  - 2\sum_{\rho=\pm 1}\cLi(x_{s_{34}},x_{t_{25}}^{\,\rho}) 
  - \Li(x_{s_{34}}^2)  
\nn\\ && \qquad {}
  - 2\ln(x_{s_{34}}) \ln(1-x_{s_{34}}^2)
  - 2 \ln(x_{s_{34}}) \ln\biggl(1-\frac{\bar t_{13}}{\Mt^2}\biggr) 
  - \ln^2(x_{t_{25}}) + \frac{\pi^2}{6} \Biggr],
\eeqar
which agrees with the result in the literature 
\cite{Beenakker:1990jr}. So, by subtracting and adding well-defined singular 
integrals we are able to switch between arbitrary regularization schemes. 
For instance, to switch to an off-shell regularization scheme, 
with some of the external particles put slightly off-shell, it is 
sufficient to know the relevant 3-point integrals in that scheme.

\section{Colour operators and correlations}
\label{app:colour}

In order to make the results in this report as compact as possible, colour
indices have been suppressed in the expressions by making use of colour 
operators. The definitions of these colour operators are given in this
appendix (see also \citeres{Catani:1996jh,Catani:2002hc,Catani:2001ef}). 

Consider the matrix element for a process involving $n$ coloured external 
particles, $\M^{a_1\,\ldots\,a_n}$, where $\,a_1,\dots,a_n\,$ are the colour
indices of the coloured particles. These colour indices range from 
1 to 8 for gluons [adjoint representation] and from 1 to 3 
for (anti)quarks [fundamental representation]. Emission of a gluon with colour 
index $c$ from the external particle $j$ is represented by a specific colour 
operator (charge) ${\bf T}_j$ acting on the matrix element:
\beq
  (\M\otimes {\bf T}_j)^{c;\,a_1\,\ldots\,a_n} 
  = 
  \sum_{b_1,\,\ldots\,,\,b_n}\delta_{a_1b_1}\cdots \,(T_j^{\,c})_{a_jb_j}
  \cdots\,\delta_{a_nb_n}\,\M^{b_1\,\ldots\,b_n}
  \ =\
  \sum_{b_j}\,(T_j^{\,c})_{a_jb_j}\,\M^{a_1\,\ldots\,b_j\,\ldots\,a_n},
\eeq
with $(T_j^{\,c})_{ab} = {}-if_{cab}\,$ for incoming/outgoing gluons,
$(T_j^{\,c})_{ab} = \lambda^c_{ab}/2\,$ for outgoing quarks or incoming
antiquarks, and $(T_j^{\,c})_{ab} = {}-\lambda^c_{ba}/2\,$ for outgoing 
antiquarks or incoming quarks. Owing to $SU(3)$ gauge invariance, the complete 
matrix element $\M$ is a colour singlet. This leads to the colour-conservation 
identity
\beq
  \sum_{j=1}^n (\M\otimes {\bf T}_j)^{c;\,a_1\,\ldots\,a_n} = 0.
\eeq

In the NLO calculations the colour operators enter as products 
$({\bf T}_j\cdot {\bf T}_k)=({\bf T}_k\cdot {\bf T}_j)$, defined according to
\beqar
  && [\M\otimes ({\bf T}_j\cdot {\bf T}_k)]^{a_1\,\ldots\,a_n} = 
  \sum_{b_j,\,b_k}\,\sum_c \,(T_j^{\,c})_{a_jb_j}\,(T_k^{\,c})_{a_kb_k}\,
  \M^{a_1\,\ldots\,b_j\,\ldots\,b_k\,\ldots\,a_n} \qquad \mbox{if \ $j\neq k$},
\nn\\
  && [\M\otimes {\bf T}_j^2\,]^{a_1\,\ldots\,a_n} =
  \sum_{b_j,\,c_j}\,\sum_{c} \,(T_j^{\,c})_{a_jc_j}\,(T_j^{\,c})_{c_jb_j}\,
  \M^{a_1\,\ldots\,b_j\,\ldots\,a_n} 
  \ =\ C_j\,\M^{a_1\,\ldots\,a_n}.
\eeqar
Here $C_j$ is the Casimir operator in the representation of particle $j$,
\ie ${\bf T}_j^2 = C_{\mathrm{A}} = 3\,$ for gluons [adjoint 
representation] and ${\bf T}_j^2 = C_{\mathrm{F}} = 4/3\,$ for 
(anti)quarks [fundamental representation].
At the cross-section level this gives rise to colour correlations of the form 
\beq
  \sum_{\mathrm{colours}}|\M|^2\otimes ({\bf T}_j\cdot {\bf T}_k) = 
  \sum_{a_1,\,\ldots\,,\,a_n} \Bigl[ \M^{a_1\,\ldots\,a_n} \Bigr]^*
  \,\sum_{b_j,\,b_k}\,\sum_c\,(T_j^{\,c})_{a_jb_j}
  \,(T_k^{\,c})_{a_kb_k}\,\M^{a_1\,\ldots\,b_j\,\ldots\,b_k\,\ldots\,a_n}
\label{eq:correlation}
\eeq  
for $j\neq k$. Needless to say that the left-hand side of 
Eq.~\refeq{eq:correlation} looks much more appealing from the point of view of 
compact expressions.

Finally, we give the explicit results for the colour correlations
in terms of the colour operators $\C^{q\bar q}$ and $\C^{gg}_m$ defined
in Eqs.~\refeq{eq:Cqq} and \refeq{eq:Cgg}, respectively. Since
these $\C$ operators represent a basis for the relevant colour structures
of the $q\bar q$ and $gg$ channels, the correlations 
$\C\otimes({\bf T}_j\cdot {\bf T}_k)$ can again be expanded in terms of 
$\C$'s:
\beq
\C^{q\bar q} \otimes ({\bf T}_j\cdot {\bf T}_k) \ =\  
\xi^{q\bar q}_{jk} \, \C^{q\bar q} + \dots,
\qquad
\C^{gg}_m \otimes ({\bf T}_j\cdot {\bf T}_k) \ =\ 
\sum_{n=1}^3 \xi^{gg,mn}_{jk} \, \C^{gg}_n,
\eeq
where the dots stand for (irrelevant) operators orthogonal to $\C^{q\bar q}$.
The coefficients $\xi$ follow from simple colour algebra. For the
$q\bar q$ channel they read
\beq
\xi^{q\bar q}_{12} = \xi^{q\bar q}_{34} = \frac{1}{6}, \quad
\xi^{q\bar q}_{13} = \xi^{q\bar q}_{24} = -\frac{7}{6}, \quad
\xi^{q\bar q}_{14} = \xi^{q\bar q}_{23} = -\frac{1}{3}.
\eeq
For $gg$ fusion it is convenient to give them in matrix form,
\beqar
\left(\xi^{gg,mn}_{12}\right) & \ = \ &
\mathrm{diag}\left(-3,-\frac{3}{2},-\frac{3}{2}\right), \qquad
\left(\xi^{gg,mn}_{34}\right) \ = \ 
\mathrm{diag}\left(-\frac{4}{3},\frac{1}{6},\frac{1}{6}\right), 
\nn\\[.5em]
\left(\xi^{gg,mn}_{13}\right) & \ = \ &
\left(\xi^{gg,mn}_{24}\right) \ = \ \pmatrix{
\vphantom{{X^X}^X} 0 & -\frac{1}{2} & 0 \cr
\vphantom{{X^X}^X} -1 & -\frac{3}{4} & -\frac{3}{4} \cr
\vphantom{{X^X}^X} 0 & -\frac{5}{12} & -\frac{3}{4} },
\nn\\[.5em]
\left(\xi^{gg,mn}_{14}\right) & \ = \ &
\left(\xi^{gg,mn}_{23}\right) \ = \ \pmatrix{
\vphantom{{X^X}^X} 0 & \frac{1}{2} & 0 \cr
\vphantom{{X^X}^X} 1 & -\frac{3}{4} & \frac{3}{4} \cr
\vphantom{{X^X}^X} 0 & \frac{5}{12} & -\frac{3}{4} }.
\eeqar
These $\xi$ coefficients have also been used in
\refse{se:UVIRdivs} in rewriting Eqs.~\refeq{eq:Mdivqq} and \refeq{eq:Mdivgg}
in terms of Eq.~\refeq{eq:Mdiv}.

\section{The Altarelli--Parisi splitting functions at NLO}
\label{app:splitting}

The so-called Altarelli--Parisi splitting functions $P^{ab}(x)$ 
\cite{Altarelli:1977zs} are a measure of the probability of finding the 
parton $b$ with fraction $x$ of longitudinal momentum inside the 
parent parton $a$.%
\footnote{We have adopted the notation of \citere{Catani:1996jh}, which
          may differ from the conventions elsewhere in the literature
          by the interchange of the parton labels $a$ and $b$.}
At NLO these splitting functions can be written in terms of a regularized 
splitting function and an IR-sensitive part according to
\beq
  P^{ab}(x) = P_{\mathrm{reg}}^{ab}(x)
     + \delta^{ab} \biggl[ 2\,{\bf T}_a^2\,\left(\frac{1}{1-x}\right)_+
                           + \gamma_a\,\delta(1-x) \biggr].
\label{eq:Pab}
\eeq
The anomalous dimensions $\gamma_a$ are given by
\beq
  \gamma_q = \gamma_{\bar q} = \frac{3}{2}\,C_{\mathrm{F}} = 2 \qquad,\qquad
  \gamma_g = \frac{11}{6}\,C_{\mathrm{A}} - \frac{2}{3}\,T_{\mathrm{R}}N_f
           = \frac{33-2N_f}{6},
\eeq
using the customary normalization $T_{\mathrm{R}}=1/2$.
The `$+$' distribution $(\dots)_+$ is defined in the usual way:
\beq
  \int_0^1\rd x\,\Big[f(x)\Big]_+ g(x) =
  \int_0^1\rd x\,f(x)\,\left[g(x)-g(1)\right]\:.
\eeq
The regularized splitting functions are given by
\beqar
  P_{\mathrm{reg}}^{q\bar q}(x) &=& P_{\mathrm{reg}}^{\bar qq}(x) 
                               \ =\ 0, \nn\\[3mm]
  P_{\mathrm{reg}}^{qq}(x)      &=& P_{\mathrm{reg}}^{\bar q\bar q}(x) 
                               \ =\ {}-C_{\mathrm{F}}\,(1+x), \nn\\[1mm]
  P_{\mathrm{reg}}^{qg}(x)      &=& P_{\mathrm{reg}}^{\bar qg}(x) 
                               \ =\ C_{\mathrm{F}}\,\frac{1+(1-x)^2}{x}, 
                               \nn\\[1mm]
  P_{\mathrm{reg}}^{gq}(x)      &=& P_{\mathrm{reg}}^{g\bar q}(x) 
                               \ =\ T_{\mathrm{R}}\,\Bigl[ x^2+(1-x)^2 \Bigr],
                               \nn\\[1mm]
  P_{\mathrm{reg}}^{gg}(x)      &=& 2\,C_{\mathrm{A}}\,\biggl[ \frac{1-x}{x}
                                    - 1 + x(1-x) \biggr].
\label{eq:Pregab}
\eeqar

If the Altarelli--Parisi splitting functions were calculated in
$D=4-2\eps$ dimensions by means of dimensional regularization, a term
${}-\eps\,\hat{P}'_{ab}(x)$ would have to be added to
Eq.~\refeq{eq:Pab}. Such a term enters the final NLO expressions
through the interference with the $1/\eps$ collinear poles. The
associated splitting functions $\hat{P}'_{ab}(x)$ amount to
\beqar
  \hat{P}'_{q\bar q}(x) &=& \hat{P}'_{\bar qq}(x) \ =\ \hat{P}'_{gg}(x) 
                            \ =\ 0, \nn\\[1mm]
  \hat{P}'_{qq}(x)      &=& \hat{P}'_{\bar q\bar q}(x) \ =\ \hat{P}'_{qg}(1-x) 
                            \ =\ \hat{P}'_{\bar qg}(1-x)
                            \ =\ C_{\mathrm{F}}\,(1-x), \nn\\[1mm] 
  \hat{P}'_{gq}(x)      &=& \hat{P}'_{g\bar q}(x) 
                            \ =\ 2\,T_{\mathrm{R}}\,x(1-x).
\label{eq:Vab}
\eeqar
Among other things, these splitting functions take into account the difference
between the number of spin degrees of freedom of the quarks ($2$) and
gluons ($D-2$) in $D$ dimensions.


\vspace*{1cm}

\begin{thebibliography}{99}
\frenchspacing

\bibitem{Higgs:1964ia}
P.~W.~Higgs,
Phys.\ Lett.\  {\bf 12} (1964) 132;
Phys.\ Rev.\ Lett.\  {\bf 13} (1964) 508
and Phys.\ Rev.\  {\bf 145} (1966) 1156; \\
F.~Englert and R.~Brout,
Phys.\ Rev.\ Lett.\  {\bf 13} (1964) 321; \\
G.~S.~Guralnik, C.~R.~Hagen and T.~W.~Kibble,
Phys.\ Rev.\ Lett.\  {\bf 13} (1964) 585; \\
T.~W.~Kibble,
Phys.\ Rev.\  {\bf 155} (1967) 1554.

\bibitem{'tHooft:rn}
G.~'t Hooft,
Nucl.\ Phys.\ B {\bf 35} (1971) 167; \\
G.~'t Hooft and M.~J.~Veltman,
Nucl.\ Phys.\ B {\bf 44} (1972) 189.

\bibitem{Carena:2000yx}
Report of the Tevatron Higgs working group,
M.~Carena, J.~S.~Conway, H.~E.~Haber, J.~D.~Hobbs  {\it et al.},
hep-ph/0010338.

\bibitem{atlas_cms_tdrs}
ATLAS Collaboration, Technical Design Report, Vols. 1 and 2, CERN--LHCC--99--14
and CERN--LHCC--99--15; \\
CMS Collaboration, Technical Proposal, CERN--LHCC--94--38.

\bibitem{Carena:2002rm}
M.~Carena, D.~W.~Gerdes, H.~E.~Haber, A.~S.~Turcot and P.~M.~Zerwas,
in {\it Proc. of the APS/DPF/DPB Summer Study on the Future 
of Particle Physics (Snowmass 2001) }, ed. R.~Davidson and C.~Quigg,
hep-ph/0203229.

\bibitem{Cabibbo:1979ay}
N.~Cabibbo, L.~Maiani, G.~Parisi and R.~Petronzio,
Nucl.\ Phys.\ B {\bf 158} (1979) 295;\\
M.~Sher,
Phys.\ Rept.\  {\bf 179} (1989) 273;\\
T.~Hambye and K.~Riesselmann,
Phys.\ Rev.\ D {\bf 55} (1997) 7255.

\bibitem{Abbaneo:2001ix}
The LEP Electroweak Working Group and the SLD Heavy Flavor and 
Electroweak Groups, D.~Abbaneo {\it et al.},
hep-ex/0112021.

\bibitem{:2001xw}
The LEP Working Group for Higgs Boson Searches,
LHWG Note/2002-01.

\bibitem{Kunszt:1984ri}
Z.~Kunszt,
Nucl.\ Phys.\ B {\bf 247} (1984) 339; \\
W.~J.~Marciano and F.~E.~Paige,
Phys.\ Rev.\ Lett.\  {\bf 66} (1991) 2433; \\
J.~F.~Gunion,
Phys.\ Lett.\ B {\bf 261} (1991) 510.

\bibitem{Beenakker:2001rj}
W.~Beenakker, S.~Dittmaier, M.~Kr\"amer, B.~Pl\"umper, M.~Spira and 
P.~M.~Zerwas,
Phys.\ Rev.\ Lett.\  {\bf 87} (2001) 201805
[hep-ph/0107081].

\bibitem{Reina:2001sf}
L.~Reina and S.~Dawson,
Phys.\ Rev.\ Lett.\  {\bf 87} (2001) 201804
[hep-ph/0107101]; \\
L.~Reina, S.~Dawson and D.~Wackeroth,
Phys.\ Rev.\ D {\bf 65} (2002) 053017
[hep-ph/0109066].

\bibitem{Goldstein:2000bp}
J.~Goldstein, C.~S.~Hill, J.~Incandela, S.~Parke, D.~Rainwater and D.~Stuart,
Phys.\ Rev.\ Lett.\  {\bf 86} (2001) 1694
[hep-ph/0006311].

\bibitem{Richter-Was:sa}
E.~Richter-Was and M.~Sapinski,
Acta Phys.\ Polon.\ B {\bf 30} (1999) 1001.

\bibitem{Drollinger:2001ym}
V.~Drollinger, T.~M\"uller and D.~Denegri,
hep-ph/0111312.

\bibitem{Maltoni:2002jr}
F.~Maltoni, D.~Rainwater and S.~Willenbrock,
Phys.\ Rev.\ D {\bf 66} (2002) 034022
[hep-ph/0202205].

\bibitem{Belyaev:2002ua}
A.~Belyaev and L.~Reina,
JHEP {\bf 0208} (2002) 041
[hep-ph/0205270].

\bibitem{Djouadi:1992tk}
A.~Djouadi, J.~Kalinowski and P.~M.~Zerwas,
Z.\ Phys.\ C {\bf 54} (1992) 255.

\bibitem{Dittmaier:1998dz}
S.~Dittmaier, M.~Kr\"amer, Y.~Liao, M.~Spira and P.~M.~Zerwas,
Phys.\ Lett.\ B {\bf 441} (1998) 383 
[hep-ph/9808433].

\bibitem{Dawson:1999ej}
S.~Dawson and L.~Reina,
Phys.\ Rev.\ D {\bf 59} (1999) 054012 
[hep-ph/9808443].

\bibitem{Abe:2001np}
T.~Abe {\it et al.},
in {\it Proc. of the APS/DPF/DPB Summer Study on the Future of Particle Physics (Snowmass 2001) }, ed. R.~Davidson and C.~Quigg,
hep-ex/0106056.

\bibitem{Juste:1999af}
A.~Juste and G.~Merino,
hep-ph/9910301; \\
A.~Gay, talk presented at the second workshop of the extended
ECFA/DESY study "Physics and Detectors for a 90 to 800 GeV Linear
Collider", April 12-15, 2002, Saint Malo, France.

\bibitem{Baer:1999ge}
H.~Baer, S.~Dawson and L.~Reina,
Phys.\ Rev.\ D {\bf 61} (2000) 013002.

\bibitem{Catani:1996jh}
S.~Catani and M.~H.~Seymour,
Phys.\ Lett.\ B {\bf 378} (1996) 287
[hep-ph/9602277] and
Nucl.\ Phys.\ B {\bf 485} (1997) 291
[Erratum-ibid.\ B {\bf 510} (1997) 291]
[hep-ph/9605323].

\bibitem{Catani:2002hc}
S.~Catani, S.~Dittmaier, M.~H.~Seymour and Z.~Tr\'ocs\'anyi,
Nucl.\ Phys.\ B {\bf 627} (2002) 189
[hep-ph/0201036].

\bibitem{Dittmaier:1999nn}
S.~Dittmaier,
Phys.\ Rev.\ D {\bf 59} (1999) 016007
[hep-ph/9805445].

\bibitem{Kublbeck:1990xc}
J.~K\"ublbeck, M.~B\"ohm and A.~Denner,
Comput.\ Phys.\ Commun.\  {\bf 60} (1990) 165; \\
H.~Eck and J.~K\"ublbeck,
{\it Guide to FeynArts 1.0}, Univ.\ W\"urzburg, 1992.

\bibitem{'tHooft:1979xw}
G.~'t Hooft and M.~Veltman,
Nucl.\ Phys.\ B {\bf 153} (1979) 365.

\bibitem{Passarino:1979jh}
G.~Passarino and M.~Veltman,
Nucl.\ Phys.\ B {\bf 160} (1979) 151.

\bibitem{Beenakker:1990jr}
W.~Beenakker and A.~Denner,
Nucl.\ Phys.\ B {\bf 338} (1990) 349. 

\bibitem{Denner:1991qq}
A.~Denner, U.~Nierste and R.~Scharf,
Nucl.\ Phys.\ B {\bf 367} (1991) 637.

\bibitem{Denner:1993kt}
A.~Denner,
Fortsch.\ Phys.\  {\bf 41} (1993) 307.

\bibitem{Vermaseren:2000nd}
J.~A.~Vermaseren,
math-ph/0010025.

\bibitem{Melrose:1965kb}
D.~B.~Melrose,
Nuovo Cim.\  {\bf 40} (1965) 181; \\
W.~L.~van Neerven and J.~A.~Vermaseren,
Phys.\ Lett.\ B {\bf 137} (1984) 241.



\bibitem{Bern:1992em}
Z.~Bern, L.~J.~Dixon and D.~A.~Kosower,
Phys.\ Lett.\ B {\bf 302} (1993) 299
[Erratum-ibid.\ B {\bf 318} (1993) 649]
[hep-ph/9212308] and
%
Nucl.\ Phys.\ B {\bf 412} (1994) 751
[hep-ph/9306240];\\
%
J.~Fleischer, F.~Jegerlehner and O.~V.~Tarasov,
Nucl.\ Phys.\ B {\bf 566} (2000) 423
[hep-ph/9907327].

\bibitem{Etensor}
A.~Denner and S.~Dittmaier,
MPI-PhT/2002-63.

\bibitem{Catani:2001ef}
S.~Catani, S.~Dittmaier and Z.~Tr\'ocs\'anyi, 
Phys.\ Lett.\ B {\bf 500} (2001) 149
[hep-ph/0011222].

\bibitem{Stelzer:tk}
T.~Stelzer and W.~F.~Long,
Nucl.\ Phys.\ Proc.\ Suppl.\  {\bf 37B} (1994) 158.

\bibitem{Murayama:1992gi}
H.~Murayama, I.~Watanabe and K.~Hagiwara,
KEK-91-11.

\bibitem{Harris:2001sx}
B.~W.~Harris and J.~F.~Owens,
Phys.\ Rev.\ D {\bf 65} (2002) 094032
[hep-ph/0102128].

\bibitem{Beenakker:1989bq}
W.~Beenakker, H.~Kuijf, W.~L.~van Neerven and J.~Smith,
Phys.\ Rev.\ D {\bf 40} (1989) 54.

\bibitem{Dittmaier:1994bj}
S.~Dittmaier,
Nucl.\ Phys.\ B {\bf 423} (1994) 384
[hep-ph/9311363].

\bibitem{Baur:1999kt}
U.~Baur, S.~Keller and D.~Wackeroth,
Phys.\ Rev.\ D {\bf 59} (1999) 013002
[hep-ph/9807417].

\bibitem{Dittmaier:2001ay}
S.~Dittmaier and M.~Kr\"amer,
Phys.\ Rev.\ D {\bf 65} (2002) 073007
[hep-ph/0109062].

\bibitem{Sommerfeld}
A.~Sommerfeld, in ``Atombau und Spektrallinien'', Vol.~II, p.~457
(F.~Vieweg und Sohn, Braunschweig 1939).

\bibitem{Nason:1988xz}
P.~Nason, S.~Dawson and R.~K.~Ellis,
Nucl.\ Phys.\ B {\bf 303} (1988) 607,
Nucl.\ Phys.\ B {\bf 327} (1989) 49; \\
W.~Beenakker, W.~L.~van Neerven, R.~Meng, G.~A.~Schuler and J.~Smith,
Nucl.\ Phys.\ B {\bf 351} (1991) 507.

\bibitem{Jersak:sp}
J.~Jersak, E.~Laermann and P.~M.~Zerwas,
Phys.\ Rev.\ D {\bf 25} (1982) 1218
[Erratum-ibid.\ D {\bf 36} (1987) 310].

\bibitem{Martin:2001es}
A.~D.~Martin, R.~G.~Roberts, W.~J.~Stirling and R.~S.~Thorne,
Eur.\ Phys.\ J.\ C {\bf 23} (2002) 73
[hep-ph/0110215],  
Phys.\ Lett.\ B {\bf 531} (2002) 216
[hep-ph/0201127], and 
hep-ph/0211080.

\bibitem{Dawson:1998im}
S.~Dawson and L.~Reina,
Phys.\ Rev.\ D {\bf 57} (1998) 5851
[hep-ph/9712400].

\bibitem{Pumplin:2002vw}
J.~Pumplin, D.~R.~Stump, J.~Huston, H.~L.~Lai, P.~Nadolsky and W.~K.~Tung,
JHEP {\bf 0207} (2002) 012
[hep-ph/0201195].

\bibitem{Altarelli:1977zs}
G.~Altarelli and G.~Parisi,
Nucl.\ Phys.\ B {\bf 126} (1977) 298.

\end{thebibliography}
\end{document}